\documentclass[a4paper,11pt]{article}
\usepackage{amsfonts}
\usepackage{latexsym}
\usepackage{amsmath}
\usepackage{amssymb}
\usepackage{amssymb}
\usepackage{slashed}
\usepackage{upgreek}
\usepackage{mathrsfs}
\usepackage{bbm}

\usepackage{amsthm}

\theoremstyle{remark}

\usepackage[utf8]{inputenc}
\usepackage[english]{babel}

\theoremstyle{definition}

\allowdisplaybreaks

\usepackage{relsize}
\usepackage{graphicx}
\usepackage{comment}

\renewcommand{\thefootnote}{\fnsymbol{footnote}}

\def\appendix#1{\addtocounter{section}{1}\setcounter{equation}{0}
\renewcommand{\thesection}{\Alph{section}}
\section*{Appendix \thesection\protect\indent \parbox[t]{11.15cm}{#1}}
\addcontentsline{toc}{section}{Appendix \thesection\ \ \ #1}}

\font\mybbbb=msbm10 at 9pt
\font\mybb=msbm10 at 11pt

\def\bb#1{\hbox{\mybb#1}}

\def\bbbb#1{\hbox{\mybbbb#1}}
\def\bZ {\bb{Z}}
\def\bbZ {\bbbb{Z}}
\def\bR {\bb{R}}

\def\bC {\bb{C}}
\def\bN {\bb{N}}

\def\RP{\bb{R}\mathrm{P}}
\def\bRP{\bbbb{R}\mathrm{P}}
\def\CP{\bb{C}\mathrm{P}}

\def\ma{\mathrm{a}}
\def\mb{\mathrm{b}}
\def\mc{\mathrm{c}}
\def\md{\mathrm{d}}

\def\h{\widehat}
\def\be{\begin{equation}}
\def\ee{\end{equation}}

\def\ii{\iota}
\def\Adj{\mathrm{Ad}}
\def\asd{\mathrm{asd}}
\def\HE{\mathrm{HE}}

\def\rsq {]\kern -2.0pt]}
\def\lsq {[\kern -2.0pt[}

\def\op {{\mathcal O}}

\newcommand{\bea}{\begin{eqnarray}}
\newcommand{\eea}{\end{eqnarray}}

\usepackage[usenames,dvipsnames,svgnames,table]{xcolor}	
\usepackage[normalem]{ulem}				
\usepackage{microtype}
\usepackage[colorlinks, citecolor=blue, linkcolor=blue, urlcolor=Maroon, filecolor=Maroon, linktocpage=true]{hyperref} 

\usepackage{chngcntr}
\counterwithout{equation}{section}

\begin{document}

\begin{center}
\vspace*{-1.0cm}
\begin{flushright}
\end{flushright}


\vspace{2.0cm} {\Large \bf Geometry and symmetries  of Hermitian-Einstein and instanton connection  moduli spaces} \\[.2cm]

\vskip 2cm
 Georgios  Papadopoulos
\\
\vskip .6cm


\begin{small}
\textit{Department of Mathematics
\\
King's College London
\\
Strand
\\
 London WC2R 2LS, UK}\\
\texttt{george.papadopoulos@kcl.ac.uk}
\end{small}
\\*[.6cm]

\end{center}

\vskip 2.5 cm

\begin{abstract}
\noindent

We investigate the geometry of the moduli spaces $\mathscr{M}_{\HE}^*(M^{2n})$ of Hermitian-Einstein irreducible connections on a vector bundle $E$  over a K\"ahler with torsion (KT) manifold $M^{2n}$  that admits  holomorphic  and $\h\nabla$-covariantly constant vector fields, where $\h\nabla$ is the connection with skew-symmetric torsion $H$.  We demonstrate that such vector fields induce an action on $\mathscr{M}_{\HE}^*(M^{2n})$ that leaves  both the metric and complex structure invariant. Moreover, if an additional condition is satisfied, the induced vector fields  are covariantly constant with respect to the connection with skew-symmetric torsion $\h{\mathcal{ D}}$ on $\mathscr{M}_{\HE}^*(M^{2n})$.  We demonstrate that in the presence of such vector fields, the geometry of $\mathscr{M}_{\HE}^*(M^{2n})$ can be modelled on that of holomorphic toric principal bundles with base space KT manifolds and give some examples.

We also extend our analysis to the moduli spaces $\mathscr{M}_{\asd}^*(M^{4})$ of instanton connections on vector bundles over  KT, bi-KT (generalised K\"ahler) and hyper-K\"ahler  with torsion (HKT) manifolds $M^4$.
 We find that the geometry of $\mathscr{M}_{\asd}^*(S^3\times S^1)$ can be modelled on that of  principal bundles with fibre $S^3\times S^1$ over  Quaternionic K\"ahler manifolds with torsion (QKT).   In addition motivated by applications to AdS/CFT, we explore the (superconformal) symmetry algebras of two-dimensional sigma models with target spaces such moduli spaces.

\end{abstract}



\newpage


\numberwithin{equation}{section}

\renewcommand{\thefootnote}{\arabic{footnote}}



\section{Introduction}\label{intro}

\underline{{\it Prologue}:}  Instantons and their moduli spaces, in addition to their  applications in  quantum field theory and differential geometry, have also been used   in the investigation of AdS/CFT correspondence of Maldacena \cite{maldacena}. Original examples include type IIB string theory on AdS$_3\times S^3\times T^4$ and AdS$_3\times S^3\times K_3$ that  are believed to be dual to the superconformal two-dimensional sigma model with target space the moduli space of instantons on $T^4$ and $K_3$, respectively.
A much explored \cite{boonstra, elitzur, deboer, gukov, eberhardt, tong, eberhardt2, eberhardt3}  extension of these dualities  is type IIB string theory on $AdS_3\times S^3\times S^1$ that is expected to be dual to a two-dimensional field theory exhibiting the large $N=4$ superconformal algebra of \cite{sevrin, schoutens} as a symmetry.

Witten  \cite{witten} has demonstrated that the two-dimensional sigma model with target space the (smooth part of the) moduli space of anti-self-dual instanton connections  over $S^3\times S^1$, $\mathscr{M}_{\asd}(S^3\times S^1)$, is invariant under (two copies of) the large $N=4$ superconformal algebra and so provided evidence for the above AdS/CFT duality.  The proof begins with the observation that $S^3\times S^1$ is a bi-HKT manifold  and as a result $\mathscr{M}_{\asd}(S^3\times S^1)$ admits a bi-HKT structure as well -- in fact $S^3\times S^1$ admits four HKT structures \cite{pw} but only two are compatible with the instanton anti-self-duality condition. Therefore, the sigma model exhibits the required number of supersymmetries as expected from the $N=4$ large superconformal algebra.  Another ingredient, as prescribed by the large $N=4$ superconformal algebra, is to demonstrate that the sigma models with target space $\mathscr{M}_{\asd}(S^3\times S^1)$ admit   $SO(4)\times U(1)$ as a symmetry group. Remarkably, this symmetry is constructed in part\footnote{The remaining three generators of the $SO(4)\times U(1)$ symmetry are associated with an R-type of transformation that  rotates the fermionic fields of the sigma model using the three complex structures of one of the two HKT structures of $\mathscr{M}_{\asd}(S^3\times S^1)$. These symmetries are not related to the isometries of $S^3\times S^1$.} by the induced action on  $\mathscr{M}_{\asd}(S^3\times S^1)$ of  left-invariant vector fields of $S^3\times S^1$, viewed as the group manifold $SU(2)\times U(1)$.    This group   acts on the complex structures  of the bi-HKT geometry  of $\mathscr{M}_{\asd}(S^3\times S^1)$, which generate the supersymmetry transformations of the sigma model, as expected.  These results, together with some key additional arguments to establish the conformal invariance at the quantum level, confirm that the two-dimensional sigma model with target space $\mathscr{M}_{\asd}(S^3\times S^1)$ exhibits all the required symmetries expected from the AdS/CFT duality.  In the same context, the moduli spaces $\mathscr{M}_{\asd}(S^3/\bZ_n\times S^1)$, $n\in\bN$,  were also investigated.

 These developments give a new impetus  to systematically   investigate the geometry and   symmetries induced on $\mathscr{M}_{\asd}(M^{4})$ from those of the underlying manifold $M^4$. These questions can be explored in the  setting of the moduli space\footnote{In what follows, we shall refer to $\mathscr{M}^*_{\HE}(M^{2n})$ as the moduli space of Hermitian-Einstein connections (over $M^{2n}$) to simplify terminology.  Furthermore, we shall assume throughout that these spaces are not empty.}, $\mathscr{M}^*_{\HE}(M^{2n})$, of irreducible Hermitian-Einstein connections on a vector bundle $E$  over a Hermitian manifold $M^{2n}$. Then, the results can be adapted to the anti-self-dual  instanton  moduli spaces.

 A Hermitian-Einstein connection  $D_A$ on a complex vector bundle $E$ over a Hermitian manifold $M^{2n}$ with metric $g$ and complex structure $I$  preserves a fibre (Hermitian) metric\footnote{Therefore, the holonomy of $D_A$ is included in $U(r)$, where $r$ is the rank of $E$.}, $h$, of $E$, $D_A h=0$, and its curvature, $F=[D_A, D_A]$, satisfies the conditions \cite{kobayashi}
\be
F^{2,0}=F^{0,2}=0~,~~~~\omega\llcorner F= -i\, \Lambda {\bf 1}_E~,
\ee
where $\Lambda$ is a constant, $\omega(X,Y)=g(X, IY)$ is the Hermitian form of $M^{2n}$ and $\llcorner$ denotes the $\omega$-trace of $F$, see also (\ref{HEcond}). Therefore, in the decomposition of forms on $M^{2n}$ along holomorphic and anti-holomorphic directions with respect to $I$, $F$ is a  $(1,1)$-form, as the $(2,0)$ and $(0,2)$ components vanish, and in addition the $(1,1)$ component obeys a $\omega$-trace condition. The existence of Hermitian-Einstein connections on compact K\"ahler and Hermitian manifolds has  extensively been investigated in \cite{Donaldson, Uhlenbeck, LiYau, Li, lubke} and a brief non-technical  outline is given in section \ref{sec:HE}.
Moreover, it has been demonstrated by L\"ubke and Teleman \cite{lubke} that the geometry of  the moduli spaces, $\mathscr{M}^*_{\HE}(M^{2n})$, of  Hermitian-Einstein connections on a vector bundle $E$ over a  compact Hermitian manifold, $M^{2n}$, is  also a Hermitian manifold with a $\partial\bar\partial$-closed Hermitian form $\Omega$, $\partial\bar\partial \Omega=0$. Equivalently, in a somewhat different terminology, $\mathscr{M}^*_{\HE}(M^{2n})$ admits a strong\footnote{The term strong refers to the condition $\partial\bar\partial \Omega=0$, which is equivalent to the closure condition of the torsion.} K\"ahler structure with torsion (KT).  Both the metric and Hermitian form on $\mathscr{M}^*_{\HE}(M^{2n})$ are induced from those\footnote{For the proof the Gauduchon metric is used on $M^{2n}$ for which the associated Lee form $\theta_i=D^k\omega_{kj} I^j{}_i$ is co-closed, $D^i \theta_i=0$, see also definition (\ref{leef}).  This is not a restriction on the Hermitian geometry of $M^{2n}$ as every Hermitian metric admits a Gauduchon representative in its conformal class.} on $M^{2n}$.  This is a generalisation of the previously known result that if $M^4$ is a K\"ahler manifold, then  $\mathscr{M}^*_{\HE}(M^{2n})$ has a K\"ahler structure as well.  Other moduli spaces that exhibit similar geometric structures include those of the heterotic vacua investigated in \cite{mcorist1, mcorist2, mcorist3}.

The above results on Hermitian-Einstein connections can be adapted to the moduli spaces, $\mathscr{M}^*_{\asd}(M^{4})$, of anti-self-dual instanton connections over a KT manifold $M^4$.  In particular, if $M^4$ is a KT manifold, then  $\mathscr{M}^*_{\asd}(M^{4})$ admits a strong KT as well. For instantons, this result has been extended in several directions. Hitchin \cite{hitchin} has demonstrated that if $M^4$ is a compact generalised K\"ahler manifold\footnote{These manifolds admit a pair of KT structures with the same Hermitian metric and torsion 3-form, i.e. bi-KT. The  two associated complex structures are covariantly constant with the connections $\h\nabla$ and $\breve \nabla$ that have torsion $H$ and $-H$, respectively. The complex structures may not commute.}  \cite{gualtieri}, i.e.  strong bi-KT, then $\mathscr{M}^*_{\asd}(M^{4})$ is also a generalised K\"ahler manifold.  Furthermore, Moraru and Verbitsky \cite{moraru} have shown that if  $M^4$ is  a hyper-K\"ahler manifold with torsion (HKT) or bi-HKT, then  $\mathscr{M}^*_{\asd}(M^{4})$ also admits a strong HKT or bi-HKT structure, respectively.

\vskip 0.2cm

\underline{{\it Description of the results}:}
One of the objectives of this article is to provide a systematic investigation of the conditions required for a symmetry of $M^{2n}$ to induce a symmetry of the geometric structure on $\mathscr{M}^*_{\HE}(M^{2n})$. Suppose that $M^{2n}$ is a KT manifold with metric $g$, complex structure $I$, Hermitian form $\omega$, and connection $\h\nabla$ with skew-symmetric torsion $H$; see section \ref{sec:kt} for more details. We show that if $X$ is a holomorphic and $\h\nabla$-covariantly constant\footnote{This assumption can be weaken somewhat and replaced with the requirement that $X$ is Killing. But for the purpose of this article, the stronger condition that $X$ is $\h\nabla$-covariantly constant will suffice. Holomorphic $\h\nabla$-covariantly constant vector fields have appeared before  in the classification of heterotic backgrounds  \cite{hetback} and more recently \cite{streets2} in generalised geometry.} vector field\footnote{If $X$ is $\h\nabla$-covariantly constant vector field, then $X$ is Killing. For Killing vector fields, $\mathcal{L}_X\omega=0$ implies that $X$ is holomorphic, $\mathcal{L}_X I=0$.}  on $M^{2n}$,
\be
\mathcal{L}_X\omega=0~,~~~\h\nabla X=0~,
\ee
then $X$ induces a vector field $\alpha_{{}_X}$  on $\mathscr{M}^*_{\HE}(M^{2n})$ that preserves both the metric $\mathcal{G}$ and the Hermitian form $\Omega$ of the KT structure on $\mathscr{M}^*_{\HE}(M^{2n})$. In other words, the induced vector field $\alpha_{{}_X}$ is both Killing and holomorphic.  It follows from the properties of the KT geometry that the torsion 3-form $\mathcal{H}$ on $\mathscr{M}^*_{\HE}(M^{2n})$ will also be invariant as it is constructed from $\mathcal{G}$ and $\Omega$.  Thus
\be
\mathcal{L}_{\alpha_{{}_X}}\mathcal{G}=0~,~~~\mathcal{L}_{\alpha_{{}_X}}\Omega=0~,~~~\mathcal{L}_{\alpha_{{}_X}}\mathcal{H}=0~.
\ee
Furthermore, we shall demonstrate that if in addition
\be
X^\flat\wedge \theta~,
\ee
is a $(1,1)$-form on $M^{2n}$, where $X^\flat$ is the 1-form associated to $X$ with respect to the metric $g$, $X^\flat(Y)=g(X,Y)$ and $\theta$, $\theta_i= D^k\omega_{kj} I^j{}_i$,  is the Lee form, see also (\ref{leef}), then the induced vector field $\alpha_{{}_X}$ is $\h{\mathcal {D}}$-covariantly constant,
\be
\h{\mathcal {D}} \alpha_{{}_X}=0
\ee
 where $\h{\mathcal {D}}$ is the connection with torsion $\mathcal{H}$ of the KT structure on $\mathscr{M}^*_{\HE}(M^{2n})$.

We use the existence of holomorphic and $\h{\mathcal {D}}$-covariantly vector fields to provide a refinement of the strong KT geometry of $\mathscr{M}^*_{\HE}(M^{2n})$. In particular, we demonstrate that the existence of one such vector field on $\mathscr{M}^*_{\HE}(M^{2n})$ implies the existence of a second.  If their orbits, or those of a linear combination of them, are closed,  then the geometry of $\mathscr{M}^*_{\HE}(M^{2n})$  can be modelled on that of a holomorphic principal $T^2$ fibration over a KT manifold.  We also present some examples of moduli spaces that include those on the manifolds $S^3\times S^3$ and $S^3\times T^3$.  According to the rigidity results of \cite{apostolov3}, see also \cite{ivanov}, these two manifolds are essentially the only compact, strong KT  manifolds in six dimensions with non-vanishing torsion,  $H\not=0$,  that the holonomy of $\h\nabla$ is included in $SU(3)$.
$S^3\times S^3$ admits several (bi-)KT structures with one of the KT structures  declared as canonical. We choose this to be one of the left-invariant KT structures on $S^3\times S^3$ viewed as the group manifold $SU(2)\times SU(2)$. For the canonical KT structure on $S^3\times S^3$, we demonstrate that $\mathscr{M}^*_{\HE}(S^3\times S^3)$ is a holomorphic $T^2$-fibration over a KT manifold with curvature $\mathcal{F}=(\mathcal{F}^1, \mathcal{F}^1)$, i.e. the two $\oplus^2\mathfrak{u}(1)$-valued components  of the curvature of the $T^2$-fibration are equal.

Similarly, $S^3\times T^3$ admits several (bi-)KT-structures. Picking a left-invariant KT structure as canonical, $\mathscr{M}^*_{\HE}(S^3\times T^3)$ is again a holomorphic $T^2$-fibration over a KT manifold but this time the curvature of the $T^2$ fibration is $\mathcal{F}=(\mathcal{F}^1, 0)$.  Thus, up to an identification with a discrete group, $\mathscr{M}^*_{\HE}(S^3\times T^3)=S^1\times \mathscr{P}_{\HE}(S^3\times T^3)$, where $\mathscr{P}_{\HE}(S^3\times T^3)$ is a circle bundle over a KT manifold with curvature $\mathcal{F}$ a $(1,1)$-form.  $\mathscr{M}^*_{\HE}(S^3\times T^3)$ also admits a $T^3$ action which leaves invariant the KT geometry on the moduli space but the associated vector fields are not $\h{\mathcal{D}}$-covariantly constant.

The results we have obtained for the moduli spaces of Hermitian-Einstein connections on general KT manifolds can be  adapted to investigate the geometry of the moduli spaces
$\mathscr{M}^*_{\asd}(M^4)$ of anti-self-dual instanton connections over a KT manifold $M^4$. As the anti-self-duality condition on $F$, ${}^*F=-F$, depends only on the choice of orientation on $M^4$ and not on the choice of complex structure, $\mathscr{M}^*_{\asd}(M^4)$ can exhibit intricate geometric structures induced by judicious choices of KT structures on $M^4$ that induce the same orientation on $M^4$. If $M^4$ admits  a KT structure\footnote{It should be noted that every compact Hermitian 4-dimensional manifold admits a strong KT structure.  This is a consequence of the Gauduchon theorem which states that given a Hermitian structure there is always another one within its conformal class such that the Lee form is divergence free. In four dimensions, the Lee form is dual to $H$ and so $H$ is closed as required for a strong KT structure. }, then the results obtained on the geometry and symmetries of $\mathscr{M}^*_{\HE}(M^{2n})$  can be adapted to $\mathscr{M}^*_{\asd}(M^4)$.    We illustrate the construction with an investigation of the geometry of $\mathscr{M}^*_{\asd}(M^4)$ for $M^4$ a torus fibration over a Riemann surface, $\Sigma_g$,  of genus $g$.

It is known that if $M^4$ is a generalised K\"ahler manifold, i.e. bi-KT,   with Hermitian forms $\h\omega$ and $\breve\omega$, and both KT structures induce the same orientation on $M^4$, i.e. $M^4$ has an {\sl oriented} bi-KT structure,  then $\mathscr{M}^*_{\asd}(M^4)$ also admits a bi-KT structure.  The proof that the metric and torsion on $\mathscr{M}^*_{\asd}(M^4)$ induced by each of the KT structures on $M^4$ are equal has been demonstrated  in \cite{hitchin}.  Furthermore, if $M^4$ admits holomorphic and $\h\nabla$-covariantly constant vector fields $\h X$,
 \be
 \mathcal{L}_{\h X} \h\omega=0~,~~~\h\nabla\h X=0~,
 \ee
 then  $\mathscr{M}^*_{\asd}(M^4)$ admits holomorphic and Killing vector fields, $\alpha_{\h X}$,  induced by those on $M^4$. The same applies for holomorphic and $\breve\nabla$-covariantly constant vector fields $\breve X$ of the other KT, where $\breve\nabla$ is the connection with torsion $-H$. The induced vector fields $\alpha_{\breve X}$ on $\mathscr{M}^*_{\asd}(M^4)$ are Killing and holomorphic. In addition,  the induced vector fields $\alpha_{\h X}$ and $\alpha_{\breve X}$ on $\mathscr{M}^*_{\asd}(M^4)$ are  $\h{\mathcal{D}}$-($\breve{\mathcal{D}}-$)covariantly constant provided that
\be
\h X^\flat\wedge \h \theta~,~~~\breve X^\flat\wedge \breve \theta~;~~~\breve \theta=-\h\theta~,
\ee
are (1,1)-forms on $M^4$ with respect to $\h I$ and $\breve I$ complex structures, respectively, where $\h{\mathcal{D}}$ and $\breve{\mathcal{D}}$ are the connections on $\mathscr{M}^*_{\asd}(M^4)$ with torsion $\mathcal{H}$ and $-\mathcal{H}$.  The condition $\breve\theta=-\h\theta$ is a consequence of the restriction\footnote{This relation does not hold for bi-KT manifolds $M^{2n}$ with $n>2$. For non-oriented bi-KT structures on $M^4$, $\breve\theta=\h\theta$.} on the bi-KT structure of $M^4$ to be oriented.  Furthermore, we express the Lie algebra of all vector fields $\alpha_{\h X}$ and $\alpha_{\breve X}$ on $\mathscr{M}^*_{\asd}(M^4)$ as well as the Lie derivatives of the complex structures $\h{\mathcal{I}}$ and $\breve{\mathcal{I}}$ with respect to $\alpha_{\h X}$ and $\alpha_{\breve X}$ in terms of those of the associated vector fields and complex structures on $M^4$.

Next we consider the case that $M^4$ admits either a HKT or an oriented bi-HKT structure. Up to an identification with a discrete group, $S^3\times S^1$ is only  compact HKT manifold in four dimensions that it is not hyper-K\"ahler. This follows from the classification of compact hyper-complex manifolds in four dimensions \cite{boyer}. In fact, it admits four HKT structures separated in two oriented bi-HKT pairs that have been explained in detail in \cite{pw}. Here, we shall focus on the oriented bi-HKT structure that is compatible with a chosen orientation of $S^3\times S^1$. Viewing $S^3\times S^1$ as the group manifold $SU(2)\times U(1)$, one of the HKT structures is left-invariant while the other is right-invariant and the associated Hermitian forms $\h\omega_r$ and $\breve\omega_r$ are self-dual in both cases. Similarly,  $\RP^3\times S^1$ also  admits a  pair of oriented bi-HKT structures\footnote{The spaces $S^3\times S^1$ and $\bRP^3\times S^1$ are the only known compact 4-dimensional manifolds that admit a bi-HKT structure. }.  These can be constructed after viewing $\RP^3\times S^1$ as the group manifold $SO(3)\times SO(2)$. It is a consequence of the results of \cite{moraru} that this induces on
$\mathscr{M}^*_{\asd}(S^3\times S^1)$  and $\mathscr{M}^*_{\asd}(\RP^3\times S^1)$ a bi-HKT structure.  The $SO(4)\times U(1)$  symmetry of $S^3\times S^1$ is generated by the left- and right-action of $SU(2)\times U(1)$  on  $S^3\times S^1$  with $SO(4)=\big(SU(2)\times SU(2)\big)/\bZ_2$.  It turns out that these vector fields satisfy all the conditions required to induce a $SO(4)\times U(1)$  action on $\mathscr{M}^*_{\asd}(S^3\times S^1)$. The associated vector fields  on $\mathscr{M}^*_{\asd}(S^3\times S^1)$  are either $\h{\mathcal{D}}$- or $\breve{\mathcal{D}}$-covariantly constant. The $SO(4)$ vector fields act non-trivially on the Hermitian forms of $\mathscr{M}^*_{\asd}(S^3\times S^1)$, while that of the $\{e\}\times U(1)$ subgroup leaves all the Hermitian forms invariant. A similar conclusion holds for the group action of $\times^2 SO(3)\times SO(2)$ on $\mathscr{M}^*_{\asd}(\RP^3\times S^1)$ and for the properties of the associated vector fields.  A detailed description of the group  action on $\mathscr{M}^*_{\asd}(S^3\times S^1)$  will be given in section \ref{sec:hktvf}. The results are in agreement  with those originally derived  in \cite{witten}.

 The left (right)  action of $SU(2)\times U(1)$ on $S^3\times S^1$ induces a free action of $\mathfrak{su}(2)\oplus \mathfrak{u}(1)$ on $\mathscr{M}^*_{\asd}(S^3\times S^1)$ as the associated vector fields are $\h{\mathcal{D}}$-($\breve{\mathcal{D}})-$covariantly constant. We use this  to show that the geometry of $\mathscr{M}^*_{\asd}(S^3\times S^1)$ and can be modelled on that of a principal
bundle with fibre $SU(2)\times U(1)$ and base space $\mathscr{B}$ a quaternionic-K\"ahler manifold with torsion (QKT) \cite{QKT}, see also \cite{ivanov2, swann2}. The $\mathfrak{u}(1)$ component of the curvature of this principal fibration vanishes and as a result $\mathscr{M}^*_{\asd}(S^3\times S^1)=S^1\times \mathscr{P}_{\asd}$ up to an identification with a discrete group. We express the metric and torsion of $\mathscr{M}^*_{\asd}(S^3\times S^1)$ in terms of those on the fibre and those of the base space $\mathscr{B}$.  The geometry of the moduli spaces $\mathscr{M}^*_{\asd}(S^3/\bZ_n\times S^1)$, $n\in \bN$,  which also  admit an HKT structure \cite{witten}, can be described in a similar way.

\vskip 0.2cm
\underline{{\sl Organisation}:}
The paper is organised as follows. In section 2, we begin with a summary of some the main tools that we use throughout this paper that include  properties of  KT structures,    aspects of the geometry on the space of connections $\mathscr{A}$,  and  the definition and some elementary  properties Hermitian-Einstein connections. The rest of the section is devoted to the proof of the statement that the moduli space of Hermitian-Einstein connections, $\mathscr{M}^*_{\HE}(M^{2n})$, admits a strong KT structure originally presented in  \cite{lubke}. We have simplified some parts and have added a few more steps in the calculations  to make the proof  easier to follow.  At the same time, we establish our notation and develop the techniques that will be useful  later to demonstrate our results. Furthermore, we stress the dependence of the torsion $\mathcal{H}$ of $\mathscr{M}^*_{\HE}(M^{2n})$ on the torsion $H$ of $M^{2n}$ and, in particular, on the Lee form $\theta$ of $M^{2n}$.

 Section 3 contains some of our main results.  We demonstrate that holomorphic and $\h\nabla$-covariantly constant vector fields $X$ on $M^{2n}$ induce Killing and holomorphic vector fields $\alpha_X$ on $\mathscr{M}^*_{\HE}(M^{2n})$. Furthermore, we show that if the vectors fields $X$ on $M^{2n}$ satisfy the condition that $X^\flat\wedge \theta$ is an $(1,1)$-form, then the induced vector fields $\alpha_X$ on $\mathscr{M}^*_{\HE}(M^{2n})$ are $\h{\mathcal{D}}$-covariantly constant, i.e. they are covariantly constant with respect to the connection with torsion $\mathcal{H}$ of  $\mathscr{M}^*_{\HE}(M^{2n})$.  Apart from the work needed to define $\alpha_X$,   this  section includes  the key lemma described in  \ref{sec:key} that is instrumental to establish the results.

 In section 4, we apply our results of section 3 to the  moduli spaces $\mathscr{M}^*_{\HE}(S^3\times S^3)$ and  $\mathscr{M}^*_{\HE}(S^3\times T^3)$.  We compute various aspects of the geometry of these moduli spaces. In particular, we point out that the geometry of both of these can be modelled on that of a holomorphic torus fibration over a KT manifold.  Furthermore, the geometry of $\mathscr{M}^*_{\HE}(S^3\times T^3)$ is a locally a product $S^1\times \mathscr{P}_{\HE}$. In addition, $\mathscr{M}^*_{\HE}(S^3\times T^3)$ admits the action of two additional vector fields that leave  both the metric and complex structure invariant but they are not
 $\h{\mathcal{D}}$-covariantly constant.

In section 5, we apply the results we have obtained for $\mathscr{M}^*_{\HE}(M^{2n})$, especially those of section 3, to the moduli spaces of anti-self-dual instantons $\mathscr{M}^*_{\asd}(M^4)$. If $M^4$ is a Hermitian manifold, then the results of section 3 carry over in a straightforward manner. As an example, we explore the geometric properties of $\mathscr{M}^*_{\asd}(M^4)$ for $M^4$ a torus fibration over a Riemann surface.  However, if $M^4$ is an (oriented) generalised K\"ahler manifold, i.e. oriented bi-KT, the proof that $\mathscr{M}^*_{\asd}(M^4)$ admits a bi-KT structure contains a subtlety.  Each KT structure on $M^4$ induces a KT structure on $\mathscr{M}^*_{\asd}(M^4)$. The metric $\mathcal{G}$  and torsion $\mathcal{H}$ of $\mathscr{M}^*_{\asd}(M^4)$, a priori, depend on the choice of KT structure on $M^4$ that has been used to define them  via  a gauge fixing  or horizontality condition.  We include the proof of \cite{hitchin}  that the metric and torsion of $\mathscr{M}^*_{\asd}(M^4)$ are actually independent from the choice of the KT structure on $M^4$.  As a result, if $M^4$ is an oriented bi-KT manifold, then $\mathscr{M}^*_{\asd}(M^4)$ admits a bi-KT structure. This is described in section \ref{sec:bkt}. In the rest of the section, we generalise our results of section 3 to take into account the bi-KT structure of $\mathscr{M}^*_{\asd}(M^4)$.  This includes a reworking of the key lemma in section \ref{sec:key}. The full analysis is presented in sections \ref{sec:bikt1} and \ref{sec:bikt2}.

In section 6, we generalise our results further to $\mathscr{M}^*_{\asd}(M^4)$ for $M^4$ an HKT or bi-HKT manifold. In section \ref{sec:hktbihkt}, we describe  the proof of \cite{moraru} that if $M^4$ is  an HKT or bi-HKT  manifold, then $\mathscr{M}^*_{\asd}(M^4)$ also admits an HKT or bi-HKT structure.    Focusing on $\mathscr{M}^*_{\asd}(S^3\times S^1)$ and $\mathscr{M}^*_{\asd}(\RP^3\times S^1)$, we demonstrate that their geometry can be modelled on those of a principal bundle fibration with fibre $S^3\times S^1$ and $\RP^3\times S^1$ over a QKT manifolds, respectively.  We also compute the curvature of this principal fibration and find that in both cases, $\mathscr{M}^*_{\asd}(S^3\times S^1)$ and $\mathscr{M}^*_{\asd}(\RP^3\times S^1)$ are products $S^1\times \mathscr{P}_{\asd}$ up to an identification with a discrete group.  We also obtain similar results for all $\mathscr{M}^*_{\asd}(S^3/\bZ_n\times S^1)$, $n\in \bN$, moduli spaces that have  been previously investigated in \cite{witten}.  Furthermore, we  comment on the geometry of $\mathscr{M}^*_{\asd}(S^3\times S^1)$ after considering a KT geometry on $S^3\times S^1$ with a squashed metric on $S^3$.

In section 7, we describe some applications. These include the use of $\mathscr{M}^*_{\HE}(M^{2n})$ and $\mathscr{M}^*_{\asd}(M^{4})$ for constructing new manifolds with strong KT, bi-KT, HKT and bi-HKT structures. We also examine the symmetries of two-dimensional sigma models with target spaces such moduli spaces.  Furthermore, we explore the superconformal properties of sigma models with target spaces $\mathscr{M}^*_{\HE}(M^{2n})$ and $\mathscr{M}^*_{\asd}(M^{4})$.

The space $S^3\times S^1$ as a group manifold can identified with either $U(2)$ or  $SU(2)\times U(1)$. In appendix A, we demonstrate that the HKT structures on $S^3\times S^1$ are independent from the group structure used to construct them up to a diffeomorphism.

\section{Geometry of the moduli space of Hermitian-Einstein connections}\label{sec:2}

\subsection{Preliminaries}

\subsubsection{KT manifolds} \label{sec:kt}

 The K\"ahler with torsion  (KT) geometries\footnote{KT and HKT  geometries are also referred to in the literature, especially for bi-KT and bi-HKT, as generalised K\"ahler and generalised hyper-K\"ahler, respectively \cite{hitchin2, gualtieri}  -- bi-KT and bi-HKT geometries contain two copies of the associated structure one with respect to the connection $\h\nabla=D+1/2 H$ and the other with respect to the connection $\breve\nabla=D-1/2 H$. Here, we use the terminology of \cite{HoweGP} to describe these geometries.   The strong condition on these geometries is also referred to as pluriclosed.}  have recently been reviewed in \cite{pw}, where a more detailed description of the KT geometry and references can be found.  The task here is to give a very brief summary of some of their properties that we use later to investigate the geometry of the moduli spaces of Hermitian-Einstein connections.

 KT manifolds are Hermitian manifolds $M^{2n}$ with metric $g$ and complex structure $I$, $I^2=-{\bf 1}$,  equipped with the unique compatible connection $\h\nabla$,
  \be
  \h\nabla_i g_{jk}=0~,~~~\h\nabla_i I^j{}_k=0~,
  \ee
  whose torsion is a 3-form $H$. Therefore, KT manifolds are complex manifolds and the metric $g$ is Hermitian with respect to $I$, i.e. $g(IX, IY)=g(X,Y)$,  where $X,Y$ are vector fields on $M^{2n}$, and $I X=(IX)^i\partial_i=I^i{}_j X^j \partial_i$ and similarly for $IY$.  In components, the connection $\h\nabla$ acts on the vector field $X$ as
 \be
 \h\nabla_i X^j=D_i X^j+\frac{1}{2} H^j{}_{ik} X^k~,
 \ee
 where $D$ is the Levi-Civita connection of $g$ and $H_{ijk}= g_{im} H^m{}_{jk}$ is a 3-form. We also define the connection $\breve \nabla$ as
 \be
 \breve \nabla_i X^j=D_i X^j-\frac{1}{2} H^j{}_{ik} X^k~,
 \ee
 which we shall use later in the description of the bi-KT geometry.

 The Hermitian form $\omega$ on $M^{2n}$ is defined as $\omega(X,Y)=g(X, I Y)$,   or equivalently in components $\omega_{ij}=g_{ik} I^k{}_j$.  As $g$ is Hermitian and $I^2=-{\bf 1}$, $\omega$ is a 2-form on $M^{2n}$. As both $g$ and $I$ are $\h\nabla$-covariantly constant, $\h\nabla g=\h\nabla I=0$,  so it is $\omega$, $\h\nabla \omega=0$.

 The conditions $\h\nabla g=\h\nabla I=0$ determine the torsion $H$ in terms of $I$ and $\omega$ uniquely as
 \be
 H=-\ii_I d\omega=-d^c \omega~,
 \ee
 where $d$ is the exterior derivative, $\ii_I$ is the inner derivation\footnote{The inner derivation of a $k$-form $L=\frac{1}{k!} L_{j_1\cdots j_k}dx^{j_1}\wedge \cdots \wedge dx^{j_{k}} $ with respect to the vector field $X$ and to the complex structure $I$ are $\ii_X L=\frac{1}{(k-1)!} X^i L_{ij_1\dots j_{k-1}} dx^{j_1}\wedge \cdots \wedge dx^{j_{k-1}}$ and $\ii_I L=\frac{1}{(k-1)!} I^i{}_{j_1} L_{ij_2\dots j_{k}} dx^{j_1}\wedge \cdots \wedge dx^{j_{k}}$, respectively. } with respect to $I$ and $d^c=\ii_I d-d \ii_I= i(\partial-\bar\partial)$ with $\partial$ ($\bar\partial$) denoting the holomorphic (anti-holomorphic)  exterior derivative on $M^{2n}$;  $M^{2n}$ is a complex manifold.  Of course if $d\omega=0$, and so $H=0$, $M^{2n}$ is a K\"ahler manifold.

 If $M^{2n}$ is a KT manifold, $\h\nabla\omega=0$ implies that
 \be
 d\omega=\ii_I H~.
 \label{doih}
 \ee
 Conversely , if $M^{2n}$ is a Hermitian manifold,  in particular $I$ is (an integrable) complex structure, and (\ref{doih}) holds for some 3-form $H$, then $\h\nabla I=0$ and $M^{2n}$ is a KT manifold  \cite{HoweGP}. This will be used below to prove that the moduli spaces of Hermitian-Einstein connections admit a KT structure.

 The integrability of the complex structure $I$ together with $\h\nabla I=0$ imply that $H$ is a $(2,1)\oplus (1,2)$-form on $M^{2n}$, where in the $(r,s)$-decomposition of a $k$-form $L$, $L=\oplus_{r+s=k} L^{r,s}$,  $r$ denotes the number of holomorphic directions while $s$ denotes the number of anti-holomorphic directions of the component $L^{r,s}$. In particular, the $(3,0)\oplus (0,3)$ component of $H$ vanishes, $H^{3,0}=H^{0,3}=0$. This condition can also be expressed either as
 \be
  I^m{}_i I^n{}_j H_{mnk}+I^m{}_k I^n{}_i H_{mnj}+I^m{}_j I^n{}_k H_{mni}-H_{ijk}=0~,
  \label{integI}
 \ee
or as
\be
  I^m{}_i  H_{mjk}+I^m{}_k  H_{mij}+I^m{}_j  H_{mki}-H_{mnp} I^m{}_i I^n{}_jI^p{}_k =0~,
  \label{integII}
 \ee
 and these formulae are useful in the calculations that will follow below.

 The Lee form $\theta$ of a Hermitian manifold is defined as
 \be
 \theta_i= D^k\omega_{kj} I^j{}_i~.
 \label{leef}
 \ee
 Using $\h\nabla I=0$, this can be re-expressed as
 \be
 \theta_i=-\frac{1}{2} H_{mnk}\, \omega^{mn} I^k{}_i~.
 \ee
 It is known that in the conformal class of every Hermitian metric $g$, there is one, the Gauduchon metric, for which $D^i\theta_i=0$ or equivalently
 \be
 d d^c\omega^{n-1}=-2i \partial \bar\partial \omega^{n-1}=0~.
 \label{gauduchon}
  \ee
  Both the Lee form $\theta$ and the Gauduchon metric enter in the investigation of the geometry of the moduli spaces $\mathscr{M}^*_{\HE}(M^{2n})$.

 The KT structure on a Hermitian manifold $M^{2n}$ is {\sl strong}, iff
 \be
 dH=0~,
 \ee
 i.e. $H$ is a closed 3-form. This is an additional condition on $H$ and it does not follow from the properties of the Hermitian structure on $M^{2n}$.  The strong condition can be stated in terms of the Hermitian form as
 \be
 d d^c\omega=0~,~~\mathrm{or~equivalently~~} \partial\bar\partial\omega=0~,
 \ee
 i.e. it implies that $\omega$ is $\partial\bar\partial$-closed.  Note also that if $dH=0$, one of the Bianchi identities of $\h\nabla$ is
 \be
\h R_{i[jk\ell]}=-\frac{1}{3} \h\nabla_i H_{jk\ell}~,
\label{bia1}
\ee
where $\h R$ is the curvature of $\h\nabla$.
This Bianchi identity holds for all connections $\h\nabla$ with torsion a 3-form $H$, provided that $dH=0$, and not just for connections with skew-symmetric torsion on Hermitian manifolds.

\subsubsection{Moduli space of connections, gauge fixing and horizontality}

Let $\mathscr{A}$ be the space connections\footnote{We shall also use $\mathscr{A}(M^d)$ to denote the space of connections $\mathscr{A}$ whenever it is required to distinguish spaces for either clarity or economy in the description. A similar notations will be adapted for the moduli spaces $\mathscr{M}$ described below.} of the principal bundle $P(M^d, G)$, where the fibre group $G$ is the gauge group and $M^d$ is the base space. The group of gauge transformations $\mathscr{G}$  are the sections of the associate bundle to $P$ with respect to the adjoint representation $\Adj$ of $G$, i.e. $\mathscr{G}=\Omega^0(P\times_{\Adj} G)$.  Locally these are maps from $M^d$ to $G$. The Lie algebra of $\mathscr{G}$ is
$\Omega^0(P\times_{\Adj} \mathfrak{g})$ and it can locally be identified with the maps from $M^d$ to $\mathfrak{g}$, where $\mathfrak{g}$ is the Lie algebra of the gauge group $G$. $\mathscr{A}$ is an affine space, where  $\Omega^1(P\times_{\Adj} \mathfrak{g})$ acts on it with translations; locally $\Omega^1(P\times_{\Adj} \mathfrak{g})$ can be identified with the $\mathfrak{g}$-valued 1-forms of $M^d$.

The action of $\mathscr{G}$ on $\mathscr{A}$ induces an infinitesimal action of $\Omega^0(P\times_{\Adj} \mathfrak{g})$ on $\mathscr{A}$ as
\be
A\rightarrow A+d_A\epsilon~,
\ee
where $\epsilon \in \Omega^0(P\times_{\Adj} \mathfrak{g})$ and
\be
d_A\epsilon\equiv d\epsilon+[A, \epsilon]_{\mathfrak{g}}~,
\ee
with $[\cdot, \cdot]_{\mathfrak{g}}$ the commutator\footnote{In what follows, we shall encounter several brackets. As it has already been mentioned $[\cdot, \cdot]_{\mathfrak{g}}$ is the commutator of Lie algebra $\mathfrak{g}$.   We reserve $[\cdot, \cdot]$ as the commutator of two vector fields on $M^d$ and $\lsq\cdot, \cdot\rsq$ as the commutator of two vector fields on the space of connections and associated moduli spaces. The latter brackets will be defined below.} of the Lie algebra  $\mathfrak{g}$.

The tangent space of $\mathscr{A}$ at any point can be identified with $\Omega^1(P\times_{\Adj} \mathfrak{g})$. The vector fields generated by the action of $\mathscr{G}$ at a point $A\in \mathscr{A}$ are
\be
a=d_A \epsilon~,
\ee
where    $\epsilon\in \Omega^0(P\times_{\Adj} \mathfrak{g})$.

Given $a_1, a_2\in\Omega^1(P\times_{\Adj} \mathfrak{g})$, an $L^2$-metric can be defined on $\mathscr{A}$ as
\begin{align}
 (a_1, a_2)_{\mathscr{A}}&\equiv \int_{M^d} d^n x \sqrt g\,\, g^{-1}\, \langle a_1, a_2\rangle_{\mathfrak{g}}=\int_{M^d} d^n x \sqrt g\,\, g^{ij} \ell_{bc}\, a^b_{1i} a^c_{2j}
\cr
&=\int_{M^d} \langle a_1\wedge {}^*a_2\rangle_{\mathfrak{g}}={1\over (d-1)!}\int_{M^d} \ell_{bc}\, a^b_{1i_1} a^c_{2k }\,\varepsilon^k{}_{i_2\dots i_{d}} dx^{i_1} \wedge\cdots \wedge dx^{d}~,
\label{Ametric}
\end{align}
where  $\langle\cdot,\cdot\rangle_{\mathfrak{g}}$ is a bi-invariant inner product on $\mathfrak{g}$ with components $(\ell_{bc})$, $g$ is a metric on $M$ and ${}^*$ denotes the Hodge duality operation\footnote{In general, the Hodge dual of a $p$-form $\chi$ is ${}^*\chi_{i_1\dots i_{d-p}}=\frac{1}{p!} \chi_{j_1\dots j_p} \epsilon^{j_1\dots j_m}{}_{i_1\dots i_{d-p}}$, where $\varepsilon$ is the volume form of $M^d$.}.  For example for $\mathfrak{g}=\mathfrak{su}(r)$ and with $\mathfrak{su}(r)$  identified with the $r\times r$ anti-Hermitian traceless matrices,  $\langle a_1, a_2\rangle_{\mathfrak{g}}=-\mathrm{tr} (a_1 a_2)$, where the sign has been added for the positivity of the inner product. From now on the gauge indices $b,c$ will be suppressed throughout.  This inner product on $\mathscr{A}$ is not gauge invariant in the sense that in general $( a_1+d_A\epsilon, a_2)_{\mathscr{A}}+( a_1, a_2+d_A\epsilon)_{\mathscr{A}}\not=0$.

 Next, postponing the discussion about smoothness properties for later, let us define a metric on $\mathscr{M}=\mathscr{A}/\mathscr{G}$.
 To define a metric on  $\mathscr{M}=\mathscr{A}/\mathscr{G}$, one has to restrict the metric (\ref{Ametric}) on $\mathscr{A}$ into directions transversal to the orbits of $\mathscr{G}$. One way to do this is to gauge fix  by imposing the condition
\be
{}^*d_A{}^* a\equiv D_A^i a_i=D^i a_i+[A^i, a_i]_{\mathfrak{g}}=0~,
\label{gaugefix1}
 \ee
 where $D$ is the Levi-Civita connection of $M^d$ and $D_A\chi \equiv D\chi+[A, \chi]_{\mathfrak{g}}$ for $\chi\in \Omega^p(P\times_{\Adj} \mathfrak{g})$. Of course $d_A=D_A$ on $\Omega^0(P\times_{\Adj} \mathfrak{g})$.

Assuming that $a\in \Omega^1(P\times_{\Adj} \mathfrak{g})$ can uniquely be written as
\be
a= a^h+d_A\epsilon~,
\label{hvdecom1}
\ee
where ${}^*d_A{}^* a^h=0$, an inner product can be defined on $\mathscr{M}$ by representing the tangent vectors of $\mathscr{M}$ with $a^h$ and taking as an inner product   $( a^h_1, a^h_2)_{\mathscr{A}}$.  For the choice of inner product (\ref{Ametric}) and gauge fixing condition (\ref{gaugefix1}), the directions tangent to $\mathscr{M}$ are orthogonal to those of the orbits of $\mathscr{G}$.

There is an alternative way to think about the gauge fixing condition (\ref{gaugefix1}). Assuming   $\mathscr{G}$ acts freely on  $\mathscr{A}$, or on a suitable open subset,  $\mathscr{A}$ can be thought as a principal bundle with fibre group $\mathscr{G}$ and  base space $\mathscr{M}$  -- it is assumed that $\mathscr{G}$ acts on the right on $\mathscr{A}$.   The projection $p: \mathscr{A}\rightarrow \mathscr{M}$  maps a connection $A$ to the orbit $[A]$ of $\mathscr{G}$ through $A$, $p(A)=[A]$. In this context, the gauge fixing condition (\ref{gaugefix1}) can be thought as defining the horizontal subspace of a principal bundle connection while the vertical directions are spanned by the vector fields generated by the group action of $\mathscr{G}$ on $\mathscr{A}$. In other words (\ref{hvdecom1}) can be seen as a decomposition of the tangent space of ${\mathscr{A}}$ at $A$ into  horizontal and vertical subspaces. This justifies the notation used in (\ref{hvdecom1}).  With this in mind, one can define an inner product on $\mathscr{M}$ as
\be
(\alpha_1, \alpha_2)_{\mathscr{M}}([A])\equiv ( a^h_1, a^h_2)_{\mathscr{A}}(A)
\ee
where $\alpha_1$ and $\alpha_2$ are tangent vectors of ${\mathscr{M}}$ at $[A]$ and $a^h_1$ and $a^h_2$ are their horizontal lifts\footnote{The horizontal lift of a vector $\alpha\in T_{[A]}{\mathscr{M}}$ with respect to a connection $\Gamma$ is a horizontal vector $a^h\in T_A{\mathscr{A}}$, $\Gamma(a^h)=0$, such that the push forward of $a^h$ is $\alpha$, i.e. $p_*(a^h)=\alpha$.}  at $A$, respectively, and $p(A)=[A]$. With this, we mean to consider $a_1$ and $a_2$ tangent vectors of ${\mathscr{A}}$ at the point $A$ such that the push-forward $p_* a_1=\alpha_1$ and $p_*a_2=\alpha_2$. As the kernel of $p_*$ is spanned by the tangent vectors of the orbit of $\mathscr{G}$ through $A$ two such choices for $a_1$  differ by a gauge transformation\footnote{From now on, the dependence of the inner products and other tensors under consideration, which will be defined later, on  the moduli space  points,   will be mostly suppressed.} and similarly for $a_2$. Given a connection, or equivalently a gauge fixing condition, a choice can be made amongst $a_1$ and $a_2$ that represent $\alpha_1$ and $\alpha_2$, respectively, on ${\mathscr{A}}$. This choice are the horizontal vector fields $a^h_1$ and $a_2^h$ that obey the gauge fixing condition  (\ref{gaugefix1}). Clearly, the inner product on ${\mathscr{M}}$ depends on the choice of connection, equivalently gauge fixing condition,  of the principal bundle.  By  construction, the pull back of $(\cdot, \cdot)_{\mathscr{M}}$ on ${\mathscr{A}}$ is
\be
p^*( a_1, a_2)_{\mathscr{M}}=( a^h_1,  a^h_2)_{\mathscr{A}}~.
\ee
This construction can be generalised. Instead of choosing the connection via the gauge fixing condition (\ref{gaugefix1}), one can choose any other principal bundle connection $\Gamma$ and write the tangent vectors of $\mathscr{A}$ as
\be
a=a^h+a^v~,
\label{hvdecom2}
\ee
where $a^h$ is the horizontal component of the tangent vector, i.e. $\Gamma(a^h)=0$, and $a^v$ is the vertical component.  As the vertical vector fields are those generated by $\mathcal{G}$, one can write $a^v=d_A \epsilon$ and so (\ref{hvdecom2}) can be rewritten as
\be
a=a^h+d_A\epsilon~,
\label{hvdecom3}
\ee
where of course $\epsilon$ will depend on both $a$ and $A$.  Using this, one can define an inner product on ${\mathscr{M}}$ by setting
\be
( \alpha_1, \alpha_2)_{\mathscr{M}, \Gamma}\equiv ( a^h_1,  a^h_2)_{\mathscr{A}}~,
\ee
where $\alpha_1$ and $\alpha_2$ are tangent vectors of $\mathscr{M}$ and $a^h_1$ and $a^h_2$ are their horizontal lifts, $\Gamma(a_1^h)=\Gamma(a_2^h)=0$. Again $p^*( a_1, a_2)_{\mathscr{M}, \Gamma}=(a^h_1,  a^h_2)_{\mathscr{A}}$.

Given a connection $\Gamma$, the decomposition of a tangent vector into horizontal and vertical components as in (\ref{hvdecom2}) is unique.  In practice, the connection $\Gamma$ in the applications that follow is determined by a gauge fixing type of condition, i.e. a differential equation on the tangent vectors of $\mathscr{A}$. In such a case, the uniqueness of the decomposition (\ref{hvdecom2}) will require a proof as it will depend on the properties of the differential equation.  As the gauge fixing condition and the horizontality condition on the tangent vectors of $\mathscr{A}$  are indistinguishable in this context,  we shall refer to the {\sl gauge fixing condition} as the {\sl horizontality condition} and vice-versa.

\subsection{Hermitian-Einstein connections}\label{sec:HE}

Suppose that $M^{2n}$ is a Hermitian manifold with complex structure $I$ and Hermitian form $\omega$, $\omega(X,Y)=g(X, IY)$, i.e. $\omega_{ij}=g_{ik} I^k{}_j$, where $IY=(I Y)^i\partial_i= I^i{}_j Y^j\partial_i$. Choose the orientation of $M^{2n}$ as $\frac{1}{n!} \wedge^n\omega$.  As it has already been stated in section \ref{sec:kt}, in the conformal class of every Hermitian metric $g$ on $M^{2n}$, there is always one, the Gauduchon metric,  that satisfies the condition (\ref{gauduchon}), i.e. $\bar\partial\partial \omega^{n-1}=0$.

Next consider a complex vector bundle $E$ on $M^{2n}$ with a connection $A$ that is compatible with a fibre Hermitian metric $h$, $D_A h=0$.  Therefore, the gauge group of $A$ is either $U(r)$ or $SU(r)$.   Therefore, $E$ should be thought as an associated vector bundle of a principal $U(r)$ or  $SU(r)$  bundle with respect to the fundamental $r$-dimensional representation of one these groups. The Hermitian-Einstein condition \cite{kobayashi} on $A$ is
\be
F^{2,0}=F^{0,2}=0~,~~~~\omega\llcorner F\equiv\frac{1}{2} \omega^{ij} F_{ij}=-i \Lambda {\bf 1}_E~,
\label{HEcond-1}
\ee
where $F$ has been expressed in the fundamental representation of $\mathfrak{u}(r)$, spanned by anti-Hermitian $r\times r$ matrices, and $\Lambda$ is real constant. Clearly, if the gauge group is $SU(r)$, $\Lambda=0$. Otherwise,
\be
\Lambda=\frac{2\pi}{(n-1)! \mathrm{Vol} (M^{2n})} \mu(E)~,
\ee
where $\mu(E)$ is the slope of $E$ that can be expressed in terms of the degree of $E$
\be
\mathrm{deg}(E)\equiv \int_{M^{2n}} c_1(E)\wedge \omega^{n-1}~,
\ee
as
\be
\mu(E)\equiv  \frac{ \mathrm{deg}(E)}{r}~,~~~
\ee
and $c_1(E)$ is the first Chern class of $E$ expressed in terms of the connection $A$. If $M^{2n}$ is K\"ahler, then $\mathrm{deg}(E)$ is a topological invariant.  Otherwise, the volume of $M^{2n}$, the slope and the degree of $E$ are all evaluated using the Gauduchon metric associated to the Hermitian metric $g$. The use of the Gauduchon metric in this context will become transparent later\footnote{There are other reasons too. If the degree of $E$ is evaluated using the Gauduchon metric, it becomes independent from  deformations of the geometry of $E$ that are not relevant in this context, like for example the conformal deformation of the fibre metric $h$ of $E$ as this does not affect the complex structure on $E$.} when the geometry of the moduli spaces  is examined.

The Hermitian-Einstein condition on a connection $A$ with gauge group $G$ of a vector bundle $E$ can also be expressed   as
\be
F^{2,0}=F^{0,2}=0~,~~~~\omega\llcorner F\equiv\frac{1}{2} \omega^{ij} F_{ij}= \lambda~,
\label{HEcond1}
\ee
after viewing the curvature $F$ as a Lie algebra valued 2-form on $M^{2n}$, where $\lambda$ is a real constant, which commutes with all the other elements of the gauge group Lie algebra $\mathfrak{g}$.  In what follows, the gauge group will  be taken as one of the following groups
\be
 U(r)~,~~~ SU(r)~,~~ \mathrm{or}~,~~~ PU(r)=PSU(r)~,
 \label{gaugeg}
 \ee
  where $PU(r)=U(r)/U(1)=SU(r)/\bZ_r=PSU(r)$ is the projective unitary group with $U(1)$ the centre of $U(r)$. Clearly $\lambda=0$ for $G=SU(r)$ or $G=PU(r)$. Most of the moduli calculation below can formally be carried out without specifying the gauge group $G$. Because of this, we shall proceed without mentioning explicitly what  $G$ is. But  as we rely on the analytic results of \cite{lubke} that have been demonstrated provided that $G$ is one of the groups in (\ref{gaugeg}), our results are valid provided $G$ is one of the groups in (\ref{gaugeg}).

Before, we proceed with a description of the geometry of the moduli space of Hermitian-Einstein connections, we shall present a brief description of these spaces.
Let $\mathscr{A}_{\HE}$ be the space of Hermitian-Einstein connections. The space of {\sl irreducible} Hermitian-Einstein connections, $\mathscr{A}^*_{\HE}$ is the subspace of $\mathscr{A}_{\HE}$ that the group of gauge transformation $\mathscr{G}^*$ acts freely. For the gauge group  $SU(r)$, $\mathscr{G}^*$ is  the group of gauge transformations of $E$ while for the gauge group $U(r)$,  $\mathscr{G}^*=\mathscr{G}/U(1)$. In the $U(r)$ case, the $U(1)$ subgroup of constant gauge transformations that lie in the centre of $\mathscr{G}$ is in the isotropy group of every connection $A$ of $E$ and this is the reason that it has to be factored out.  The associated moduli space $\mathscr{M}^*_{\HE}=\mathscr{A}^*_{\HE}/\mathscr{G}^*$ of irreducible Hermitian-Einstein connections can be given a smooth structure and it is the smooth part of the moduli space of Hermitian-Einstein connections whose geometry we shall describe later.

It is clear from the definition of $\mathscr{A}^*_{\HE}$ that the only solution to $d_A\epsilon =0$ is that $\epsilon=0$. It can be shown that if the equation $d_A\epsilon =0$ admits an non-trivial solution, then $E$ decomposes as an $h$-orthogonal sum of subbundles.  The proof utilises the eigenspaces of the endomorphism generated by $\epsilon$ on $E$ \cite{lubke}.

It remains to explore the existence of Hermitian-Einstein connections \cite{Donaldson, Uhlenbeck, LiYau}. Taking either $G=U(r)$ or $G=SU(r)$,  this has been investigated in the context of the correspondence\footnote{The proof of this correspondence is rather involved. One of the difficulties is that the group of gauge transformations on holomorphic structures is the complexified group $\mathscr{G}\otimes \bC$ of $\mathscr{G}$.  The proof also contains a substantial analytic side which is essential to determine a manifold structure on  $\mathscr{M}^*_{\HE}$.} between Hermitian-Einstein connections  and stable holomorphic structures in $E$. An elementary  way to explain the construction, see \cite{lubke} for the detailed proof,  is to solve locally the $F^{0,2}=0$ condition and express the anti-holomorphic component of the connection as
\be
A_{\bar p}=\bar V^{-1} \partial_{\bar p} \bar V~,
\ee
where we have introduced complex coordinates $(z^p, z^{\bar p} ; p=1, \cdots, n)$ on $M^{2n}$ and $V$ is an element of the complexified gauge group. Such a $V$ is not unique as the transformations
\be
\bar V\rightarrow \kappa(z) \bar V~,
\ee
where $\kappa$ are holomorphic, $\bar \partial \kappa=0$, leave $A_{\bar p}$ invariant. Therefore, it is expected that at the intersection of two trivialisations of $E$, the two solutions for $V$ will  patch together with a holomorphic  $\kappa$ transition functions. Performing a $\bar V^{-1}$ gauge transformations, which lies in the complexified group of gauge transformations,  the $D_A$ covariant derivative transforms as
\be
\bar V D_{\bar p} \bar V^{-1}= \bar\partial_{\bar p}~,~~~\bar V D_{ p} \bar V^{-1}= \partial_p+ \bar V A_p\bar V^{-1}+\bar V \partial_p\bar V^{-1}~.
\ee
This is the Chern connection  of $E$, which has the property that  $\bar D_A=\bar\partial$ and it is compatible with its holomorphic structure.  This is well defined as the holomorphic transition functions preserve its form. Given a Hermitian fibre metric $h$ on $E$, a compatible Chern connection, $D^C$, is uniquely determined. Indeed,
\be
D^C_p h_{\alpha\bar \beta}=\partial_p h_{\alpha\bar \beta}- A_p{}^\gamma{}_\alpha h_{\gamma\bar \beta}=0 \Longrightarrow A_p{}^\alpha{}_\beta=h^{\alpha\bar\gamma}\partial_{p} h_{\beta\bar\gamma}~,
\ee
where $\alpha, \beta=1,\cdots, r$.  The theorem demonstrates that there always exist a fibre metric $h$ such that the Chern connection satisfies the Hermitian-Einstein condition (\ref{HEcond-1}). Stability in this context means that for every subbundle $F$ of $E$, $F\subset E$, the slope of $F$ is less than that of $E$, $\mu(F)<\mu(E)$ -- it is required for solving the differential equation for the fibre metric.

\subsection{The Hermitian structure on $\mathscr{M}^*_{\HE}$}

The complex structure on $M^{2n}$ induces a complex structure $\mathcal{I}$ on $\mathscr{A}$ as
\be
\mathcal{I} a\equiv -\ii_I a=-i a^{1,0}+i a^{0,1}~,
\label{actI}
\ee
where we have used $I$ to decompose $a=a^{1,0}+a^{0,1}$ as an 1-form on $M^{2n}$  into holomorphic $a^{1,0}$ and anti-holomorphic $a^{0,1}$ components. One of the reasons in the minus sign that appears in the definition of $\mathcal{I}$ is that  $(I X)_i =g_{ij} I^j{}_k X^k=- I^k{}_i X_k$. With this convention  $\bar\partial_A=\bar\partial+A^{0,1}$ depends holomorphically  on $A^{0,1}$ and it is used in the proof of the correspondence mentioned in the previous section.
Next, one can define a Hermitian 2-form on $\mathscr{A}$ as
\be
\Omega_{\mathscr{A}}(a_1, a_2)\equiv ( a_1, \mathcal{I} a_2)_{\mathscr{A}} =\frac{1}{(n-1)!}\, \int_M \omega^{n-1}\wedge  \langle a_1\wedge  a_2\rangle_{\mathfrak{g}}~,
\label{Aform}
\ee
which demonstrates that the metric $( \cdot, \cdot)_{\mathscr{A}}$ is Hermitian with respect $\mathcal{I}$ as $\Omega$ is skew-symmetric in the exchange of $a_1$ and $a_2$.

The tangent space of $\mathscr{A}^*_{\HE}$ at a point $A$ is characterised by the elements of $\Omega^1(P\times_{\Adj} \mathfrak{g})$ that satisfy the conditions
\be
(d_A a)^{2,0}=(d_A a)^{0,2}=0~,~~~\omega\llcorner (d_A a)=0~.
\label{HEcond}
\ee
For $\mathscr{M}^*_{\HE}$ to be a complex manifold, it is required that if $a\in T_{[A]}\mathscr{M}_{\HE}$, then $\mathcal{I}a$ is also an element of $T_{[A]}\mathscr{M}^*_{\HE}$. As $\mathcal{I}a$ solves the first two conditions of (\ref{HEcond}), if $a$ does, it remains to make sure that it also solves the last condition in (\ref{HEcond}).   This is achieved
by choosing the gauge fixing condition that specifies  the tangent space of $\mathscr{M}^*_{\HE}$ as a subspace of the tangent space of $\mathscr{A}^*_{\HE}$.  This is equivalent to choosing a connection in $\mathscr{A}^*_{\HE}$ for which the horizontal vectors are characterised by the condition \cite{lubke}
\be
\omega\llcorner (d_A \ii_I a)=-\omega\llcorner (d_A \mathcal{I} a)= - \omega\llcorner d_A^c a=0~,
\label{HEcone}
\ee
where $d_A^c\equiv \ii_I d_A -d_A \ii_I=i(\partial_A-\bar\partial_A)$. As
\be
\omega\llcorner (d_A \ii_I a)=D^i_A a_i+\theta^i a_i~,
\ee
where $\theta$ is the Lee form (\ref{leef}), $\theta_i=D^j \omega_{jk} I^k{}_i$,  of the Hermitian manifold $M^{2n}$,  the horizontality condition (\ref{HEcone}) on the tangent vector $a$ can also be expressed as
\be
D^i_A a_i+\theta^i a_i=0~.
\label{horizon}
\ee
In general, this condition depends on the choice of the complex structure $I$ on $M^{2n}$ as the Lee form $\theta$ depends on $I$.  However, we shall demonstrate later that there are occasions that this is not the case and this is a significant observation that will be used to determine the geometry of the moduli space of instantons.

As we aim to define a connection on $\mathscr{A}^*_{\HE}$,  the tangent vectors should decompose uniquely into horizontal and vertical components
as
\be
a=a^h+ a^v~,
\label{HEhv}
\ee
where $a^h$ satisfies both (\ref{HEcond}) and (\ref{HEcone}) and of course $a^v=d_A \epsilon$. This requires a proof.  Before, we proceed with it, we point out that the if $a$ satisfies the Hermitian-Einstein conditions (\ref{HEcond}), then $a^h$ will also satisfy these conditions. As $a$ and $a^h$ differ by a gauge transformation, there is $\eta\in \Omega^0(P\times_{\Adj} \mathfrak{g})$, such that $a=a^h+d_A \eta$.  Then
\be
d_A a=d_Aa^h+ d_A^2\eta=d_A a^h+[F, \eta]_{\mathfrak{g}}~.
\label{HEcomp}
\ee
Thus if $d_A a$ satisfies the Hermitian-Einstein conditions (\ref{HEcond}) so does $d_A a^h$ because $F$ is a $(1,1)$-form and $\lambda$ commutes with all the other elements of $\mathfrak{g}$.  In other words, the Hermitian-Einstein conditions are compatible with the horizontality condition and so $a^h$ can be taken to satisfy both (\ref{HEcond}) and the gauge fixing condition (\ref{HEcone}).

Returning to the proof of (\ref{HEhv}) given a tangent vector that satisfies (\ref{HEcond}), one has to show that there is an $\epsilon$ such that (\ref{HEhv}) is satisfied with $a^v=d_A\epsilon$.  Applying the operator
$\omega\llcorner d_A^c$ on (\ref{HEhv}), one finds that
\be
\omega\llcorner d_A^c a=\omega\llcorner d_A^c d_A\epsilon
\ee
So one has to show that the operator
\be
\op=-\omega\llcorner d_A^c d_A=D_A^iD_{Ai}+\theta^i  D_{Ai}~,
\label{oop}
 \ee
has an inverse and it  is onto.  For the existence of an inverse, the kernel of the operator must be trivial -- incidentally this implies the uniqueness of the solution. Indeed consider
\begin{align}
&\int_M \sqrt{g} d^{2n}x \langle \epsilon, \op \epsilon\rangle_{\mathfrak{g}} =\int_M \sqrt{g} d^{2n}x \langle \epsilon, (D_A^iD_{Ai}+\theta^i  D_{Ai}) \epsilon\rangle_{\mathfrak{g}}
\cr
&
\qquad =-\int_M\sqrt{g} d^{2n}x \Big(\langle D_A^i\epsilon, D_{Ai}\epsilon \rangle_{\mathfrak{g}}+{1\over2} D_i\theta^i \langle\epsilon,  \epsilon\rangle_{\mathfrak{g}}\Big)
\cr
&
\qquad =-\int_M\sqrt{g} d^{2n}x \langle D_A^i\epsilon, D_{Ai}\epsilon \rangle_{\mathfrak{g}}~,
\end{align}
where in the last step we have used that the Lee form is co-closed,
\be
D_i\theta^i=0~.
\ee
If $\epsilon$ is in the kernel of the operator $\op$, then the identity above implies $D_A\epsilon=d_A\epsilon=0$. But for irreducible connections, $A\in \mathscr{A}^*_{\HE}$, $D_A\epsilon=d_A\epsilon=0$ implies that $\epsilon=0$. Thus the kernel of  $\op$ is trivial.  The co-closure  condition on the Lee form  $\theta$ is another expression for the Gauduchon condition (\ref{gauduchon}).  Therefore, this does not imply a restriction on the Hermitian geometry of $M^{2n}$ as in the conformal class of every Hermitian metric, there is always one that satisfies the Gauduchon condition.

Let us now turn to the proof that $\op$ is onto. Suppose that it is not and there is $\eta\in \Omega^0(P\times_{\Adj} \mathfrak{g})$, $\eta\not=0$, such that $\eta$ is orthogonal to the image of $\op$. This implies that
\be
\int_M \sqrt{g} d^{2n}x \langle \eta, \op \epsilon\rangle_{\mathfrak{g}}=0~,
\ee
for every $\epsilon \in \Omega^0(P\times_{\Adj} \mathfrak{g})$.  Integrating by parts, this gives
\be
\int_M \sqrt{g} d^{2n}x \langle D_A^i D_{Ai}\eta-(D_{Ai} (\theta^i  \eta),  \epsilon\rangle_{\mathfrak{g}}=0~.
\ee
Thus
\be
D_A^i D_{Ai}\eta-D_{Ai} (\theta^i  \eta)=D_A^iD_{Ai}\eta-\theta^i D_{Ai}\eta=0~,
\ee
where in the last step we have used the Gauduchon condition $D_i\theta^i=0$.  The operator
\be
\op^\dagger\equiv D_A^i(D_A)_i\eta-\theta^i (D_A)_i~,
 \ee
 is the formal adjoint of $\op$. It can be shown that its kernel is trivial for $A\in \mathscr{A}_{\HE}$. The  proof is similar to that for $\op$ above and so $\eta=0$.  Thus $\op$ is onto.  Therefore the splitting (\ref{HEhv}) always exists and it is unique as expected from the definition of a connection.

The metric on $\mathscr{M}^*_{\HE}$ is defined by restricting that on $\mathscr{A}^*_{\HE}$ on the horizontal subspace.  In particular
\be
 (\alpha, \beta)_{\mathscr{M}^*_{\HE}}\equiv ( a^h, b^h)_{\mathscr{A}^*_{\HE}}=\int_{M^{2n}} d^{2n}x \sqrt{g} g^{-1} \langle a^h, b^h\rangle_{\mathfrak{g}}~,
\label{mmetric}
\ee
where $a^h$ and $b^h$ are the horizontal lifts of $\alpha$ and $\beta$, respectively, and $( a, b)_{\mathscr{A}^*_{\HE}}$ is the restriction of (\ref{Ametric}) on $\mathscr{A}^*_{\HE}$. Of course the metric $g$ used in the definition of $( a, b )_{\mathscr{A}^*_{\HE}}$ is that of the Hermitian manifold $M^{2n}$.  It is straightforward to prove that the metric $(\cdot, \cdot)_{\mathscr{M}^*_{\HE}}$ is Hermitian. This follows from the fact that if $a^h$ is a horizontal vector field so is $\mathcal{I} a^h$. Then
\be
(\mathcal{I}\alpha, \mathcal{I}\beta)_{\mathscr{M}^*_{\HE}}\equiv (\mathcal{I}a^h, \mathcal{I}b^h)_{\mathscr{A}^*_{\HE}}=( a^h, b^h)_{\mathscr{A}^*_{\HE}}=(\alpha, \beta)_{\mathscr{M}^*_{\HE}}~,
\ee
where  the Hermiticity of the metric $g$ of $M$ and (\ref{actI}) have been used.

Similarly, one can defined the 2-form
\be
\Omega_{\mathscr{M}^*_{\HE}}(\alpha, \beta)\equiv \Omega_{\mathscr{A}^*_{\HE}}(a^h, b^h)=\frac{1}{(n-1)!}\, \int_M \omega^{n-1}\wedge  \langle a^h\wedge  b^h \rangle_{\mathfrak{g}}~,
\label{mherm}
\ee
where $\Omega$ is given in (\ref{Aform}) but now restricted on $\mathscr{A}^*_{\HE}$ and $a^h$ and $b^h$ are the horizontal lifts of $\alpha$ and $\beta$, respectively. Using again that if $a^h$ and $b^h$ are horizontal vectors so are $\mathcal{I} a^h$ and $\mathcal{I} b^h$, one can verify that $\Omega_{\mathscr{M}^*_{\HE}}$ is the Hermitian form of the metric
$(\cdot, \cdot)_{\mathscr{M}^*_{\HE}}$. Note that the complex structure is integrable as it has ``constant'' components, i.e. it is invariant under the action of
$\Omega^1(P\times_{\Adj} \mathfrak{g})$ on the space of connections.

\subsection {Strong KT structure on $\mathscr{M}^*_{\HE}$}

\subsubsection{The exterior derivative of Hermitian form}

After defining a metric and a Hermitian form, we have prepared the ground  to show that $\mathscr{M}^*_{\HE}$ is a strong KT manifold that has  originally been demonstrated by L\"ubke and Teleman in \cite{lubke}.
To compute the exterior derivative of $\Omega_{\mathscr{M}^*_{\HE}}$ recall that the exterior derivative of   a 3-form $\chi$ on a manifold  is
\be
d\chi(a_1, a_2, a_3)=a_1 \chi(a_2, a_3)-\chi([a_1, a_2], a_3)+ \mathrm{cyclic~ in~} (a_1, a_2, a_3)~,
\label{dform}
\ee
where $a_1, a_2, a_3$ are vector fields.

Note that $\Omega_{\mathscr{A}}$ in (\ref{Aform}) is closed as there is no dependence of either the components of the form or of the tangent vectors on the base point $A$ that it is evaluated. Pulling this back with the inclusion of $\mathscr{A}^*_{\HE}$ in $\mathscr{A}$ it remains closed as the exterior derivative commutes with the pullback map. To continue, let us denote with $\cdot$ the action\footnote{Typically to compute a directional derivative, one has to consider all curves with tangent $a$ that pass through $A$.  However, $\mathscr{A}$ is an affine space and the restriction of the computation to straight lines suffices.} of a vector field $a$ on a function $f$ on ${\mathscr{A}^*_{\HE}}$, i.e.
\be
a\cdot f(A)\equiv \frac{d}{dt} f(A+t a)\vert_{t=0}~.
\label{actvf}
\ee
Then, the commutator $\lsq\cdot, \cdot\rsq$  of two vector fields on ${\mathscr{A}^*_{\HE}}$ is
\be
 \lsq a_1, a_2\rsq f\equiv a_1\cdot(a_2\cdot f)-a_2 \cdot (a_1\cdot f)~.
 \ee
Using (\ref{dform}), we find
\begin{align}
d\Omega_{\mathscr{M}^*_{\HE}}(\alpha_1, \alpha_2, \alpha_3)&\equiv \alpha_1\cdot \Omega_{\mathscr{M}^*_{\HE}}( \alpha_2, \alpha_3)- \Omega_{\mathscr{M}^*_{\HE}}(\lsq\alpha_1,\alpha_2\rsq,  \alpha_3)+ \mathrm{cyclic~ in~} (a^h_1, a^h_2, a^h_3)
\cr
&=a^h_1\cdot \Omega_{\mathscr{A}^*_{\HE}}(a^h_2, a^h_3)-\Omega_{\mathscr{A}^*_{\HE}}(\lsq a^h_1, a_2^h\rsq^h, a_3^h)+ \mathrm{cyclic~ in~} (a^h_1, a^h_2, a^h_3)
\cr
&=a^h_1\cdot \Omega_{\mathscr{A}^*_{\HE}}(a^h_2, a^h_3)-\Omega_{\mathscr{A}^*_{\HE}}(\lsq a^h_1, a_2^h\rsq-\lsq a^h_1, a_2^h\rsq^v, a_3^h)+ \mathrm{cyclic~ in~} (a^h_1, a^h_2, a^h_3)
\cr
&=d\Omega_{\mathscr{A}^*_{\HE}}(a_1^h, a_2^h, a^h_3)+\big(\Omega_{\mathscr{A}^*_{\HE}}(\lsq a^h_1, a_2^h\rsq^v, a_3^h)+ \mathrm{cyclic~ in~} (a^h_1, a^h_2, a^h_3)\big)
\cr
&=\Omega_{\mathscr{A}^*_{\HE}}(\lsq a^h_1, a_2^h\rsq^v, a_3^h)+ \mathrm{cyclic~ in~} (a^h_1, a^h_2, a^h_3)~,
\end{align}
where we have used that $\Omega_{\mathscr{A}^*_{\HE}}$ is closed. As $\lsq a^h_1, a_2^h\rsq^v$ is the vertical component of the commutator, there is a $\Theta$ such that
\be
\lsq a^h_1, a_2^h\rsq^v=- d_A\Theta(a^h_1, a^h_2)~.
\ee
$\Theta(\alpha_1, \alpha_2)\equiv \Theta(a^h_1, a^h_2)$ is  the curvature\footnote{By definition, the curvature of a principal bundle is the exterior derivative of the connection  evaluated along the horizontal directions and so it can be written as $\Theta(a_1, a_2)$. However, we shall maintain the notation $\Theta(a^h_1, a^h_2)$ for uniformity and as an explicit reminder of the horizontal projection.} of the principal bundle connection defined on  ${\mathscr{A}^*_{\HE}}$ with horizontality condition (\ref{horizon}).
$\Theta$ is skew-symmetric in the exchange of $a^h_1$ and  $a^h_2$ and takes values in the Lie algebra of the gauge group $\mathcal{G}^*$ as expected.

Continuing with the computation of the exterior derivative, we find that
\begin{align}
d\Omega_{\mathscr{M}^*_{\HE}}&(\alpha_1, \alpha_2, \alpha_3)=\Omega_{\mathscr{A}^*_{\HE}}(\lsq a^h_1, a_2^h\rsq^v, a_3^h)+ \mathrm{cyclic~ in~} (a^h_1, a^h_2, a^h_3)
\cr
&=\frac{1}{(n-1)!} \,\int_{M^{2n}} \omega^{n-1}\wedge \langle \lsq a^h_1, a_2^h\rsq^v\wedge  a^h_3\rangle_{\mathfrak{g}}+ \mathrm{cyclic~ in~} (a^h_1, a^h_2, a^h_3)
\cr
&=-\frac{1}{(n-1)!} \,\int_{M^{2n}} \omega^{n-1}\wedge \langle d_A\Theta (a_1^h, a_2^h) \wedge  a^h_3\rangle_{\mathfrak{g}}+ \mathrm{cyclic~ in~} (a^h_1, a^h_2, a^h_3)
\cr
&=\frac{1}{(n-1)!} \,\Big(\int_{M^{2n}} d\omega^{n-1}\wedge \langle \Theta (a_1^h, a_2^h),   a^h_3\rangle_{\mathfrak{g}}
\cr &\qquad\qquad+  \,\int_{M^{2n}} \omega^{n-1}\wedge \langle \Theta (a_1^h, a_2^h),   d_Aa^h_3\rangle_{\mathfrak{g}}\Big) + \mathrm{cyclic~ in~} (a^h_1, a^h_2, a^h_3)
\cr
&=\frac{1}{(n-1)!} \,\int_{M^{2n}} d\omega^{n-1}\wedge \langle \Theta (a_1^h, a_2^h),   a^h_3\rangle_{\mathfrak{g}} + \mathrm{cyclic~ in~} (a^h_1, a^h_2, a^h_3)~.
\label{domega}
\end{align}
The second term after the fourth equality    vanishes because $\omega^{n-1}\wedge d_A a^h_3=\omega^{n-1}\wedge d_A a_3=0$ as a consequence of (\ref{HEcond}) and the analysis below (\ref{HEcomp}).  Clearly if $d\omega^{n-1}=0$, then $d\Omega_{\mathscr{M}^*_{\HE}}=0$ and $\mathscr{M}^*_{\HE}$ will be a K\"ahler manifold.

\subsubsection{The KT structure on $\mathscr{M}^*_{\HE}$}

To make further progress, let us find an equation for $\Theta$.  Assuming that both $a_1$ and $a_2$ satisfy the horizontality condition (\ref{HEcone}), we have  $a_1\cdot \omega\llcorner d^c_A a_2-a_2\cdot \omega\llcorner d^c a_1=0$.  Evaluating this expression, we find that
\be
\omega \llcorner d_A^c \lsq a^h_1, a^h_2\rsq =2 g^{-1} [a^h_1, a^h_2]_{\mathfrak{g}}~.
\ee
 The component $\lsq a^h_1, a^h_2\rsq^h$ of the commutator  is in the kernel of the operator $\omega \llcorner d_A^c$. Thus one has that $\omega \llcorner d_A^c \lsq a^h_1, a^h_2\rsq^v =2 g^{-1} [a^h_1, a^h_2]_{\mathfrak{g}}$ or equivalently
\be
\omega \llcorner d_A^c d_A \Theta(a^h_1, a^h_2)=-2g^{-1} [a^h_1, a^h_2]_{\mathfrak{g}}\Longleftrightarrow \op \Theta(a^h_1, a^h_2)=2 g^{ij}[a^{h}_{1i}, a^h_{2j}]_{\mathfrak{g}}~.
\label{diftheta}
\ee
 The above equation uniquely  determines $\Theta$ because the operator $\op$ is invertible.  Observe also that $\Theta$ is an $(1,1)$-form on  $\mathscr{M}^*_{\HE}$ as the equations that determine $\Theta(\mathcal{I}a^h_1, \mathcal{I}a^h_2)$
and $\Theta(a^h_1, a^h_2)$ have the same source term $2 g^{-1} [a^{h}_1, a^h_{2}]_{\mathfrak{g}}$ and so they must be equal.

Using that $\Theta$ is an $(1,1)$-form, one also finds that
\begin{align}
d^c\Omega_{\mathscr{M}^*_{\HE}}&(\alpha_1, \alpha_2, \alpha_3)=\ii_{\mathcal{I}} d\Omega_{\mathscr{M}^*_{\HE}}(\alpha_1, \alpha_2, \alpha_3)
\cr
&=\frac{1}{(n-1)!} \,\int_{M^{2n}} d\omega^{n-1}\wedge \Big(\langle \Theta (a_1^h, a_2^h),   \mathcal{I} a^h_3\rangle_{\mathfrak{g}} + \mathrm{cyclic~ in~} (a^h_1, a^h_2, a^h_3)\Big)
\cr
&=-\frac{1}{(n-1)!} \,\int_{M^{2n}} d\omega^{n-1}\wedge \Big(\langle \Theta (a_1^h, a_2^h),  \ii_I a^h_3\rangle_{\mathfrak{g}} + \mathrm{cyclic~ in~} (a^h_1, a^h_2, a^h_3)\Big)
\cr
&=\frac{1}{(n-1)!} \,\int_{M^{2n}} d^c\omega^{n-1}\wedge \Big(\langle \Theta (a_1^h, a_2^h),    a^h_3\rangle_{\mathfrak{g}} + \mathrm{cyclic~ in~} (a^h_1, a^h_2, a^h_3)\Big)~.
\end{align}
Next define the 3-form $\mathcal {H}_{\mathscr{M}^*_{\HE}}$ on  $\mathscr{M}^*_{\HE}$ as
\begin{align}
\mathcal {H}_{\mathscr{M}^*_{\HE}}&(\alpha_1, \alpha_2, \alpha_3)=-d^c\Omega_{\mathscr{M}^*_{\HE}}(\alpha_1, \alpha_2, \alpha_3)
\cr
&
=-\frac{1}{(n-1)!} \,\int_{M^{2n}} d^c\omega^{n-1}\wedge \Big(\langle \Theta (a_1^h, a_2^h),   a^h_3\rangle_{\mathfrak{g}} + \mathrm{cyclic~ in~} (a^h_1, a^h_2, a^h_3)\Big)
\cr
&
=\frac{1}{(n-2)!} \,\int_{M^{2n}} H\wedge\omega^{n-2}\wedge \Big(\langle \Theta (a_1^h, a_2^h),   a^h_3\rangle_{\mathfrak{g}} + \mathrm{cyclic~ in~} (a^h_1, a^h_2, a^h_3)\Big)~.
\label{mtorsion}
\end{align}
It is clear that $\mathcal {H}_{\mathscr{M}^*_{\HE}}$ is a $(2,1)\oplus (1,2)$-form as  expected, because $\mathcal{I}$ is integrable, and follows from construction that $d \Omega_{\mathscr{M}^*_{\HE}}=\ii_{\mathcal{I}} \mathcal{H}_{\mathscr{M}^*_{\HE}}$. These imply, see \ref{sec:kt} and (\ref{doih}),  that the  complex structure $\mathcal {I}$ is covariantly constant with respect to the connection
$\h {\mathcal{D}}$ on $\mathscr{M}_{\HE}$ with torsion $\mathcal{H}_{\mathscr{M}^*_{\HE}}$, i.e.
\be
\h {\mathcal{D}}\mathcal {I}=0~.
\label{modkt}
\ee
Therefore,  $\mathscr{M}^*_{\HE}$ is a KT manifold.

It is worth noticing that $\mathcal {H}_{\mathscr{M}^*_{\HE}}$ depends only on the Lee form of $M^{2n}$. Indeed, (\ref{mtorsion}) can be rewritten as
\be
\mathcal {H}_{\mathscr{M}^*_{\HE}}(\alpha_1, \alpha_2, \alpha_3)=- \,\int_{M^{2n}}d^{2n}x\,\sqrt{g}\,\theta_i \, g^{ij}\,  \Big(\langle \Theta (a_1^h, a_2^h),   a^h_{3j}\rangle_{\mathfrak{g}} + \mathrm{cyclic~ in~} (a^h_1, a^h_2, a^h_3)\Big)~.
\label{ttorsion}
\ee
Therefore, if $\theta=0$, then $\mathscr{M}^*_{\HE}$ will be a K\"ahler manifold. The $\theta=0$ condition is equivalent to $d\omega^{n-1}=0$ found for the closure
$\Omega_{\mathscr{M}^*_{\HE}}$. This is a restriction on the Hermitian structure of $M^{2n}$. Nevertheless, there is a large class of Hermitian manifolds, which are not K\"ahler,  with $\theta=0$.  For all these manifolds,
$\mathscr{M}^*_{\HE}$ has a K\"ahler structure.

\subsubsection{The strong  KT structure on  $\mathscr{M}^*_{\HE}$}

To show that the KT structure defined in the previous section on $\mathscr{M}^*_{\HE}$ is strong, one has to demonstrate that $\mathcal {H}_{\mathscr{M}^*_{\HE}}$ is closed.  For this,  one can use the usual expression for the exterior derivative of a 3-form, as that of a 2-form in (\ref{dform}),
to find that
\begin{align}
d \mathcal {H}_{\mathscr{M}^*_{\HE}}&(\alpha_1, \alpha_2, \alpha_3, \alpha_4)=-   dd^c\Omega_{\mathscr{M}^*_{\HE}}(\alpha_1, \alpha_2, \alpha_3, \alpha_4)
\cr &
=-\frac{1}{(n-1)!} \,\int_{M^{2n}} d^c\omega^{n-1}\wedge \Big(\langle \Theta (a_1^h, a_2^h),  \lsq a^h_3, a^h_4\rsq -\lsq a^h_3, a^h_4\rsq^h\rangle_{\mathfrak{g}} + \dots\Big)
\cr
&=-\frac{1}{(n-1)!} \,\int_{M^{2n}} d^c\omega^{n-1}\wedge \Big(\langle \Theta (a_1^h, a_2^h),  -d_A\Theta(a_3^h, a_4^h)\rangle_{\mathfrak{g}} + \dots\Big)
\cr
&=\frac{1}{(n-1)!} \,\int_{M^{2n}} d^c\omega^{n-1}\wedge \Big(\big(\langle \Theta (a_1^h, a_2^h),  d_A\Theta(a_3^h, a_4^h)\rangle_{\mathfrak{g}} +\langle \Theta (a_3^h, a_4^h),  d_A\Theta(a_1^h, a_2^h)\rangle_{\mathfrak{g}}\big)
\cr
&\qquad\qquad\qquad+ \mathrm{cyclic~ in~} (a^h_1, a^h_2, a^h_3)\Big)
\cr
&=\frac{1}{(n-1)!} \,\int_M d^c\omega^{n-1}\wedge \Big(d \big(\langle \Theta (a_1^h, a_2^h),  \Theta(a_3^h, a_4^h)\rangle_{\mathfrak{g}}\big) + \mathrm{cyclic~ in~} (a^h_1, a^h_2, a^h_3)\Big)
\cr
&=\frac{1}{(n-1)!} \,\int_M d d^c\omega^{n-1}\Big(\langle \Theta (a_1^h, a_2^h),  \Theta(a_3^h, a_4^h)\rangle_{\mathfrak{g}} + \mathrm{cyclic~ in~} (a^h_1, a^h_2, a^h_3)\Big)~.
\label{dhm}
\end{align}
where the expressions with the dots contain  six terms in total and arise from the skew-symmetrisation in $(a^h_1, a^h_2, a^h_3, a^h_4)$ and the Bianchi identity $d\Theta(a_1^h, a_2^h, a_3^h)=0$ has been used for $\Theta$.  Furthermore after the second equality, the first term in $\lsq a^h_3, a^h_4\rsq -\lsq a^h_3, a^h_4\rsq^h$ arises from acting
on $\langle \Theta (a_1^h, a_2^h),   a^h_4\rangle_{\mathfrak{g}}$ with $a_3^h$ and after considering the skew-symmetrisation in the vector fields while the second term arises from the definition of the exterior derivative of the 3-form.

It is clear from (\ref{dhm}) that if $d d^c\omega^{n-1}=0$, i.e. the metric $g$ is Gauduchon, $\Omega_{\mathscr{M}^*_{\HE}}$ is $\partial\bar\partial$-closed. This in turn implies that $d\mathcal{H}_{\mathscr{M}^*_{\HE}}=0$ and so $\mathscr{M}^*_{\HE}$ is a strong KT manifold.

\section{Covariantly constant vector fields on  $\mathscr{M}_{\HE}$}\label{sec:3}

\subsection{ KT manifolds with holomorphic and covariantly constant vector fields}\label{sec:ktvf}

Before we proceed with the investigation of $\h{\mathcal{D}}$-covariant vector fields on $\mathscr{M}_{\HE}$, it is useful to
summarise some of the properties of $\h\nabla$-covariantly constant vector fields on a KT manifold $M^{2n}$, see also \cite{gp1}.  Suppose that  $M^{2n}$ is a KT manifold, see \ref{sec:kt}, that admits a $\hat \nabla$-covariantly constant  vector field $X$
\be
\h\nabla_i X^j=0~,
\ee
usually written as $\h\nabla X=0$ for shorthand.
The covariant constancy condition $\h\nabla X=0$ implies that
\be
\mathcal{L}_X g=0~,~~~dX^\flat=\ii_X H~,
\ee
i.e. $X$ is Killing and the exterior derivative $dX^\flat$ of the associated 1-form $X^\flat$ to $X$, $X^\flat(Z)\equiv g(X,Z)$, is given in terms of the inner product of $X$ with torsion of $H$. These conditions do not imply that $H$ is invariant under the action of $X$.  However, if the KT structure is strong, $dH=0$, then
$\mathcal{L}_X H=d\ii_X H+\ii_X d H=0$ and so $H$ is invariant under the action of $X$.

 The $\hat \nabla$-covariant constancy condition on $X$ does not imply that $X$ is holomorphic, i.e. ${\mathcal L}_X I$ may not vanish.  However, the conditions $\h\nabla X=0$ and $\h\nabla I=0$ imply  that $X$ is holomorphic, ${\mathcal L}_X I=0$, provided that in addition $\ii_X H$ is an $(1,1)$-form. Indeed, as $X$ is Killing, it suffices to evaluate the Lie derivative of the Hermitian form as
\begin{align}
{\mathcal{L}}_X\omega_{ij}&= X^k D_k\omega_{ij}+D_i X^k \omega_{kj}-D_j X^k \omega_{ki}\cr
&=X^k \h \nabla_k \omega_{ij}+\h\nabla_i X^k \omega_{kj}-\h\nabla_j X^k \omega_{ki}+X^k H_{k\ell j} I^\ell{}_i-X^k H_{k\ell i} I^\ell{}_j
\cr
& =(\ii_I\ii_X H)_{ij}~.
\end{align}
This vanishes, iff $\ii_X H$ is a (1,1)-form\footnote{On strong KT manifolds $M^{2n}$, $dH=0$,  with a $\h\nabla$-covariantly constant vector field $X$, the $\ii_X H^{(2,0)\oplus (0,2)}$ components of $\ii_XH$ are $\h\nabla$-covariantly constant.  This follows from the Bianchi identity (\ref{bia1}) together with
$\h R_{ijk\ell} X^\ell=0$, which arises as integrability condition of $\h\nabla X=0$, and the restriction on the holonomy of $\h\nabla$ to be included in $U(n)$.} on $M^{2n}$.

If a KT manifold admits a $\h\nabla$-covariantly constant vector field $X$, it also admits a second one\footnote{The minus sign in the expression of $Y$ in terms of $X$ has been added so that the Hermitian form in  (\ref{gktdec}) below decomposes as stated. Otherwise, one has to add a minus sign in the first term involving $X^\flat$ and $Y^\flat$ in the decomposition of $\omega$.} $Y\equiv - I X$ as both $I$ and $X$ are $\h\nabla$-covariantly constant.  $Y$  is linearly independent from $X$. Furthermore, the commutator of any two $\h\nabla$-covariantly constant  vector fields $Z, W$ is given by
\be
[Z, W]=\ii_Z \ii_W H^\flat~.
\label{zwcom}
\ee
For strong KT manifolds, the commutator $[X, Y]$ is again a $\h\nabla$-covariantly constant vector field as a consequence of the Bianchi identity (\ref{bia1})
  and the integrability conditions $\h R_{ijk\ell} X^\ell=\h R_{ijk\ell} Y^\ell=0$.
Provided that the repeated commutators do not vanish and the resulting vector fields are linearly independent, this can continue till $M^{2n}$ becomes locally isometric to a group manifold.

 {\sl Suppose that $X$ is $\h\nabla$-covariantly constant and holomorphic}\footnote{It should be pointed out that if $X$ is Killing and holomorphic, and so it leaves invariant $H$,  it is not necessarily $\h\nabla$-covariantly constant.  An example of such (strong) HKT geometry has been presented in \cite{gp2, tod}.  We shall encounter this phenomenon in the investigation of vector fields on $\mathscr{M}_{\HE}$ below.}. {\sl In such a case,    $Y=-IX$ is also holomorphic and $[X,Y]=0$.}   First, for $Y$ to be holomorphic $\ii_Y H$ must be an $(1,1)$-form.  Indeed, we have
\be
Y^\ell H_{\ell ij}= -I^\ell{}_k X^k H_{\ell ij}= I^\ell{}_j X^k H_{\ell k i}+I^\ell{}_i X^k H_{\ell j k }-  X^k H_{mnp} I^m{}_k I^n{}_i I^p{}_j~,
\ee
where the second equality follows as a consequence of the integrability of the complex structure $I$, which implies that $H$ is a $(2,1)\oplus (1,2)$-form, see (\ref{integII}).
The above identity can be rearranged as
\be
Y^\ell H_{\ell ij}- Y^\ell H_{\ell np} I^n{}_i I^p{}_j= -(\ii_I \ii_X H)_{ij}=0~,
\ee
where the last equality is a consequence of the holomorphicity of $X$.  Therefore, $\ii_Y H$ is an $(1,1)$-form and so $Y$ is holomorphic. This also follows from simply observing that in complex coordinates the components of $Y$ are appropriately holomorphic.

Next  to prove the second part of the statement  that  $[X,Y]=0$ , notice that
\be
[X, Y]^\flat_i= H_{i jk} Y^j X^k=-H_{ijk} I^j{}_\ell X^\ell X^k=- H_{\ell jk} I^j{}_i X^\ell X^k=0~,
\ee
where the third equality follows because $X$ is holomorphic and so $\ii_X H$ is an $(1,1)$-form.

The geometry of KT manifolds that admit two holomorphic $\h\nabla$-covariantly constant vector fields $X,Y$, $(Y=-IX)$, can be modelled on that of a principal $T^2$ fibration. As $X,Y$ are orthogonal and their length is constant, after a possible rescaling, one can choose them to have length 1. Since, they commute $\ii_X \ii_Y H=0$.  Moreover,  $\ii_X H$ and $\ii_Y H$ are $(1,1)$-forms as $X$ and $Y$ are holomorphic. The KT structure on $M^{2n}$ can be written as
\begin{align}
&g=(X^\flat)^2+(Y^\flat)^2+ g^\perp~,
\cr
&H=X^\flat\wedge dX^\flat+Y^\flat\wedge dY^\flat+ H^\perp~,
\cr
&\omega=X^\flat\wedge Y^\flat+ \omega^\perp~,
\label{gktdec}
\end{align}
where $g^\perp$ is orthogonal to the plane spanned by $X$ and $Y$, $\ii_X H^\perp =\ii_Y H^\perp=0$ and similarly for $\omega^\perp$.  Furthermore, $g^\perp$, $H^\perp$ and $\omega^\perp$ are invariant under the action of both $X$ and $Y$.  Therefore, they project ``down'' on the orbit space $B^{2n-2}$ of the $T^2$ fibration, especially whenever $X$ and $Y$, or a linear combination of them, has closed orbits in $M^{2n}$. In fact $B^{2n-2}$ is a KT manifold with these data. As, the exterior derivative of $H^\perp$ is
\be
d H^\perp=- dX^\flat\wedge dX^\flat-dY^\flat\wedge dY^\flat~,
\label{dhktdec}
\ee
$B^{2n-2}$ may not admit a strong KT structure.  If $M^{2n}$ is a torus fibration over $B^{2n-2}$, the consistency of the above equation requires that $c_1^2=0$, where $c_1$ is the first Chern class.

So far we have investigated the geometry of the KT manifold $M^{2n}$ in the presence of one holomorphic $\h\nabla$-covariantly constant vector field. This  can be easily generalised to describe the geometry of $M^{2n}$ in the presence of several holomorphic $\h\nabla$-covariantly constant vector fields $( X_\alpha; \alpha=1, \dots, 2q)$; their number is even as their Lie algebra $\mathfrak{k}$ must be closed under the operation $X_\alpha\rightarrow I(X_\alpha)$.  It turns out that these vector fields must commute under Lie brackets. To prove this,  without loss of generality, choose $( X_\alpha; \alpha=1, \dots, 2q)$   to be orthonormal $g(X_\alpha, X_\beta)=\delta_{\alpha\beta}$. The structure constants of $\mathfrak{k}$ are given by $H_{\alpha\beta\gamma}= H(X_\alpha, X_\beta, X_\gamma)$.  The requirement that $X_\alpha$ are holomorphic implies the condition
\be
{\mathcal L}_{ X_\alpha} \omega_{ \beta\gamma }=H^\delta{}_{\alpha\beta}  \omega_{  \delta \gamma}-H^\delta{}_{\alpha\gamma}  \omega_{  \delta\beta}=0~,
\ee
see also (\ref{complexlie}) below for more explanation, where $\omega_{\alpha\beta}=\omega(X_\alpha, X_\beta)$ and can be thought as the Hermitian form  of $\mathfrak{k}$ with $(I^\alpha{}_\beta)$ the associated complex structure, $\omega_{\alpha\beta}=\delta_{\alpha\gamma} I^\gamma{}_\beta$. The condition above implies that $(I^\alpha{}_\beta)$ must be bi-invariant.  Lie algebras of compact groups\footnote{This is because the existence of such complex structure on the Lie algebra would have implied that the group is a K\"ahler manifold that cannot be the case for non-abelian compact groups.} do not admit such complex structures unless they are abelian.    Thus, the results obtained in this section can be generalised only to holomorphic  $\h\nabla$-covariantly constant vector fields that commute under Lie brackets.  In such a case, the geometry of $M^{2n}$ can modelled after  that of a holomorphic principal $T^{2p}$ fibration\footnote{Perhaps a more appropriate notion to describe $M^{2n}$ is that of a foliation with $B^{2(n-p)}$ identified as the space of leaves. This is also the case later in the investigation of the geometry of moduli spaces. But as it is more intuitive to think in terms of bundles, we shall keep the bundle terminology to describe the local geometry.} over a KT manifold $B^{2(n-p)}$. The decomposition of metric, Hermitian form and torsion in (\ref{gktdec}) and the expression for the exterior derivative (\ref{dhktdec}) generalise in a straightforward way to this more general case.

\subsection{Vector fields on the moduli space}\label{vsec1}

Given a vector field $X$ on $M^{2n}$, one can define the tangent vector
\bea
a_X(A)= \ii_X F(A)-D_A \epsilon~,
\eea
 on $\mathscr{A}$ at the point $A\in  \mathscr{A}$  up to a  gauge transformation with infinitesimal parameter $\epsilon$.

Here, {\sl  we shall demonstrate that if $X$ is holomorphic and $\h\nabla$-covariantly constant, and $A$ is a Hermitian-Einstein connection over the KT manifold $M^{2n}$, then $\ii_X F$ is tangent to
$\mathscr{A}_{\HE}$ at $A$ and horizontal.}

In what follows below, we shall focus on KT manifolds $M^{2n}$ that admit a single holomorphic $\h\nabla$-covariantly constant vector field $X$ and therefore another given by $Y=-IX$. The generalisation of the results below to several  holomorphic $\h\nabla$-covariantly constant vector fields, which necessarily commute under Lie brackets, is straightforward.

 \subsubsection{$\ii_X F$ is tangent to $\mathscr{A}_{\HE}$}

 For $\ii_X F$ to be tangent to $\mathscr{A}_{\HE}$, it has to satisfy the conditions (\ref{HEcond}).  This means that $d_A\ii_X F$  has to be an $(1,1)$-form on $M^{2n}$ and $\omega\llcorner d_A\ii_XF=0$.

 Before we proceed with the proof, let us set
 \be
 \h\nabla_{Ai} F_{jk}\equiv \h\nabla_i F_{jk}+ [A_i, F_{jk}]_\mathfrak{g}~.
 \ee
 It is convenient to use this derivative in the calculations below as both $X$ and $I$ are $\h\nabla$-covariantly constant.

 To prove that $d_A\ii_X F$ is  (1,1)-form, we have
\begin{align}
 \big(d_A\ii_X F\big)_{mn} I^m{}_i I^n{}_j&=\big((d_A\ii_X+\ii_X d_A) F\big)_{mn} I^m{}_i I^n{}_j
 =\big(d_A\ii_X+\ii_X d_A\big)(F_{mn}  I^m{}_i I^n{}_j)
 \cr
 &=
 \big((d_A\ii_X+\ii_X d_A)F\big)_{ij}=\big(d_A\ii_X F\big)_{ij}~,
 \end{align}
 where we have used repeatedly the Bianchi identity of $F$, the holomorphicity of $X$, $\mathcal{L}_X I=0$,  and that $F$ is an $(1,1)$-form. Therefore, $d_A \ii_X F$ is an $(1,1)$-form.

 It remains to demonstrate the second condition in (\ref{HEcond}).  For this, we have
 \begin{align}
 \omega\llcorner (d_A \ii_X F)&= \omega\llcorner \big((d_A \ii_X+\ii_X d_A) F\big)
 \cr
 &
 =-\frac{1}{2} {\mathcal{L}}_X \omega^{ij} F_{ij}+\frac{1}{2}X^kD_{Ak}(\omega^{ij} F_{ij})=0~,
 \end{align}
 as $X$ is Killing and holomorphic, $\mathcal{L}_X \omega=0$,  and $\omega^{ij} F_{ij}$ satisfies the Hermitian-Einstein condition (\ref{HEcond1}) with $\lambda$  (covariantly) constant because it commutes with all elements of $\mathfrak{g}$.

 \subsubsection {$\ii_X F$ is horizontal}

It remains to demonstrate that $\ii_X F$ is horizontal.  A consequence of the Bianchi identity for $F$  and the Hermitian-Einstein conditions (\ref{HEcond1}) is that
\be
\h\nabla_A^i F_{ij}+\theta^i F_{ij}=0~.
\ee
Contracting this equation  with $X$, one has that
\be
\h\nabla_A^i (F_{ij} X^j)+\theta^i F_{ij} X^j=0 \Longleftrightarrow  D_A^i(X^j F_{ji})+ \theta ^i (X^j F_{ji})=0~,
\label{horiz}
\ee
where we have used that $X$ is $\h\nabla$-covariantly constant. This clearly implies the horizontality condition for $\ii_X F$.

Therefore, we have demonstrated that if $X$ is a holomorphic $\h\nabla$-covariantly constant\footnote{A similar statement can be shown under weaker assumptions. In particular,  if $X$ is a holomorphic Killing vector field on $M^{2n}$, i.e. $X$ is not necessarily $\h\nabla$-covariantly constant, then there is a gauge transformation $\eta$ such that $\ii_X F-d_A\eta$ is tangent to $\mathscr{A}_{\HE}$ and horizontal, where $\eta$ is determined by the equation  $\mathcal{O}\eta=-F_{ij} \h\nabla^i X^j$. } vector field on $M^{2n}$, it induces a vector field $\alpha_{{}_X}$ on
$\mathscr{M}^*_{\HE}$ such that its horizontal lift on the tangent space of $\mathscr{A}^*_{\HE}$ is
\be
a_{{}_X}^h=\ii_X F~.
\ee
Sometimes is more convenient below to use the notation $a_{{}_X}^h$ instead of $\ii_X F$ for uniformity.

To conclude this section, let $X,Z$  be holomorphic $\h\nabla$-covariantly constant vector fields on $M^{2n}$.  The associated vector fields $\alpha_{{}_X}$ and $\alpha_{{}_Z}$ on the moduli space $\mathscr{M}^*_{\HE}$ have horizontal lifts $a_{{}_X}^h=\ii_X F$ and $a_{{}_Z}^h=\ii_Z F$. Clearly, the commutator vector field $[X,Z]$ is holomorphic. Moreover, if $M^{2n}$ is a strong KT manifold, $[X,Z]$ will be $\h\nabla$-covariantly constant. As a result, $\ii_{[X,Z]}F$ is tangent to $\mathscr{A}^*_{\HE}$ and horizontal.  Furthermore, the commutator $\lsq \alpha_{{}_X}, \alpha_{{}_Z}\rsq$    has  horizontal lift
\be
\lsq \ii_X F, \ii_Z F\rsq^h=(\ii_Z d_A \ii_X F-\ii_X d_A \ii_Z F)^h=(\mathcal {L}_Z \ii_X F -\ii_X \mathcal {L}_Z F-d_A \ii_Z \ii_X F)^h=-\ii_{[X,Z]} F~,
\label{commutxy}
\ee
where we have used one of the Lie derivative identities.
Therefore, the vector field $\lsq \alpha_{{}_X}, \alpha_{{}_Z}\rsq$ has horizontal lift $-\ii_{[X,Z]} F$ and so the commutators  of such vector fields on $\mathscr{M}^*_{\HE}$ are determined from those of the associated vector fields on $M^{2n}$.

\subsection{Covariantly constant and KT invariant vector fields on the moduli space}\label{vsec2}

The main objective of this section is to show that if $X$ is a holomorphic $\h\nabla$-covariantly constant vector field on $M^{2n}$, then the induced vector field $\alpha_{{}_X}$ on ${\mathscr{M}}^*_{\HE}$ is Killing and holomorphic.  Moreover, it will also be  $\h{\mathcal {D}}$-covariantly constant  provided that the 2-form
\be
X^\flat\wedge \theta~,
\ee
is (1,1) with respect to the complex structure $I$ on $M^{2n}$, where $\h{\mathcal {D}}$ is the connection on ${\mathscr{M}}^*_{\HE}$ with torsion $\mathcal{H}$, see (\ref{modkt}).  This statement will be proven in the subsection \ref{ccvf} below.

\subsubsection{A key lemma}\label{sec:key}

To prove the main result of this section, we require the formula
\be
\Theta (a^h, a^h_{{}_X})=-\Theta (a^h_{{}_X}, a^h)=\ii_X a^h~,
\label{TF}
\ee
 where $X$ is a holomorphic $\h\nabla$-covariantly constant vector field\footnote{It should be noted that if $X$ is holomorphic and Killing,  but not necessarily $\h\nabla$-covariantly constant,  so that $a_X^h=\ii_X F-d_A\eta$ with $\mathcal{O}\eta=F_{ij}\h\nabla^i X^j$, then $\Theta(a^h, a_X^h)=\ii_X a^h+ a^h\cdot \eta$. We shall not pursue this further here but most of the properties of the moduli spaces associated with holomorphic $\h\nabla$-covariant constant vector fields $X$ can be generalised to holomorphic Killing vector fields.} and so  $a^h_{{}_X}=\ii_X F$.  A similar formula has been derived in \cite{witten}   for  instantons   using the special properties of vector fields on $S^3\times S^1$.
To prove (\ref{TF}), act with the tangent vector $a$ of $\mathscr{A}^*$ on (\ref{horiz}) to find
\be
[a^j, X^i F_{ij}]_{\mathfrak{g}}+ D_A^j(X^i (d_A a)_{ij})+ \theta^j (d_A a)_{ij} X^i=0~.
\label{com1}
\ee
First notice that
\be
\theta^j (d_A a)_{ij} X^i= X^i D_{Ai} (\theta^j a_j)-\theta^j D_{Aj}(a_i X^i)~,
\ee
where we have used that $\mathcal{L}_X\theta=0$.
Then,
\begin{align}
D_A^j(X^i (d_A a)_{ij})&=X^i D_{Aj} D_{Ai} a^j- D_A^2 (a_i X^i)+ a_i D^2 X^i
\cr
&
=  X^i D_{Ai} D_{Aj} a^j +X^i R_{ij} a^j+ [a^j, F_{ij} X^i]_{\mathfrak{g}}- D_A^2 (a_i X^i)- a^i R_{ij} X^j
\cr
&
= X^i D_{Ai} D_{Aj} a^j+[a^j, F_{ij} X^i]_{\mathfrak{g}}-  D_A^2 (a_i X^i)~,
\end{align}
where we have used that $X$ is Killing and so $D^2 X_i= -R_{ij} X^j$. Putting these terms in (\ref{com1}), we find that
\be
-D_A^2 (a_i X^i)-\theta^j D_{Aj} (a_i X^i)+ 2 [a^j, X^i F_{ij}]_{\mathfrak{g}}+ X^i D_{Ai} (D_{Aj} a^j+\theta^j a_j)=0~.
\label{hconxf}
\ee
If the tangent vector $a$ satisfies the horizontality condition (\ref{horizon}), $D_{Aj} a^j+\theta^j a_j=0$,  then
\be
{\mathcal O} (a^h_i X^i)= 2 [a^{hj}, X^i F_{ij}]_{\mathfrak{g}}=2 g^{-1} [a^h, a^h_{{}_X}]_{\mathfrak{g}}~.
\ee
As the operator ${\mathcal O}$ is invertible,  this together with the differential equation that determines  $\Theta$ in (\ref{diftheta}) imply (\ref{TF}).

\subsubsection{Covariantly constant vector fields on ${\mathscr{M}}^*_{\HE}$}\label{ccvf}

From the properties  of $\h\nabla$-covariantly constant vector fields on KT manifolds  in section \ref{sec:ktvf}, one deduces that in order to show that $\alpha_{{}_X}$ is $\h{\mathcal{D}}$-covariantly constant on ${\mathscr{M}}^*_{\HE}$, it suffices to show that $\alpha_{{}_X}$ is Killing and $d\alpha^\flat_{{}_X}=\ii_{\alpha_{{}_X}} \mathcal{H}$.

Let us begin with the proof of the second condition first. It is clear from (\ref{mtorsion}) that
\begin{align}
\mathcal {H}&(\alpha_{{}_X}, \alpha_2, \alpha_3)=-\frac{1}{(n-1)!} \,\int_M d^c\omega^{n-1}\wedge \big(\langle \Theta (\ii_X F, a_2^h),   a^h_3\rangle_{\mathfrak{g}} -
\langle \Theta (\ii_X F, a_3^h),   a^h_2\rangle_{\mathfrak{g}}+ \langle \Theta ( a^h_2, a_3^h),  \ii_X F \rangle_{\mathfrak{g}}\big)
\cr
&
=-\frac{1}{(n-1)!} \,\int_M d^c\omega^{n-1}\wedge \big(-\langle \ii_Xa_2^h,   a^h_3\rangle_{\mathfrak{g}} +
\langle \ii_X  a_3^h,   a^h_2\rangle_{\mathfrak{g}}+ \langle \Theta ( a^h_2, a_3^h),  \ii_X F \rangle_{\mathfrak{g}}\big)
\cr
&
=\frac{1}{(n-1)!} \,\int_M \omega^{n-1}\wedge \big(\langle d_A\ii_Xa_2^h\wedge \ii_I   a^h_3\rangle_{\mathfrak{g}} -
\langle d_A\ii_X  a_3^h\wedge \ii_I  a^h_2\rangle_{\mathfrak{g}}-\langle d_A\Theta ( a^h_2, a_3^h)\wedge  \ii_I\ii_X F \rangle_{\mathfrak{g}}\big)
\cr
&= \int_{M^{2n}} d^{2n}x \sqrt g g^{-1}\big( \langle D_{A} (\ii_X a_2^h), a_3^h\rangle_{\mathfrak{g}}-\langle D_{A} (\ii_X a_3^h), a_2^h\rangle_{\mathfrak{g}}-\langle d_A\Theta ( a^h_2, a_3^h),  \ii_X F \rangle_{\mathfrak{g}}\big)
\cr
&= \int_{M^{2n}}d^{2n}x \sqrt g  \big( (X\wedge \theta^\flat)^{ij} \langle a^h_{2i}, a^h_{3j}\rangle_{\mathfrak{g}}   -g^{-1} \langle d_A\Theta ( a^h_2, a_3^h),  \ii_X F \rangle_{\mathfrak{g}}\big)
\cr
&= \int_{M^{2n}}d^{2n}x \sqrt g  \big( (X\wedge \theta^\flat)^{ij} \langle a^h_{2i}, a^h_{3j}\rangle_{\mathfrak{g}}   -\langle\Theta ( a^h_2, a_3^h),  \ii_{\theta^\flat}\ii_X F \rangle_{\mathfrak{g}}\big)~,
\label{HF}
\end{align}
where $\theta^\flat=g^{ij} \theta_i \partial_j$.
On the other hand the associated 1-form $\mathcal{F}_X$ to the vector field $\ii_X F$ with respect to the metric $\mathcal{G}$ on the moduli space ${\mathscr{M}}^*_{\HE}$ is
\begin{align}
\alpha_{{}_X}^\flat(\alpha)&\equiv \mathcal{F}_X(a^h)\equiv \int_{M^{2n}} \sqrt{g} g^{-1} \langle \ii_X F, a^h\rangle_{\mathfrak{g}}
\cr
&=\frac{1}{(n-2)!}\int_{M^{2n}} \omega^{n-2}\wedge \langle F\wedge a^h\rangle_{\mathfrak{g}}\wedge X-\frac{1}{(n-1)!}\int_{M^{2n}} \omega^{n-1}\wedge \langle \Lambda, a^h\rangle_{\mathfrak{g}}\wedge X~,
\end{align}
where we have used that $F$ is a (1,1)-form.
Thus the exterior derivative is
\begin{align}
d\mathcal{F}_X&(a^h_2, a^h_3)=\int_{M^{2n}} d^{2n}x \sqrt{g} g^{-1}\big( a_2^h \cdot \langle \ii_X F, a_3^h\rangle_{\mathfrak{g}}-a_3^h \cdot \langle \ii_X F, a_2^h\rangle_{\mathfrak{g}}- \langle \ii_X F, \lsq a_2^h, a_3^h\rsq^h \rangle_{\mathfrak{g}}\big)
\cr
&=\int_{M^{2n}} d^{2n}x \sqrt{g} g^{-1}\big(   \langle \ii_X d_A a_2^h, a_3^h\rangle_{\mathfrak{g}}-  \langle \ii_X d_Aa_3^h, a_2^h\rangle_{\mathfrak{g}}+ \langle \ii_X F, \lsq a_2^h, a_3^h\rsq^v \rangle_{\mathfrak{g}}\big)
\cr
&=\int_{M^{2n}} d^{2n}x \sqrt{g} g^{-1}\big(   \langle \ii_X d_A a_2^h, a_3^h\rangle_{\mathfrak{g}}-  \langle \ii_X d_Aa_3^h, a_2^h\rangle_{\mathfrak{g}}- \langle \ii_X F, d_A\Theta(a_2^h, a_3^h) \rangle_{\mathfrak{g}}\big)~.
\label{interm}
\end{align}
To continue, let us evaluate the first two terms in the last  equation above. Indeed suppressing the horizontality label on the vector fields $a_2$ and $a_3$ to simplify the expressions, we have
\begin{align}
&\langle\ii_X d_A a_{2i}, a_3^i\rangle_{\mathfrak{g}}- (a_3, a_2)=\langle(d_A a_2)_{mn} I^m{}_i X^i I^n{}_j, a_3^j \rangle_{\mathfrak{g}} -(a_3,a_2)
\cr
&
= \langle Y^i (d_A a_2)_{in} I^n{}_j, a_3^j\rangle_{\mathfrak{g}}- (a_3,a_2)
\cr
&
= \Big(Y^i D_{i} \langle I^{jk} a_{2j}, a_{3k}\rangle_{\mathfrak{g}}- \langle a_{2n}, Y^i D_i I^n{}_j a_3^j\rangle_{\mathfrak{g}}-\langle  I^n{}_j a_{2n},Y^i  D_{Ai} a_3^j
\rangle_{\mathfrak{g}}\cr
&
- \langle Y^i D_{An} a_{2i},  I^n{}_j a_3^j\rangle_{\mathfrak{g}}\Big)- (a_3,a_2)
\cr
&=Y^i D_{i} \langle I^{jk} a_{2j}, a_{3k}\rangle_{\mathfrak{g}}- Y^iD_n \langle a_{2i}, I^n{}_j a_3^j\rangle_{\mathfrak{g}}+Y^iD_n \langle a_{3i}, I^n{}_j a_2^j\rangle_{\mathfrak{g}}
\cr
&\qquad\qquad+ \langle Y^i a_{2i},  D_n I^n{}_j a_3^j\rangle_{\mathfrak{g}}-\langle Y^i a_{3i}, D_n I^n{}_j a_2^j\rangle_{\mathfrak{g}}
\cr
&
=Y^i D_{i} \langle I^{jk} a_{2j}, a_{3k}\rangle_{\mathfrak{g}}-D_n \langle Y^i a_{2i}, I^n{}_j a_3^j\rangle_{\mathfrak{g}} +D_n \langle Y^i a_{3i},  I^n{}_j a_2^j \rangle_{\mathfrak{g}}
\cr
&\qquad\qquad+ \langle Y^i a_{2i},  D_n I^n{}_j a_3^j\rangle_{\mathfrak{g}}-\langle Y^i a_{3i},  D_n I^n{}_j a_2^j\rangle_{\mathfrak{g}}~,
\end{align}
where we have used that $d_A a^h_2$ and $d_A a^h_3$ are (1,1)-forms,  $\omega \llcorner d_A a_2=\omega \llcorner d_A a_3=0$ and that $dY$ is a (1,1)-form as $Y=-IX$ is holomorphic.  The divergence of $I$ can be expressed in terms of the Lee form using (\ref{leef}). Moreover, note that $Y^iD_i I^j{}_k= Y^i \h\nabla_i I^j{}_k=0$  because $\ii_Y H$ is a (1,1)-form and $\h\nabla I=0$.

Substituting the expression above into (\ref{interm}), after reinstating the horizontality labels on $a_2$ and $a_3$, and   integrating out the surface terms, we find that
\begin{align}
d\mathcal{F}_X(a^h_2, a^h_3)&=\int_{M^{2n}} d^{2n}x \sqrt{g} \big(  (IX\wedge I\theta^\flat)^{ij} \langle a_{2i}^h, a_{3j}^h\rangle_{\mathfrak{g}}- g^{-1} \langle \ii_X F, d_A\Theta(a_2^h, a_3^h) \rangle_{\mathfrak{g}}\big)
\cr
& =\int_{M^{2n}} d^{2n}x \sqrt{g} \big(  (IX\wedge I\theta^\flat)^{ij} \langle a_{2i}^h, a_{3j}^h\rangle_{\mathfrak{g}}- \langle\Theta ( a^h_2, a_3^h),  \ii_\theta\ii_X F \rangle_{\mathfrak{g}}\big)~,
\label{dF}
\end{align}
where $(I\theta^\flat)^i=I^i{}_j\theta^j$.  On comparing (\ref{dF}) with (\ref{HF}), we conclude that
\be
d\alpha^\flat_{{}_X}=\ii_{\alpha_{{}_X}} \mathcal{H}~,
\ee
 provided that
\be
X^\flat\wedge\theta=\ii_I X^\flat\wedge \ii_I \theta~.
\label{xtii}
 \ee
 This is the condition for  $X^\flat\wedge\theta$ to be  an (1,1)-form with respect to $I$. Therefore, there is an additional condition on the vector field $X$ and on the Lee form $\theta$ of the KT structure of $M^{2n}$ for the induced vector field $\alpha_{{}_X}$ on ${\mathscr{M}}_{\HE}^*$ to be $\h{\mathcal{D}}$-covariantly constant.
 The condition (\ref{xtii}) together with the condition that $\Theta$ is an $(1,1)$-form on the moduli space ${\mathscr{M}}_{\HE}^*$ imply that $d\mathcal{F}_X$, and so $\ii_{\alpha_{{}_X}} \mathcal{H}$,  is (1,1)-form on ${\mathscr{M}}_{\HE}^*$ as it may have been expected.

 It remains to demonstrate that $\alpha_{{}_X}$ is Killing. For this, we use the formula
\begin{align}
\mathcal{L}_{\alpha_{{}_X}} ( \alpha_1 ,& \alpha_2)_{{\mathscr{M}}^*_{\HE}}=\alpha_{{}_X} \cdot ( \alpha_1, \alpha_2)_{{\mathscr{M}}^*_{\HE}}-(\lsq \alpha_{{}_X}, \alpha_1\rsq, \alpha_2)_{{\mathscr{M}}^*_{\HE}}-( \alpha_1, \lsq \alpha_{{}_X}, \alpha_2\rsq)_{{\mathscr{M}}^*_{\HE}}
\cr
&
=\ii_XF\cdot ( a^h_1, a^h_2)_{{\mathscr{A}}^*_{\HE}}-(\lsq \ii_XF, a^h_1\rsq^h, a^h_2)_{{\mathscr{A}}^*_{\HE}}-( a^h_1, \lsq \ii_XF, a^h_2\rsq^h)_{{\mathscr{A}}^*_{\HE}}~,
\end{align}
for the Lie derivative.  This gives
\begin{align}
\mathcal{L}_{\alpha_{{}_X}} ( \alpha_1,& \alpha_2)_{{\mathscr{M}}^*_{\HE}}=\int_{M^{2n}} d^{2n}x \sqrt{g} g^{-1} \big(\ii_X F\cdot\langle a_1^h, a_2^h\rangle_{\mathfrak{g}}-\langle\lsq \ii_XF, a_1^h\rsq^h, a_2^h\rangle_{\mathfrak{g}}-
\langle a_1^h, \lsq\ii_XF , a_2^h\rsq^h\rangle_{\mathfrak{g}}\big)
\cr
&
=\int_{M^{2n}} d^{2n}x \sqrt{g} g^{-1} \big(\langle\ii_X F\cdot a_1^h, a_2^h\rangle_{\mathfrak{g}}+\langle a_1^h, \ii_X F\cdot a_2^h\rangle_{\mathfrak{g}}
\cr
&\qquad\qquad-\langle\lsq \ii_XF, a_1^h\rsq^h, a_2^h\rangle_{\mathfrak{g}}-
\langle a_1^h, \lsq\ii_XF, a_2^h\rsq^h\rangle_{\mathfrak{g}}\big)
\cr
&
=\int_{M^{2n}} d^{2n}x \sqrt{g} g^{-1} \big(\langle\lsq\ii_X F, a_1^h\rsq+ d_A \ii_Xa_1^h, a_2^h\rangle_{\mathfrak{g}}+\langle a_1^h, \lsq\ii_X F, a_2^h\rsq+ d_A \ii_Xa_2^h\rangle_{\mathfrak{g}}
\cr
&\qquad\qquad
-\langle\lsq \ii_X F, a_1^h\rsq+d_A \Theta(\ii_XF, a_1^h), a_2^h\rangle_{\mathfrak{g}}-
\langle a_1^h, \lsq \ii_XF, a_2^h\rsq+d_A \Theta(\ii_XF, a_2^h)\rangle_{\mathfrak{g}}\big)
\cr
&
=\int_{M^{2n}} d^{2n}x \sqrt{g} g^{-1} \big(\langle (\ii_X d_A+d_A \ii_X) a_1^h, a_2^h\rangle_{\mathfrak{g}}+\langle a_1^h,  (\ii_X d_A+d_A \ii_X) a_2^h\rangle_{\mathfrak{g}}\big)
\cr
&
=\int_{M^{2n}} d^{2n}x \sqrt{g} g^{-1}  \mathcal{L}_X \langle a_1^h, a_2^h\rangle_{\mathfrak{g}}=0~,
\label{kproof}
\end{align}
as $X$ is a Killing vector field. Therefore, $\alpha_{{}_X}$ is a Killing vector field on the moduli space ${\mathscr{M}}^*_{\HE}$.

 To summarise, we have demonstrated that provided the condition (\ref{xtii}) holds the vector field $\alpha_{{}_X}$ on the moduli space ${\mathscr{M}}^*_{\HE}$ is both Killing and satisfies $d\alpha^\flat_{{}_X}=\ii_{\alpha_{{}_X}}\mathcal{H}$. Therefore, it is $\h{\mathcal{D}}$-covariantly constant.

 \subsubsection{Holomorphic vector fields on ${\mathscr{M}}^*_{\HE}$}\label{sec:hol}

 As in the previous section, let us assume that the vector field  $X$ on $M^{2n}$ is holomorphic and $\h\nabla$-covariantly constant. We shall demonstrate that
 the induced vector field $\alpha_{{}_X}$ on ${\mathscr{M}}^*_{\HE}$ is holomorphic as well.  For this, we have to demonstrate that $\mathcal{L}_{\alpha_{{}_X}}\mathcal{I}=0$.  However, we have already demonstrated that $\alpha_{{}_X}$ is Killing.  As a result, it suffices to show that
 the Hermitian form $\Omega$ is invariant under the action of $\alpha_{{}_X}$.  We can easily demonstrate this using the expression for $d\Omega$ in (\ref{domega}) and the expression for $\Theta(\ii_X F,a)$ in  (\ref{TF}) as
\begin{align}
\mathcal{L}_{\alpha_{{}_X}} \Omega(\alpha_1,& \alpha_2)=\ii_{\alpha_{{}_X}} d\Omega(\alpha_1, \alpha_2)+d \ii_{\alpha_{{}_X}} \Omega(\alpha_1, \alpha_2)
\cr
&
=\frac{1}{(n-1)!} \int_{M^{2n}} \omega^{n-1}\wedge \Big(-\langle d_A\Theta (\ii_X F, a_1^h)\wedge a_2^h\rangle_{\mathfrak{g}}-\langle d_A\Theta ( a_2^h, \ii_X F)\wedge a_1^h\rangle_{\mathfrak{g}}
\cr
&\qquad -\langle d_A\Theta ( a_1^h, a_2^h)\wedge \ii_X F\rangle_{\mathfrak{g}}+ a_1^h \cdot \langle \ii_XF\wedge a_2^h\rangle_{\mathfrak{g}}- a_2^h \cdot \langle \ii_XF\wedge a_1^h\rangle_{\mathfrak{g}}
-\langle \ii_XF\wedge \lsq a_1^h, a_2^h\rsq^h \rangle_{\mathfrak{g}}\Big)
\cr
&
=\frac{1}{(n-1)!} \int_{M^{2n}} \omega^{n-1}\wedge \Big(\langle d_A\ii_Xa_1^h\wedge a_2^h\rangle_{\mathfrak{g}}-\langle d_A \ii_X a_2^h\wedge a_1^h\rangle_{\mathfrak{g}}-\langle d_A\Theta ( a_1^h, a_2^h)\wedge \ii_X F\rangle_{\mathfrak{g}}
\cr
&\qquad +   \langle \ii_Xd_A a_1^h\wedge a_2^h\rangle_{\mathfrak{g}}-  \langle \ii_Xd_A a_2^h \wedge a_1^h\rangle_{\mathfrak{g}}
-\langle \ii_XF\wedge d_A\Theta(a_1^h, a_2^h)  \rangle_{\mathfrak{g}}\Big)
\cr
&
=\frac{1}{(n-1)!} \int_{M^{2n}} \omega^{n-1}\wedge \big(\langle \mathcal{L}_X a_1^h\wedge a^h_2\rangle_{\mathfrak{g}}+\langle  a_1^h\wedge \mathcal{L}_X a^h_2\rangle_{\mathfrak{g}}\big)
\cr
&
=-\frac{1}{(n-1)!} \int_{M^{2n}} \mathcal{L}_X \omega^{n-1}\wedge \langle  a_1^h\wedge a^h_2\rangle_{\mathfrak{g}}=0~,
\label{holx}
\end{align}
 where the last equality follows because $X$ is Killing and holomorphic and so  $\mathcal{L}_X\omega=0$. Thus $\alpha_{{}_X}$ is also holomorphic.

Let us denote the Lie algebra of holomorphic and $\h \nabla$-covariant constant vector fields that satisfy (\ref{xtii}) on $M^{2n}$ with $\mathfrak{k}$. One of the application of the results above is that the action of $\mathfrak{k}$ on $\mathscr{M}_{\HE}$ is free. Indeed, as for all $X\in \mathfrak {k}$ the associated vector field $\alpha_{{}_X}$ on the moduli space $\mathscr{M}^*_{\HE}$ is covariantly constant with respect to the connection $\h{\mathcal{D}}$, it implies that if the vector field is non-vanishing at the point, it is non-vanishing everywhere on $\mathscr{M}_{\HE}$. In particular, this  implies that the action of $\mathfrak{k}$ on ${\mathscr{M}}^*_{\HE}$ has no fixed points.  Therefore, it is free.

 It remains to demonstrate that if $X\not=0$, then $\alpha_{{}_X}$  does not vanish on ${\mathscr{M}}^*_{\HE}$.  For this to be the case, every vector field $X\not=0$ in $\mathfrak{k}$ must induce a non-trivial vector field $\alpha_{{}_X}$ on $\mathscr{M}_{\HE}$. We shall provide an argument for this by contradiction. Suppose there is an $X\not=0$ in $\mathfrak{k}$ such that $\alpha_{{}_X}=0$ and so
\be
\ii_X F=d_A\eta~,
\ee
for some $\eta\in \Omega^0(P\times_{\Adj} \mathfrak{g})$.  As  $\ii_X F$ is horizontal
\be
D_A^i\ii_X F_i+\theta^i\ii_X F_i=0\Longrightarrow {\mathcal O}\eta=0~.
\ee
As the kernel of $\mathcal{O}$ is trivial $\eta=0$ and so $\ii_X F=0$.  Thus $\ii_X F$ must vanish identically on ${\mathscr{A}}^*_{\HE}$ for every connection $A$. However, the only connections that satisfy $\ii_X F=0$ are those that are invariant under the action of $\mathfrak{k}$ on ${\mathscr{A}}^*_{\HE}$. It is expected that ${\mathscr{A}}^*_{\HE}$ contains non-invariant connections under the action of $\mathfrak{k}$ -- this is indeed the case in many examples including instanton moduli spaces\footnote{For instantons, $\ii_X F=0$ together with the anti-self-duality condition on $F$ imply that $F=0$. As a result, the instanton number of $E$ that admits such connections vanishes. Therefore for bundles with non-trivial instanton number, such connections do not exist.}. Thus, there must be a connection $A$ for which $\alpha_{{}_X}\not=0$.  But as $\alpha_{{}_X}$ is $\h{\mathcal{D}}$-covariantly constant, it must be non-vanishing everywhere on ${\mathscr{M}}^*_{\HE}$.

\subsubsection{KT invariant vector fields on the moduli space}\label{kvf}

There is a refinement of the results demonstrated in the previous section for $\h\nabla$-covariantly constant and holomorphic vector fields $X$ on $M^{2n}$ that $X^\flat\wedge \theta$ is not a (1,1)-form on  $M^{2n}$, i.e.
\be
(X^\flat\wedge \theta)^{2,0}\not=0~.
\label{nonoo}
\ee
As a result, the induced vector fields $\alpha_{{}_X}$ on ${\mathscr{M}}^*_{\HE}$ are not $\h{\mathcal  {D}}$-covariantly constant.  However, it turns out that they are both Killing and holomorphic. As a result, they also leave both the torsion $\mathcal{H}$ and the Hermitian form $\Omega$ of the moduli space invariant, i.e. they leave invariant the whole KT structure on ${\mathscr{M}}^*_{\HE}$.  From here on, we shall refer to such vector fields $\alpha_{{}_X}$ as KT invariant vector fields.

The proof that  $\alpha_{{}_X}$,  with  $X$ satisfying (\ref{nonoo}), is Killing  has already been given in (\ref{kproof}).  For this it suffices to notice that the computation described in (\ref{kproof}) does not use the property that $X^\flat\wedge \theta$ is an $(1,1)$-form. Therefore, it is valid for all holomorphic $\h\nabla$-covariantly constant  vector fields $X$. The same applies for the proof described in (\ref{holx}) on the invariance of the Hermitian form $\Omega$ under the action of $\alpha_{{}_X}$. Therefore, $\alpha_{{}_X}$ is both Killing and holomorphic, but not necessarily  $\h{\mathcal{D}}$-covariantly constant,  without requiring additional conditions on $X$.  There are examples of KT manifolds admitting KT invariant vector fields, which are not covariantly constant.  Such examples have been constructed in \cite{gp2, tod} in the context of HKT manifolds.

As the torsion $\mathcal{H}$ of ${\mathscr{M}}^*_{\HE}$ depends on the metric and complex structure of ${\mathscr{M}}^*_{\HE}$, which is the case for all KT manifolds, we conclude that $\mathcal{H}$
is also invariant under the action of $\alpha_{{}_X}$.  We can directly   confirm this  by taking the exterior derivative of $\ii_{\alpha_{{}_X}} \mathcal{H}$ given in (\ref{HF}) to find
\begin{align}
d\mathcal{H}(\ii_XF, &a_1^h, a_2^h, a_3^h)= \int_{M^{2n}}d^{2n}x \sqrt g  \biggl(\Big( (X\wedge \theta^\flat)^{ij} \big(a_1^h\cdot \langle a^h_{2i}, a^h_{3j}\rangle_{\mathfrak{g}} -\langle \lsq a_1^h,a^h_{2}\rsq^h_i, a^h_{3j}\rangle_{\mathfrak{g}}\big)
\cr
&\qquad
-\langle\Theta ( a^h_1, a_2^h),  a_3^h\cdot \ii_{\theta^\flat}\ii_X F \rangle_{\mathfrak{g}}\Big) +\mathrm{cyclic~in~} (a_1^h, a_2^h, a_3^h)\biggr)
\cr
&
= \int_{M^{2n}}d^{2n}x \sqrt g  \biggl(\Big( (X\wedge \theta^\flat)^{ij} \big(a_1^h\cdot \langle a^h_{2i}, a^h_{3j}\rangle_{\mathfrak{g}} -\langle \lsq a_1^h,a^h_{2}\rsq_i+D_{Ai} \Theta(a_1^h, a_2^h), a^h_{3j}\rangle_{\mathfrak{g}}\big)
\cr
&\qquad
-\langle\Theta ( a^h_1, a_2^h),  (X\wedge \theta^\flat)^{ij} D_{Ai} a_{3j}^h \rangle_{\mathfrak{g}}\Big) +\mathrm{cyclic~in~} (a_1^h, a_2^h, a_3^h)\biggr)
\cr
&
=- \int_{M^{2n}}d^{2n}x \sqrt g \Big(\big( (X\wedge \theta^\flat)^{ij} D_{i}\langle\Theta(a_1^h, a_2^h), a_{3j}^h\rangle_{\mathfrak{g}}\big)+\mathrm{cyclic~in~} (a_1^h, a_2^h, a_3^h)\Big)
\cr
&=0~,
\end{align}
where we have used that the 2-form
\be
\int_{M^{2n}}d^{2n}x \sqrt g  \big( (X\wedge \theta^\flat)^{ij}  \langle a_{1i}, a_{2j}\rangle_{\mathfrak{g}}\big)~,
\ee
 on ${\mathscr{A}}^*_{\HE}$ is closed\footnote{The argument for this is similar to that we have used to demonstrate the closure of $\Omega(a_1, a_2)$ on ${\mathscr{A}}^*_{\HE}$.}, the Gauduchon condition  $D^i\theta_i=0$ on the metric $g$ of $M^{2n}$,  $D^i X_i=0$ as $X$ is a Killing vector field on $M^{2n}$   and
 $\mathcal{L}_X \theta=0$ because  $X$ is Killing and holomorphic.  Using the closure of  $\mathcal{H}$, we have
 \be
 \mathcal{L}_{\alpha_{{}_X}} \mathcal{H}=\ii_{\alpha_{{}_X}} d \mathcal{H}+ d\ii_{\alpha_{{}_X}}\mathcal{H}=0~,
 \ee
 which proves the result.

\section{Examples}

There is a vast  choice of Hermitian manifolds that we can use to illustrate with examples  the moduli space computations we have presented  in the previous section.  Here, we shall choose  $S^3\times S^3$ and $S^3\times T^3$, viewed as group manifolds, for three reasons. One is that such manifolds appear as parts of backgrounds in AdS/CFT duality. The second reason is that they are  group manifold and so they give rise   to WZW models. The third reason is the recent rigidity result of \cite{apostolov3}.  This  states that $S^3\times S^3$ and $S^3\times T^3$ are, up to an identification with a discrete group, the only compact 6-dimensional strong KT manifolds with the holonomy of $\h\nabla$ contained in $SU(3)$. So let us focus on these two manifolds and investigate some aspects of the geometry of their  Hermitian-Einstein connection moduli spaces.   Before, we proceed with this, we shall describe some  KT structures on these two manifolds.

\subsection{The geometry of $\h{\mathscr{M}}^*_{\HE}(S^3\times S^3)$}

\subsubsection{A KT geometry on $S^3\times S^3$}

Viewing $S^3\times S^3$ as the group manifold $SU(3)\times SU(3)$, we denote with $(L_r,\tilde L_r; r=1,2,3)$ the left-invariant vector fields on $S^3\times S^3$, where $(L_r; r=1,2,3)$ are the left-invariant vector fields of the $SU(2)\times \{e\}$ subgroup while $(\tilde L_r; r=1,2,3)$ are the left-invariant vector fields of the $\{e\}\times SU(2)$ subgroup. We choose $L_3$ ($\tilde L_3$) to be the vector field tangent to the Hopf fibres of $S^3\times \{e\}$  ($\{e\}\times S^3$). The commutators are
 \be
 [L_r, L_s]=-\epsilon_{rs}{}^t L_t~,~~~[\tilde L_r, \tilde L_s]=-\epsilon_{rs}{}^t \tilde L_t~,~~~[L_r, \tilde L_s]=0~.
 \label{Lcom}
 \ee
 Similarly, we denote with  $(R_r,\tilde R_r; r=1,2,3)$, the right-invariant vector fields with commutators
 \be
 [R_r, R_s]=\epsilon_{rs}{}^t R_t~,~~~[\tilde R_r, \tilde R_s]=\epsilon_{rs}{}^t \tilde R_t~,~~~[R_r, \tilde R_s]=0~.
 \label{Rcom}
 \ee
 It follows from these that the dual left-invariant $(L^r, \tilde L^r; r=1,2,3)$ and  right-invariant  1-forms $(R^r, \tilde R^r; r=1,2,3)$  satisfy the
 differential relations
 \begin{align}
&dL^r=\frac{1}{2}\epsilon^r{}_{st} L^s\wedge L^t~,~~~d\tilde L^r=\frac{1}{2}\epsilon^r{}_{st} \tilde L^s\wedge \tilde L^t~,
\cr
&dR^r=-\frac{1}{2}\epsilon^r{}_{st} R^s\wedge R^t~,~~~d\tilde R^r=-\frac{1}{2}\epsilon^r{}_{st} \tilde R^s\wedge \tilde R^t~,
\end{align}
where $L^r(L_s)=\delta^r_s$ and $R^r(R_s)=\delta^r_s$ and similarly for $\tilde L^r$ and $\tilde R^r$.

 A left-invariant KT structure on $S^3\times S^3$ is defined as
\be
g=  \delta_{rs} L^r L^s+  \delta_{rs} \tilde L^r \tilde L^s = \delta_{rs} R^r R^s+ \delta_{rs} \tilde R^r \tilde R^s~,~~~\h\omega=L^1\wedge L^2+\tilde L^1\wedge \tilde L^2+L^3\wedge \tilde L^3~,
\label{s3s3kt}
\ee
where  as indicated the metric $g$ is bi-invariant, i.e. it is invariant under both the left and right actions of the group on itself. As such it can be written in terms of both the left- and the right-invariant 1-forms.  We have denoted the Hermitian form with $\h\omega$ instead of $\omega$ as, for example, in section \ref{sec:kt}. The reason for this is to avoid confusion, as it will become shortly apparent below.  The connection with torsion $\h\nabla$ is identified with the left-invariant connection on $SU(2)\times SU(2)$, which is characterised by the property that the left-invariant vector fields $(L_r,\tilde L_r; r=1,2,3)$ are $\h\nabla$-covariantly constant. As $\h\nabla L_r=\h\nabla\tilde L_r=0$, the curvature of $\h\nabla$ vanishes and this is a parallelisable connection. Thus, $g$ and $\h\omega$ are $\h\nabla$-covariantly constant, $\h\nabla g=\h\nabla\h\omega=0$.

The torsion $H$ of $\h\nabla$ can be easily evaluated from $\h\nabla L_r=\h\nabla\tilde L_r=0$ to find that
\be
H=L^1\wedge L^2\wedge L^3+ \tilde L^1\wedge \tilde L^2\wedge \tilde L^3~. \label{hs3s3}
\ee
Thus, $H$ is the sum of the volume forms of the two $S^3$ subspaces of $S^3\times S^3$ and as such $H$ is bi-invariant and closed $dH=0$. In particular, it can be rewritten in terms of the right-invariant forms as
\be
H= R^1\wedge R^2\wedge R^3+ \tilde R^1\wedge \tilde R^2\wedge \tilde R^3~.
\ee
The complex structure $\h I$  associated with the metric $g$ and Hermitian form $\h\omega$ in (\ref{s3s3kt}) as  $\h\omega_{ij}=g_{ik} \h I^k{}_j$ is integrable.  This can be easily seen by observing that $H$ is a $(2,1)\oplus (1,2)$-form on $S^3\times S^3$, see (\ref{integI}). As $\h I$ is integrable and $\h\nabla g=\h\nabla\h\omega=0$, $S^3\times S^3$ is a strong KT manifold. One can easily also verify that $H=-\ii_{\h I} \h\omega$.   The Lee form of this KT structure is
\be
\h\theta=L^3-\tilde L^3~.
\ee
As $\h\theta$ is $\h\nabla$-covariantly constant, it obeys the Gauduchon condition $D^i \h\theta_i=0$.
We shall refer to (\ref{s3s3kt}) as the standard left-invariant KT structure on $S^3\times S^3$.

However, such a KT structure is not unique.   For example fixing the orientation of $S^3\times S^3$ as that given by $\h\omega^3$ as well as the metric $g$ and torsion $H$, another KT structure\footnote{There are more possibilities of commuting KT structures. For example, one can also consider KT structures on $S^3\times S^3$ with Hermitian forms $\h\omega_2=-L^1\wedge L^2+\tilde L^1\wedge \tilde L^2-L^3\wedge \tilde L^3$ and $\h\omega_3=L^1\wedge L^2-\tilde L^1\wedge \tilde L^2-L^3\wedge \tilde L^3$.  For another non-commuting KT structure, one can consider the Hermitian form $\h\omega_4= L_3\wedge L_1+ L_2\wedge \tilde L_3+ \tilde L_1\wedge \tilde L_2$.} can be defined on $S^3\times S^3$ associated with  Hermitian form
\begin{align}
&\h\omega_1=-L^1\wedge L^2-\tilde L^1\wedge \tilde L^2+L^3\wedge \tilde L^3~.~~~
\label{oherm}
\end{align}
The complex structures $\h I$ and $\h I_1$ associated to the Hermitian forms $\h\omega$ and $\h\omega_1$ commute. Therefore, $S^3\times S^3$ exhibits a commuting pair of KT structures.

There is a corresponding standard right-invariant  KT structure on $S^3\times S^3$ with Hermitian form
\be
\breve \omega=R^1\wedge R^2+\tilde R^1\wedge \tilde R^2+R^3\wedge \tilde R^3~.
\ee
This Hermitian form is covariantly constant with respect to the
right-invariant connection $\breve \nabla$ on $S^3\times S^3$, which is characterised by the property that $\breve\nabla R_r=\breve\nabla \tilde R_r=0$ and whose torsion is $-H$, while that of $\h\nabla$ is $H$ (\ref{hs3s3}). The proof that $\breve\omega$ induces a KT structure on $S^3\times S^3$ with respect to $\breve \nabla$ is similar to that presented for the left-invariant KT structure above.

Again, the right-invariant standard KT-structure on $S^3\times S^3$ is not unique.   There are also several other right-invariant commuting and non-commuting KT structures on $S^3\times S^3$, as the left-invariant ones we have discussed above.
We shall not pursue this aspects of the KT geometry of $S^3\times S^3$ here.  The reason is that they do not have an application\footnote{However, they have an application on supersymmetric sigma models with target space $S^3\times S^3$. Such sigma models will admit several pairs of commuting and non-commuting (2,2) worldsheet supersymmetries induced by each  bi-KT structure on $S^3\times S^3$.} on the geometry  of $\h{\mathscr{M}}^*_{\HE}(S^3\times S^3)$.  As the Hermitian-Einstein conditions depend on the choice of complex structure on the underlying manifold, the moduli spaces of Hermitian-Einstein connections also depends on that choice. As a result a different choice of a complex structure leads to a different moduli space.  Though of course, all these moduli spaces share the same type of geometry. Nevertheless, they   are different as spaces and so the existence of additional KT structures on $S^3\times S^3$ does not induce additional KT structures on the same moduli space.  This is unlike the case of the moduli spaces of anti-self dual connections on KT and HKT 4-dimensional manifolds $M^4$.  This is  because the anti-self-duality condition does not depend on the choice of complex structure(s) of the underlying 4-dimensional manifold  provided that the  orientation of  $M^4$ is preserved.

Considering $S^3\times S^3$ with the standard KT structure, one can notice that
 the vector field $X=L_3$ is $\h\nabla$-covariantly constant and holomorphic. $L_3$ is $\h\nabla$-covariantly constant because it is left-invariant. It is also holomorphic because
\be
dX^\flat=dL^3=\ii_{L_3} H=L^1\wedge L^2~,
\ee
which is a (1,1)-form with respect to $\h I$.   $Y=-\h I X=\tilde L_3$ is also a holomorphic and $\h\nabla$-covariantly constant vector field.  The commutator of $X$ and $Y$, as it is required by the general argument in  section \ref{sec:ktvf}, vanishes $[X,Y]=0$.   The remaining $\h\nabla$-covariantly constant vector fields $L_1, L_2, \tilde L_1, \tilde L_2$, although $\h\nabla$-covariantly constant, are not holomorphic.  For example, consider the vector field $L_1$ and observe that $dL^1=L^2\wedge L^3$  is not an (1,1)-form on $S^3\times S^3$ with respect to $\h I$.  Therefore,  $L_1$ is not holomorphic.

\subsection{The KT structure on $\h{\mathscr{M}}^*_{\HE}(S^3\times S^3)$}

 Let us consider $S^3\times S^3$ with standard left-invariant KT structure (\ref{s3s3kt}) and denote  the corresponding  moduli space of Hermitian-Einstein connections with $\h{\mathscr{M}}^*_{\HE}=\h{\mathscr{M}}^*_{\HE}(S^3\times S^3)$.  It has already been demonstrated that $\h{\mathscr{M}}^*_{\HE}$ admits a strong KT structure, where the  Hermitian form and torsion are given in (\ref{Aform}) and (\ref{mtorsion}), respectively.  The metric $g$ and Hermitian form $\omega$ in these formulae are those in (\ref{s3s3kt}) of the standard KT structure on $S^3\times S^3$.

The existence of holomorphic $\h\nabla$-covariantly constant vector fields on $S^3\times S^3$ leads to a refinement of the geometry of $\h{\mathscr{M}}_{\HE}$. To see this, observe first that
\be
X\wedge \theta=- L^3\wedge \tilde L^3~,~~Y\wedge \theta=\tilde L^3\wedge L^3=-L^3\wedge \tilde L^3~,
\ee
are (1,1) forms on $S^3\times S^3$ with respect to $\h I$.

From the general theory developed in sections \ref{vsec1} and \ref{vsec2}, one concludes that the vectors fields $\alpha_{{}_{L_3}} $ and $\alpha_{{}_{\tilde L_3}} $ on the moduli space $\h{\mathscr{M}}_{\HE}$ are holomorphic and $\h{\mathcal{D}}$-covariantly constant, where $\h{\mathcal{D}}$ is the KT connection on  $\h{\mathscr{M}}^*_{\HE}$ with  torsion $\mathcal{H}$. In addition, the two vector fields on  $\h{\mathscr{M}}^*_{\HE}$ commute
\be
\lsq\alpha_{{}_{L_3}}, \alpha_{{}_{\tilde L_3}}\rsq=0~.
\ee
This follows from (\ref{commutxy}) and  $[L_3, \tilde L_3]=0$.

Furthermore,  the exterior derivative of the associated 1-forms to these two vector fields on $\h{\mathscr{M}}^*_{\HE}$, see eqn (\ref{dF}), are
\begin{align}
d\mathcal{F}_{L_3}(a^h_2, a^h_3) &= d\mathcal{F}_{\tilde L_3}(a^h_2, a^h_3)=-\int_{M^{2n}} d^{2n}x \sqrt{g} \big(  (L^3\wedge \tilde L^3)^{ij} \langle a_{2i}^h, a_{3j}^h\rangle_{\mathfrak{g}}
\cr
&
- \langle\Theta ( a^h_2, a_3^h),   F(L_3, \tilde L_3) \rangle_{\mathfrak{g}}\big)~.
\label{dF2}
\end{align}
  Notice that $d\mathcal{F}_{L_3}$ and $d\mathcal{F}_{\tilde L_3}$ are equal and, as expected, $(1,1)$-forms on $\h{\mathscr{M}}^*_{\HE}$.

Such a geometry on  $\h{\mathscr{M}}^*_{\HE}$ can be modelled after that of the $T^2$ holomorphic principal fibration over KT base space $\h{\mathscr{B}}_{\HE}$ with $\mathcal{F}_{ L_3}$ and $\mathcal{F}_{\tilde L_3}$ interpreted as the connections of the fibration.  Working with the horizontal lifts of $\alpha_{{}_{L_3}} $ and $\alpha_{{}_{\tilde L_3}} $, after a re-scaling of $\ii_{L_3} F$ and $\ii_{\tilde L_3} F$ with a constant, one can take them to have unit length with respect to the metric $\mathcal{G}$ of $\h{\mathscr{M}}_{\HE}$.  As $\ii_{\tilde L_3} F= \mathcal{I} \ii_{ L_3} F$, the hermiticity of $\mathcal{G}$ with respect to $\mathcal{I}$ implies that $\ii_{\tilde L_3} F$ and $\ii_{ L_3} F$ are orthogonal.  The $\h{\mathcal{D}}$-covariant constancy of $\ii_{\tilde L_3} F$ and $\ii_{ L_3} F$ also implies that they are Killing.

As $\alpha_{{}_{L_3}} $ and $\alpha_{{}_{\tilde L_3}} $ commute, the  $\h{\mathcal{D}}$-covariant constancy of $\alpha_{{}_{L_3}} $ and $\alpha_{{}_{\tilde L_3}} $ expresses their commutator in terms of $\mathcal{H}$, which in turn implies that
\be
\ii_{\alpha{{}_{L_3}}} \ii_{\alpha{{}_{\tilde L_3}}} \mathcal{H}=0~.
\ee
Moreover,  $\mathcal{H}$ is invariant under the action of $\alpha_{{}_{L_3}} $ and $\alpha_{{}_{\tilde L_3}} $  that again follows from the  $\h{\mathcal{D}}$-covariant constancy of $\alpha_{{}_{L_3}} $ and $\alpha_{{}_{\tilde L_3}} $ and $d\mathcal{H}=0$.

Putting all these together, the metric $\mathcal{G}$ and $\mathcal {H}$ of $\h{\mathscr{M}}^*_{\HE}$ can be decomposed as
\begin{align}
\mathcal{G}&=\mathcal{F}_{L_3}\otimes \mathcal{F}_{L_3}+\mathcal{F}_{\tilde L_3}\otimes \mathcal{F}_{\tilde L_3}+ \mathcal{G}^\perp
\cr
\mathcal{H}&=(\mathcal{F}_{L_3}+\mathcal{F}_{\tilde L_3})\wedge  d\mathcal{F}_{L_3}+\mathcal{H}^\perp
\end{align}
where $\mathcal{G}^\perp$ is orthogonal to the vector space spanned by $\alpha_{{}_{L_3}} $ and $\alpha_{{}_{\tilde L_3}} $ at each point in $\h{\mathscr{M}}^*_{\HE}$ and $\mathcal{H}^\perp$ has the property that $\ii_{\alpha_{{}_{L_3}}} \mathcal{H}^\perp=\ii_{\alpha_{{}_{\tilde L_3}}} \mathcal{H}^\perp=0$.

The base space $\h{\mathscr{B}}_{\HE}$, i.e. the space of orbits of the $T^2$ action, admits a KT structure with metric $\mathcal{G}^\perp$ and torsion $\mathcal{H}^\perp$ as both of these are invariant under the action of $\alpha_{{}_{L_3}} $ and $\alpha_{{}_{\tilde L_3}} $  vector fields. It turns out that $\h{\mathscr{B}}_{\HE}$ admits a complex structure with Hermitian form $\Omega^\perp$, where $\Omega= \mathcal{F}_{L_3}\wedge \mathcal{F}_{\tilde L_3}+\Omega^\perp$.
Using that $d\mathcal{H}=0$, one also finds that
\be
d\mathcal{H}^\perp=- 2 d\mathcal{F}_{L_3}\wedge d\mathcal{F}_{L_3}~.
\ee
A priori the right hand side of the expression above does not vanish. Thus, $\h{\mathscr{B}}_{\HE}$ admits a weak KT structure with respect to $\mathcal{G}^\perp$ and $\Omega^\perp$.

\subsection{The geometry of ${\mathscr{M}}^*_{\HE}(S^3\times T^3)$}

\subsubsection{A KT geometry on $S^3\times T^3$}

The manifold $S^3\times T^3$ can again be viewed as the group manifold $SU(2)\times U(1)^3$. The description of the KT structure on $S^3\times T^3$ is similar to that of the  previous $S^3\times S^3$ case, with some differences, and because of this most details will be neglected.
As in the previous case, one can introduce on $S^3\times T^3$ the left- and right-invariant $(L_r, V_r, r=1,2,3)$ and $(R_r, V_r; r=1,2,3)$ vector fields, respectively, where the left- and right-invariant vector field on $T^3$ coincide as this is an abelian group -- this is the reason that in both cases they are denoted\footnote{We set $V_r=\partial_{\tau^r}$, $r=1,2,3$ and $0\leq \tau^r<2\pi$ the standard angular coordinates of $T^3$.} with $V_r$. The commutators of $L_r$ and $R_r$ are as those in (\ref{Lcom}) and (\ref{Rcom}), respectively, with all the remaining commutators to vanish.  Considering again the corresponding dual left- and right-invariant 1-forms  on $S^3\times T^3$, one has that
\be
d L^r=\frac{1}{2} \epsilon^r{}_{st} L^s\wedge L^t~,~~~d V^r=0~,~~~d R^r=-\frac{1}{2} \epsilon^r{}_{st} R^s\wedge R^t~.
\ee
The standard KT structure\footnote{A more general bi-invariant metric can be put on $S^3\times T^3$, especially along $T^3$, but the one stated here suffices for our purpose.} on $S^3\times T^3$ can be defined as
\be
g=\delta_{rs} L^r L^s+\delta_{rs} V^r V^s=\delta_{rs} R^r R^s+\delta_{rs} V^r V^s~,~~~\hat\omega=L^1\wedge L^2+ V^1\wedge V^2+ L^3\wedge V^3~.
\label{s3t3kt}
\ee
The connection $\h\nabla$ is the left-invariant connection on $S^3\times T^3$ and has zero curvature and $\h\nabla g=\h\nabla \hat\omega=0$. The torsion of $\h\nabla$ is given by
\be
H=L^1\wedge L^2\wedge L^3=R^1\wedge R^2\wedge R^3~,
\label{s3t3tor}
\ee
which is the volume form of $S^3$.  The proof that the metric and Hermitian form in  (\ref{s3t3kt}) define a KT structure on $S^3\times T^3$ is similar to that for $S^3\times S^3$ and it will not be repeated.  Both the metric $g$ and torsion $H$ are bi-invariant but not the Hermitian form $\h\omega$, which is only left-invariant.
The Lee form of the standard KT structure on $S^3\times T^3$ is
\be
\hat\theta= -V^3~.
\ee
There is also a standard right-invariant KT structure on $S^3\times T^3$ with metric given as in (\ref{s3t3kt}) and Hermitian form
\be
\breve\omega=R^1\wedge R^2+ V^1\wedge V^2+ R^3\wedge V^3~.
\ee
The compatible connection with torsion $\breve\nabla$ is the right-invariant connection on $S^3\times T^3$ which has torsion $-H$, where $H$ is given in (\ref{s3t3tor}).  There are several other\footnote{Writing $S^3\times T^3$  as $S^3\times S^1\times T^2$, one can put the self-dual and anti-self-dual Hermitians form on $S^3\times S^1$, described in \cite{pw}, and add to them the Hermitian form of $T^2$. This will generate 6 left-invariant Hermitian forms and 6 right-invariant Hermitian forms. Every pair of a left-invariant and right-invariant KT structure with generate an either commuting or non-commuting  bi-KT structure on $S^3\times T^3$ and in turn the associated sigma model will exhibit a (2,2) worldsheet supersymmetry. However, as in the previous example, these additional KT structures will not be induced on the moduli space of Hermitian-Einstein connections as the latter depends on the choice of the complex structure on $S^3\times T^3$. So, we do not explore them further.} KT structures that one can put on $S^3\times T^3$.

Let us consider $S^3\times T^3$ equipped with the standard left-invariant KT structure. It turns out   that the vector fields $L_3$ and $V_r$ are all $\h\nabla$-covariantly constant, by construction, and holomorphic. For the latter  observe that $dL^3=L^1\wedge L^2$ is an $(1,1)$-form on $S^3\times T^3$ and as $dV^r=0$, they  are trivially $(1,1)$-forms. Then, from the results of section \ref{vsec1}, we conclude that the vector fields  $\ii_{L_3} F$ and $\ii_{V_r} F$ on $\mathscr{A}_{\HE}(S^3\times T^3)$ are the horizontal lifts of the tangent vector fields $\alpha_{{}_{L_3}}$ and $\alpha_{{}_{V_r}}$ of the moduli space $\h{\mathscr{M}}^*_{\HE}(S^3\times T^3)$ of Hermitian-Einstein connections on $S^3\times T^3$.

\subsection{The KT structure on $\h{\mathscr{M}}^*_{\HE}(S^3\times T^3)$}

As $L_3$ and $V_r$ are holomorphic $\h\nabla$-covariantly constant vector fields on $S^3\times T^3$, we conclude from the results of section \ref{kvf} that $\alpha_{{}_{L_3}}$ and $\alpha_{{}_{V_r}}$ are holomorphic Killing  vector fields on $\h{\mathscr{M}}^*_{\HE}=\h{\mathscr{M}}^*_{\HE}(S^3\times T^3)$, i.e. they leave invariant the KT structure of the moduli space.
However not all of them are  $\h{\mathcal{D}}$-covariantly constant on  $\h{\mathscr{M}}^*_{\HE}$. This is because from the results of section \ref{vsec2}, the $\h{\mathcal{D}}$-covariant constancy condition   requires,  in addition,  that  $X^\flat\wedge \theta$ is an $(1,1)$-form on $M^{2n}$.   Clearly for $X=L_3$ and $X=V_3$, $X^\flat\wedge\h\theta$ is an (1,1)-form, but  for $X=V_1$ and $X=V_2$ this is not the case. Therefore, $\alpha_{{}_{L_3}}$ and $\alpha_{{}_{V_3}}$ are $\h{\mathcal{D}}$-covariantly constant on $\h{\mathscr{M}}_{\HE}$ but  $\alpha_{{}_{V_1}}$ and $\alpha_{{}_{V_2}}$, although holomorphic and Killing, are not $\h{\mathcal{D}}$-covariantly constant.

The two  $\h{\mathcal{D}}$-covariantly constant vector fields $\alpha_{{}_{L_3}}$ and $\alpha_{{}_{V_3}}$  are orthogonal and commute under Lie brackets as a consequence of (\ref{commutxy}) and $[L_3, V_3]=0$.     Furthermore, one finds from (\ref{dF}) that
\begin{align}
 d\mathcal{F}_{\tilde L_3}(a^h_2, a^h_3)&=-\int_{M^{2n}} d^{2n}x \sqrt{g} \big(  (L^3\wedge  V^3)^{ij} \langle a_{2i}^h, a_{3j}^h\rangle_{\mathfrak{g}}
- \langle\Theta ( a^h_2, a_3^h),   F(L_3,  V_3) \rangle_{\mathfrak{g}}\big)~,
\cr
d\mathcal{F}_{V_3}(a^h_2, a^h_3) &=0~.~~~
\label{dF2}
\end{align}
Therefore,  $\alpha_{{}_{V_3}}$ is covariantly constant with respect to the Levi-Civita connection in $\h{\mathscr{M}}^*_{\HE}$.  The moduli space, up to an identification with a  discrete group, is a product
\be
\h{\mathscr{M}}^*_{\HE}=S^1\times \h{\mathscr{P}}_{\HE}~,
\ee
where the vector field $\alpha_{{}_{V_3}}$   is tangent to $S^1$.
The geometry of $\h{\mathscr{P}}_{\HE}$ can be modelled on  that of a principal $S^1=U(1)$ bundle, where the action of the $U(1)$ on $\h{\mathscr{P}}_{\HE}$ is generated the vector field $\alpha_{{}_{L_3}}$. The metric $\mathcal{G}$ and torsion $\mathcal{H}$ of the moduli space $\h{\mathscr{M}}^*_{\HE}$  can be written as
\begin{align}
\mathcal{G}&=\mathcal{F}_{V_3}\otimes \mathcal{F}_{V_3}+\mathcal{F}_{L_3}\otimes \mathcal{F}_{L_3}+ \mathcal{G}^\perp~,
\cr
\mathcal{H}&=\mathcal{F}_{L_3}\wedge  d\mathcal{F}_{L_3}+\mathcal{H}^\perp~,
\end{align}
where $\mathcal{G}^\perp$ is orthogonal to the vector space spanned by $\alpha_{{}_{L_3}}$ and $\alpha_{{}_{V_3}}$ at each point in $\h{\mathscr{M}}^*_{\HE}$ and $\mathcal{H}^\perp$ has the property that $\ii_{\alpha_{{}_{L_3}}} \mathcal{H}^\perp=\ii_{\alpha_{{}_{V_3}}} \mathcal{H}^\perp=0$. We have also re-scaled  $\alpha_{{}_{L_3}}$ and $\alpha_{{}_{V_3}}$ to have length one. As $d\mathcal{F}_{V_3} =0$, one can adapt a coordinate $0\leq\tau<2\pi R$ such that $\mathcal{F}_{V_3}=d\tau$.  The metric $\mathcal{G}^\perp$ and torsion $\mathcal{H}^\perp$ together with $\Omega^\perp$ induce a KT structure on the space of orbits  $\h{\mathscr{B}}_{\HE}$ of  $\alpha_{{}_{L_3}}$ and $\alpha_{{}_{V_3}}$. This is a weak KT structure as
\be
d\mathcal{H}^\perp=-d\mathcal{F}_{L_3}\wedge  d\mathcal{F}_{L_3}~.
\ee

As it has already been mentioned, $\h{\mathscr{P}}_{\HE}$ admits the action of two more Killing vector fields $\alpha_{{}_{V_1}}$ and $\alpha_{{}_{V_2}}$, which leave both the metric $\mathcal{G}$ and Hermitian form $\Omega$ on $\h{\mathscr{M}}^*_{\HE}$ invariant. The vector fields $\ii_{L_3} F$ and $\ii_{V_r} F$ commute and so $\h{\mathscr{M}}^*_{\HE}$ admits a (holomorphic) $T^4$ action.  This can be used to decompose the metric $\mathcal{G}^\perp$ and $\mathcal{H}^\perp$ further.  However, as the length of $\alpha_{{}_{V_1}}$ and $\alpha_{{}_{V_2}}$ may not be constant and potentially these vector fields may have fixed points, we shall refrain of decomposing the metric and torsion further.  Though, it is straightforward to describe such a local geometry.

\section{Geometry of instanton moduli spaces}\label{instanton}

\subsection{Instantons on KT manifolds}

On oriented 4-manifolds $M^4$, one can impose on the curvature $F$ of a principal  bundle $P(G, M^4)$ the anti-self-duality condition
\be
{}^* F=-F~;~~~\frac{1}{2} \epsilon^{mn}{}_{ij} F_{mn}=-F_{ij}~,
\label{asd}
\ee
where $\epsilon$ is the volume form of $M^4$. This  condition depends on the choice of orientation of $M^4$ and  on the conformal class of the metric $g$ on $M^4$. The solutions of this equation are the anti-self-dual instantons whose  moduli space\footnote{For $G=SU(r)$, the (virtual) dimension of $\mathscr{M}^*_{\asd}(M^4)$ is $4rk-(r^2-1) (1-b^1+b^2_+)$, where $k$ is the second Chern class of $P$, $k=c_2(P)$, that is identified with the instanton number,  $b^1$ and $b^2$ are the Betti numbers of $M^4$ and $b^2_+$ is the dimension of the space of self-dual harmonic forms \cite{atiyah}.  The moduli spaces $\mathscr{M}^*_{\asd}(M^4)$ are non-empty for $G=SU(r)$ or $PU(r)$, especially for large enough instanton number \cite{taubes1, taubes2, witten}.}
we denote with $\mathscr{M}_{\asd}(M^4)$.

On a Hermitian manifold $M^4$ with form $\omega$ and volume form chosen as  $\epsilon=\frac{1}{2} \omega^2$, the instanton condition becomes the Hermitian-Einstein condition (\ref{HEcond})  with $\lambda=0$, i.e.
\be
F^{2,0}=F^{0,2}=0~,~~~~\omega\llcorner F\equiv\frac{1}{2} \omega^{ij} F_{ij}= 0~.
\label{HEcond2}
\ee
The Hermitian form $\omega$ is self-dual, ${}^*\omega=\omega$. As the degree of any associated vector bundle to  $P(G, M^4)$ vanishes, we shall take the gauge group $G$ to be either $SU(r)$ or $PU(r)$.

The anti-self-duality condition depends only on the orientation of $M^4$. Any other Hermitian structure $\tilde \omega$ that induces on $M^4$ the same orientation as $\omega$, i.e. $\epsilon=\frac{1}{2} \tilde \omega^2$ up to a rescaling with a strictly positive function,   will lead to the same anti-self-duality condition (\ref{asd}).
The implications of this are twofold. The first one is that the geometry of anti-self-dual instanton moduli spaces for Hermitian manifolds $M^4$ can be investigated as a special case of those of Hermitian-Einstein connections.  Therefore, all the results we have obtained so far on  geometry of the moduli spaces of Hermitian-Einstein connections can be adapted to investigate the geometry of instanton moduli spaces. The other is that anti-self-dual instanton moduli spaces enjoy a much more intricate geometric structure induced as a consequence of judicious choices of Hermitian structures on $M^4$.

The results described in section  \ref{sec:2} for the moduli spaces of Hermitian-Einstein connections $\mathscr{M}_{\HE}(M^4)$ can be extended  to instanton moduli spaces $\mathscr{M}_{\asd}(M^4)$. A concise description how $\mathscr{M}_{\asd}(M^4)$ and $\mathscr{M}_{\HE}(M^4)$ are related for $M^4$ a Hermitian manifold has been given by L\"ubke and Teleman in \cite{lubke} chapter 4. Our results, when applied  to $\mathscr{M}_{\asd}(S^3\times S^1)$,  are in agreement with those  of Witten \cite{witten} derived using in parts topological field theory \cite{witten2} methods. The various formulae can be related upon  replacing the fields  $\sigma$ and $\psi$   of twisted $N=2$ super-Yang-Mills  field theory with the curvature $\Theta$ and the tangent vectors $a$ of the principal bundle $\mathscr{A}^*_{\asd}$ over $\mathscr{M}^*_{\asd}$, respectively, i.e.  $\sigma$ is identified with the curvature of the fibration.

There are  simplifications in the description of geometry of instanton moduli spaces for 4-dimensional manifolds. In particular, observe that the Lee form $\theta$ of a KT structure on $M^4$ can also be written as
\be
\h\theta=-{}^* H~;~~~\h\theta_i=-\frac{1}{6} H_{mnj} \epsilon^{mnj}{}_i~.
\label{4lee}
\ee
One consequence of this is that the Gauduchon condition on the metric $g$, $D^i\theta_i=0$, is equivalent to the strong condition on the KT structure, $dH=0$,  of $M^4$, i.e.
\be
D^i\theta_i=0 \Longleftrightarrow dH=0~.
\ee
Another consequence is that any two KT structures on $M^4$ with same  metric $g$  and torsion $H$, but different Hermitian forms $\omega$ and $\tilde \omega$ that induce the same orientation on $M^4$, give rise to two different (strong) KT structures on   $\mathscr{M}^*_{\asd}(M^4)$.  This is because the anti-self-duality condition on $F$ is the same for both of KT structures as the orientation of $M^4$ induced by $\omega$ and $\tilde \omega$ is the same.   Moreover, the horizontal subspace of the tangent space ${\mathscr{A}}^*_{\asd}$ is the same for both KT structures. This is because both the anti-self-duality condition and the operator $\mathcal{O}$ that identifies the horizontal subspace in the tangent space of ${\mathscr{A}}^*_{\asd}$ do not depend on the choice of complex structure on $M^4$.  In particular, the operator $\mathcal{O}$ in (\ref{oop}) can be written as
\be
\mathcal{O} a=D^2 a-{}^*(H\wedge a)~,
\ee
and it depends only on the metric $g$ and torsion $H$; the sign difference in the $H$ term with  the $W$ operator of \cite{witten} is conventional.  As a result both KT structures on $M^4$ will induce the same metric $\mathcal{G}$ and torsion $\mathcal{H}$ on ${\mathscr{M}}^*_{\asd}(M^4)$, see either (\ref{mtorsion}) for $n=2$ or (\ref{ttorsion}).
However, the Hermitian forms of the KT structures on ${\mathscr{M}}^*_{\asd}(M^4)$ depend on the choice of KT structure on $M^4$. Thus, two different KT structures on $M^4$ with the same metric $g$ and torsion $H$, and so Lee form $\theta$,  can induce two different KT structures on $\mathscr{M}^*_{\asd}(M^4)$.

Our results in section \ref{sec:3} can be easily adapted to the moduli space ${\mathscr{M}}^*_{\asd}(M^4)$. Choosing a KT structure on $M^4$ that admits holomorphic $\h\nabla$-covariantly constant vector fields $X$, these vector fields will induce vector fields $\alpha_{{}_X}$ on ${\mathscr{M}}^*_{\asd}(M^4)$ that will leave invariant the metric $\mathcal{G}$, Hermitian form $\Omega$ and torsion $\mathcal{H}$ of the associated KT structure on ${\mathscr{M}}^*_{\asd}(M^4)$.  If, in addition, $X^\flat\wedge \h\theta$ are (1,1)-forms on $M^4$,  the vector fields $\alpha_{{}_X}$ on ${\mathscr{M}}^*_{\asd}$ will also be $\h{\mathcal{D}}$-covariantly constant.

\subsection{Examples}\label{sec:ktex}

A class of 4-dimensional manifolds $M^4$ that admit a strong KT structure can be constructed as principal torus, $T^2$, fibrations over a Riemann surface $\Sigma_g$ of genus $g$.  These torus fibrations are classified by two integers $(m,n)$, their first Chern classes. Given a metric $g_{{}_{\Sigma_g}}$ and a K\"ahler form $\omega_{{}_{\Sigma_g}}$ on ${\Sigma_g}$, one can consider several KT structures on $M^4$. Here, we shall describe the simplest example and focus instead on the geometry of $\mathscr{M}^*_{\asd}$. In particular, a KT structure on $M^4$ is given by
\begin{align}
g&=(\kappa^1)^2+(\kappa^2)^2+g_{{}_{\Sigma_g}}~,~~~\omega=\kappa^1\wedge \kappa^2+\omega_{{}_{\Sigma_g}}~,
\cr
H&=\kappa^1\wedge d\kappa^1+\kappa^2\wedge d\kappa^2~,
\end{align}
where $(\kappa^1, \kappa^2)$ is a principal bundle connection of the torus fibration. The vector fields $X_r=\partial_{\tau^r}$, $r=1,2$, with $\kappa^r(\partial_{\tau^s})=\delta^r_s$, are generated by the $T^2$ action on $M^4$. Thus
\be
\kappa^1=d\tau^1+ C^1~,~~~\kappa^2=d\tau^2+ C^2~,
\ee
where $C^1$ and $C^2$ are locally defined 1-forms on $\Sigma_g$. If the Chern numbers $m,n\not=0$, $ d\kappa^1$ and $d\kappa^2$ do not vanish.  With these choice of metric, Hermitian form and torsion, $M^4$ has a strong KT structure as $dH=0$.  The Lee form is
\be
\theta=(\omega_{{}_{\Sigma_g}} \llcorner d\kappa_2) \kappa_1-(\omega_{{}_{\Sigma_g}} \llcorner d\kappa_1) \kappa_2~.
\ee
Furthermore, the vector fields $X_r$ commute, are holomorphic and $\h\nabla$-covariantly constant. Both
\be
 \kappa_1\wedge \theta=-(\omega_{{}_{\Sigma_g}} \llcorner d\kappa_1) \kappa_1\wedge \kappa_2~,~~~\kappa_2\wedge \theta=-(\omega_{{}_{\Sigma_g}} \llcorner d\kappa_2) \kappa_1\wedge \kappa_2~,
\ee
are (1,1)-forms on $M^4$

From the general results of section \ref{sec:2}, ${\mathscr{M}}^*_{\asd}$ admits a strong KT structure with metric, Hermitian form and torsion given in equations
(\ref{mmetric}), (\ref{mherm}) and (\ref{mtorsion}), for $n=2$,  respectively.

The vector fields $X_r$ fulfill all the requirements for $\alpha_{{}_{X_r}}$ to be holomorphic and $\hat{\mathcal{D}}$-covariantly constant on the moduli space ${\mathscr{M}}^*_{\asd}$.  The exterior derivatives of the associated 1-forms  given in (\ref{dF}) can be expressed as
\begin{align}
d\mathcal{F}_{X_1}&(\alpha_2, \alpha_3)=\int_{M^{4}} d^{4}x \sqrt{g} \big(  (X_1\wedge \theta)^{ij} \langle a_{2i}^h, a_{3j}^h\rangle_{\mathfrak{g}}- \langle\Theta ( a^h_2, a_3^h),  \ii_\theta\ii_{X_1} F \rangle_{\mathfrak{g}}\big)
\cr
&=-\int_{M^{4}} d^{4}x \sqrt{g} (\omega_{{}_{\Sigma_g}} \llcorner d\kappa_1) \big( \langle a_{2\tau_1}^h, a_{3\tau_2}^h\rangle_{\mathfrak{g}}-\langle a_{2\tau_2}^h, a_{3\tau_1}^h\rangle_{\mathfrak{g}}+ \langle\Theta ( a^h_2, a_3^h),  F_{\tau_1\tau_2}\rangle_{\mathfrak{g}} \big)~,
\cr
d\mathcal{F}_{X_2}&(\alpha_2, \alpha_3)=\int_{M^{4}} d^{4}x \sqrt{g} \big(  (X_2\wedge \theta)^{ij} \langle a_{2i}^h, a_{3j}^h\rangle_{\mathfrak{g}}- \langle\Theta ( a^h_2, a_3^h),  \ii_\theta\ii_{X_2} F \rangle_{\mathfrak{g}}\big)
\cr
&=-\int_{M^{4}} d^{4}x \sqrt{g} (\omega_{{}_{\Sigma_g}} \llcorner d\kappa_2) \big( \langle a_{2\tau_1}^h, a_{3\tau_2}^h\rangle_{\mathfrak{g}}-\langle a_{2\tau_2}^h, a_{3\tau_1}^h\rangle_{\mathfrak{g}}- \langle\Theta ( a^h_2, a_3^h), F_{\tau_1\tau_2}\rangle_{\mathfrak{g}}\big) ~.
\label{dF4}
\end{align}
These do not a priori vanish unless $M^4$ is a trivial $T^2$ fibration.  As a result, the geometry of $\mathscr{M}^*_{\asd}(M^4)$ can be modelled on that of a twisted principal $T^2$ fibration with a base space $\mathscr{B}_{\asd}$ a manifold with a KT structure. The metric, Hermitian form and torsion 3-form can be decomposed as
\begin{align}
\mathcal{G}&=(\mathcal{F}_{X_1})^2+(\mathcal{F}_{X_2})^2+\mathcal{G}^\perp~,
\cr
\Omega&=\mathcal{F}_{X_1}\wedge  \mathcal{F}_{X_2}+ \Omega^\perp~,
\cr
\mathcal{H}&=\mathcal{F}_{X_1}\wedge d\mathcal{F}_{X_1}+\mathcal{F}_{X_2}\wedge d\mathcal{F}_{X_2}+\mathcal{H}^\perp~,
\end{align}
where $\mathcal{G}^\perp$, $\Omega^\perp$ and $\mathcal{H}^\perp$ induces a KT structure on the space of orbits $\mathscr{B}_{\asd}$ of the $T^2$ fibration and we have normalised the vector fields $\alpha_{{}_{X_r}}$ to have length one.

\subsection{Instantons on bi-KT manifolds}

\subsubsection{Geometry of bi-KT manifolds or generalised K\"ahler manifolds}

A bi-KT, or equivalently generalised K\"ahler,  manifold $M^{2n}$ is a KT manifold with Hermitian metric $g$ and complex structure $\h I$ that are covariantly constant with respect to the connection $\h \nabla$ with skew-symmetric torsion $H$. In addition, such manifold admits a second KT structure with the same Hermitian metric $g$ but with different complex structure $\breve I$ that are covariantly constant with respect to the connection $\breve \nabla$ that has torsion $-H$.  In general, $\h I$ and $\breve I$ can be independent.  However, if the complex structures commute, $\h I \breve I=\breve I \h I$, then the bi-KT structure on $M^{2n}$ is called commuting.  Similarly, if the complex structures $\h I \breve I=\breve I \h I$ induce the same orientation on $M^{2n}$, then the bi-KT structure is called {\it oriented}.

 A consequence of the above definition of the bi-KT structure on a manifold $M^{2n}$  is that
\be
H=-d_{\h I} \h \omega=d_{\breve I} \breve \omega~,
\ee
where $\h \omega$ is the Hermitian form of the complex structure $\h I$ while $\breve \omega$ is the Hermitian form of the complex structure $\breve I$.

For manifolds $M^{2n}$ with $n>2$ and with a generic bi-KT structure, the associated Lee forms $\h \nabla$ and $\breve \nabla$ may not be related.  However, if the bi-KT structure is commuting, then it has been shown in \cite{pw} that
\be
\h\theta=\breve \theta~.
\ee
The case of 4-dimensional bi-KT manifolds is special. If $\h I$ and $\breve I$ induce an oriented bi-KT structure on $M^4$, regardless on whether they commute or not,
\be
\h\theta=-\breve \theta~.
\ee
This is because the Lee form of a KT structure on $M^4$ does not depend on the underlying complex structure but on the orientation that it induces on $M^4$.
As the Lee form $\h\theta=-{}^* H$, see (\ref{4lee}),  and the Lee form $\breve \theta=-{}^*(-H)={}^* H$, one arrives at the equation above.

On the other hand if $\h I$ and $\breve I$ induce opposite orientations on $M^4$, then a similar argument reveals that
\be
\h\theta=\breve \theta~.
\ee
The Gauduchon condition on the metric $g$ of the bi-KT manifold $M^4$, i.e. $D^i\h\theta_i=0$ and/or $D^i\breve\theta_i=0$,  is equivalent to the closure
of $H$, $dH=0$. There are many other aspects of bi-KT geometry, especially in relation to generalised geometry, but the summary provided here suffices for the applications below on the moduli space of anti-self-dual instantons.

There are many examples of compact bi-KT manifolds in dimensions more than four.  For example any group manifold  with a left-invariant KT structure investigated in \cite{Spindel, OP} also admits a corresponding right-invariant KT structure. These give a bi-KT structure to all these group manifolds.  However, in four dimensions the problem is more subtle as there is a limited supply of such group manifolds with $SU(2)\times U(1)$ be the main example.  But  this group manifold admits a bi-HKT structure and so it is not strictly a bi-KT manifold. This problem has been investigated in both \cite{grantcharov} and \cite{hitchin} and examples have been constructed for which both complex structures induce the same orientation on $M^4$. Such examples of oriented bi-KT manifolds are the connected sums $S^3\times S^1\sharp_k \bar \CP^2$ for $k>0$.

\subsubsection{Moduli spaces of instatons on bi-KT manifolds}\label{sec:bkt}

In order  the anti-self-duality condition on $F$ to be compatible with the bi-KT structure on $M^4$, the complex structures $\h I$ and $\breve I$ must induce the same orientation on $M^4$. In such a case, the anti-self-duality condition on $F$ will imply that $F$ satisfies the Hermitian-Einstein condition with both $\h I$ and $\breve I$ complex structures on $M^4$. This means that the anti-self-duality condition on $F$ will imply that $F$ is  an (1,1)-form\footnote{Suppose that $\h I$ and $\breve I$ induce opposite orientations on $M^4$.   Then anti-self-duality condition can be arranged such that  $F$ is an (1,1)-form with respect to $\h I$ and $\h\omega\llcorner F=0$.  However, the same condition will imply that $F$ is $(2,0)\oplus (0,2)$-form with respect to $\breve I$ and $\breve\omega \llcorner F\not=0$.  Therefore, $F$ will not satisfy the Hermitian-Einstein condition with respect to $\breve I$.} with respect to both $\h I$ and $\breve I$ and $\h\omega\llcorner F=\breve\omega\llcorner F=0$.  However, in turn this requires, as it has been explained in the previous section, that
\be
\h\theta=-\breve \theta~.
\ee
This changes the horizontality condition that identifies the representative for the tangent vectors of the moduli space $\mathscr{M}^*_{\asd}$ in the tangent space of
$\mathscr{A}^*_{\asd}$.  In particular, the horizontality condition with respect to the  KT-structure with complex structure $\h I$ is
\be
D_A^i \h a_i^h+\h \theta^i \h a^h_i=0~,
\label{ha}
\ee
while the horizontality condition for the  KT structure with complex structure $\breve I$ is
\be
D_A^i \breve a_i^h+\breve \theta^i \breve a^h_i=D_A^i \breve a_i^h-\h \theta^i \breve a^h_i=0~.
\label{baha}
\ee
As $\h a^h$ and $\breve a^h$ satisfy different differential equations, the horizontal subspace is different in each case. This  can  potentially  induce two different geometries on $\mathscr{M}^*_{\asd}$ that have a different metric and torsion.

However, it was demonstrated by Hitchin  \cite{hitchin} that this is not the case. The metric and torsion induced on $\mathscr{M}^*_{\asd}$ by  both KT structures is the same.  To show this, as both $\h a^h$ and $\breve a^h$ represent the same tangent vector $\alpha$ on $\mathscr{M}^*_{\asd}$, there is an $\eta$ such that
\be
\breve a^h=\h a^h+d_A\eta~.
\ee
Substituting this in (\ref{baha}) and using (\ref{ha}), $\eta$ satisfies
\be
\mathcal{O}^\dagger\eta\equiv D_A^2 \eta-\h \theta^i D_{Ai}\eta=2\h\theta^i\h{a}^h_i~,
\label{etaeqn}
\ee
where $\mathcal{O}^\dagger$ is the adjoint of $\mathcal{O}$.
Then to demonstrate that the two metric are equal, we have
\begin{align}
(\breve a^h,& \breve a^h)_{\mathscr{A}^*_{\asd}}=(\h a^h+d_A\eta, \h a^h+d_A\eta)_{\mathscr{A}^*_{\asd}}=(\h a^h, \h a^h)_{\mathscr{A}^*_{\asd}}+2 ( d_A\eta, \h a^h)_{\mathscr{A}^*_{\asd}}+( d_A\eta, d_A\eta)_{\mathscr{A}^*_{\asd}}
\cr
&
=(\h a^h, \h a^h)_{\mathscr{A}^*_{\asd}}+ \int_{M^4} d^4x \sqrt{g} g^{-1}\Big(2\langle d_A\eta, \h a^h\rangle_{\mathfrak{g}}+\langle d_A\eta,  d_A\eta\rangle_{\mathfrak{g}}\Big)
\cr
&
=(\h a^h, \h a^h)_{\mathscr{A}^*_{\asd}}+ \int_{M^4} d^4x \sqrt{g} \Big(-2\langle \eta, D_A^i\h a_i^h\rangle_{\mathfrak{g}}-\langle \eta,D^2_A\eta\rangle_{\mathfrak{g}}\Big)
\end{align}
where in the last step we have integrated by parts. We have also used both the horizontality condition for $\h a$ and (\ref{etaeqn}). The second expression in the formula above vanishes as
\begin{align}
&\int_{M^4} d^4x \sqrt{g} \Big(-2\langle \eta, D_A^i\h a_i^h\rangle_{\mathfrak{g}}-\langle \eta,D^2_A\eta\rangle_{\mathfrak{g}}\Big)
\cr & \qquad
=-\frac{1}{2}\int_{M^4} d^4x \sqrt{g} \h\theta^i D_i\langle \eta,\eta\rangle_{\mathfrak{g}}=0~,
\end{align}
where we have used both the horizontality of $\h a$ and (\ref{etaeqn}).  Therefore, $(\breve a^h, \breve a^h)_{\mathscr{A}^*_{\asd}}=(\h a^h, \h a^h)_{\mathscr{A}_{\asd}}$  and so $\breve a^h$ and $\h a^h$ have the same length. Therefore, the length of $\alpha$  does not depend on the choice of representatives $\breve a^h$ and $\h a^h$.  As a result, the metric on $\mathscr{M}^*_{\asd}$ does not depend on the choice of horizontal lift.

It remains to show that the torsion satisfies a similar condition. For this use (\ref{mtorsion}) and define
\begin{align}
\h{\mathcal {H}}(\h a^h_1, \h a_2^h, \h a^h_3)&\equiv  \,\int_{M^4} H\wedge \Big(\langle \h\Theta (\h a_1^h, \h a_2^h),   \h a^h_3\rangle_{\mathfrak{g}} + \mathrm{cyclic~ in~} (\h a^h_1, \h a^h_2, \h a^h_3)\Big)~,
\cr
\breve{\mathcal {H}}(\breve a^h_1, \breve a_2^h, \breve a^h_3)&\equiv - \,\int_{M^4} H\wedge \Big(\langle \breve\Theta (\breve a_1^h, \breve a_2^h),   \breve a^h_3\rangle_{\mathfrak{g}} + \mathrm{cyclic~ in~} (\breve a^h_1, \breve a^h_2, \breve a^h_3)\Big)~.
\label{mtorsion2}
\end{align}
We shall demonstrate that
\be
\mathcal {H}(\alpha_1, \alpha_2, \alpha_3)\equiv \h{\mathcal {H}}(\h a^h_1, \h a_2^h, \h a^h_3)=-\breve{\mathcal {H}}(\breve a^h_1, \breve a_2^h, \breve a^h_3)~.
\label{onetorsion}
\ee
Indeed
\begin{align}
\h{\mathcal {H}}(\h a^h_1,& \h a_2^h, \h a^h_3)= \,-\int_M  \Big(\langle \h\Theta (\h a_1^h, \h a_2^h),   \h\theta^i\h a^h_{3i}\rangle_{\mathfrak{g}} + \mathrm{cyclic~ in~} (\h a^h_1, \h a^h_2, \h a^h_3)\Big)
\cr
&
=-\frac{1}{2} \,\int_M d^4x \sqrt{g}  \Big(\langle \mathcal{O}\h\Theta (\h a_1^h, \h a_2^h),   \eta_3\rangle_{\mathfrak{g}} + \mathrm{cyclic~ in~} (1, 2, 3)\Big)
\cr
&
= \,\int_M d^4x \sqrt{g}  \Big( g^{-1}\langle [\h a_1^h, \h a_2^h]_{\mathfrak{g}},   \eta_3\rangle_{\mathfrak{g}} + \mathrm{cyclic~ in~} (1, 2, 3)\Big)
\end{align}
where in the second step we have used that $\mathcal{O}^\dagger$ is the adjoint of $\mathcal{O}$ and (\ref{etaeqn}).  In particular, we have used $\mathcal{O}^\dagger \eta_r=2\h a^{hi}_r\h\theta_i$ for $r=3$, where $r=1,2,3$.

A similar calculation reveals that
\be
\breve{\mathcal {H}}(\breve a^h_1, \breve a_2^h, \breve a^h_3)=\,-\int_M d^4x \sqrt{g}  \Big( g^{-1}\langle [\breve a_1^h, \breve a_2^h]_{\mathfrak{g}},   \eta_3\rangle_{\mathfrak{g}} + \mathrm{cyclic~ in~} (1, 2, 3)\Big)
\ee
where note that $\mathcal{O}\eta_r=-2\breve h_i \breve a_r^{hi}$.
Therefore, the proof can be completed provided that we can demonstrate the identity
\begin{align}
\int_M d^4x \sqrt{g}\Big(&\langle [D_A^i \eta_1, \h a_{2i}^h]_{\mathfrak{g}},   \eta_3\rangle_{\mathfrak{g}}+\langle [\h a_{1i}^h, D_A^i \eta_2]_{\mathfrak{g}},   \eta_3\rangle_{\mathfrak{g}}
\cr
& \qquad+\langle [D_A^i \eta_1, D_{Ai} \eta_2]_{\mathfrak{g}},   \eta_3\rangle_{\mathfrak{g}}+ \mathrm{cyclic~ in~} (1, 2, 3)\Big)=0
\label{biktid}
\end{align}
Collecting the terms that contain $\h a^h_1$ in the above expression, we have
\begin{align}
\int_M d^4x& \sqrt{g}\Big(\langle [D_A^i \eta_3, \h a_{1i}^h]_{\mathfrak{g}},   \eta_2\rangle_{\mathfrak{g}}+\langle [\h a_{1i}^h, D_A^i \eta_2]_{\mathfrak{g}},   \eta_3\rangle_{\mathfrak{g}}\Big)
\cr
&
=-\int_M d^4x \sqrt{g}\Big(\langle [\eta_2, D_A^i \eta_3 ]_{\mathfrak{g}},   \h a_{1i}^h\rangle_{\mathfrak{g}}+\langle [ D_A^i \eta_2, \eta_3]_{\mathfrak{g}},    \h a_{1i}^h\rangle_{\mathfrak{g}}\Big)
\cr
&
=- \int_M d^4x \sqrt{g}\Big( \langle D_A^i ([ \eta_2, \eta_3]_{\mathfrak{g}}),   \h a_{1i}^h\rangle_{\mathfrak{g}}\Big)=-\int_M d^4x \sqrt{g}\Big(\langle  ([ \eta_2, \eta_3]_{\mathfrak{g}}),    \h\theta^i \h a_{1i}^h\rangle_{\mathfrak{g}}\Big)
\cr
&
=\frac{1}{2} \int_M d^4x \sqrt{g} \langle  [ \eta_2, \eta_3]_{\mathfrak{g}},    \mathcal{O}^\dagger \eta_1\rangle_{\mathfrak{g}}~.
\end{align}
Considering all the cyclic permutations of the above expression, we have that
\begin{align}
&\frac{1}{2} \int_M d^4x \sqrt{g}\Big(\langle  [ \eta_2, \eta_3]_{\mathfrak{g}},    \mathcal{O}^\dagger \eta_1\rangle_{\mathfrak{g}}+ \mathrm{cyclic~ in~} (1, 2, 3)\Big)
\cr
&\qquad = \frac{1}{2} \int_M d^4x \sqrt{g}\Big(\langle  [ \eta_2, \eta_3]_{\mathfrak{g}},    (D_A^2- \h \theta^i D_{Ai}) \eta_1\rangle_{\mathfrak{g}}+ \mathrm{cyclic~ in~} (1, 2, 3)\Big)
\cr
&\qquad
=-\int_M d^4x \sqrt{g}\Big(\langle  [ D_A^i\eta_1, D_{Ai}\eta_3]_{\mathfrak{g}},    \eta_3\rangle_{\mathfrak{g}}+ \mathrm{cyclic~ in~} (1, 2, 3)\Big)~,
\end{align}
where we have perform a partial integration in the terms containing $D_A^2$ and have used that the inner product $\langle\cdot, \cdot\rangle_{\mathfrak{g}}$ is bi-invariant.  In addition, the terms containing $\h\theta$ vanish upon integrated over $M^4$ as a consequence of $D^i\h\theta_i=0$. Substituting this result in (\ref{biktid}), we establish that it vanishes. This concludes the proof of (\ref{onetorsion}).

\subsubsection{Covariantly constant and KT invariant vector fields on $\mathscr{M}^*_{\asd}$}\label{sec:bikt1}

 Let us first explore the properties of covariantly constant vector fields. For this suppose that $M^4$ is an oriented bi-KT manifold with complex structures $\h I$ and $\breve I$,  and $\mathscr{M}^*_{\asd}=\mathscr{M}^*_{\asd}(M^4)$ is an instanton moduli space of connections  on $M^4$ with gauge group $SU(r)$.  Suppose that $M^4$ admits covariantly constant vector fields $\h X$ and $\breve X$ such that
 \be
 \h\nabla \h X=0~,~~~\breve \nabla \breve X=0~,
 \ee
 which are also holomorphic
 \be
 \mathcal{L}_{\h X} \h I=0~,~~~\mathcal{L}_{\breve X} \breve I=0~.
 \ee
 It is easy to see that the vector fields $\h Y=-\h I \h X$ and $\breve Y=-\breve I\breve X$ are also covariantly constant, $\h \nabla \h Y=\breve\nabla\breve Y=0$ and holomorphic $\mathcal{L}_{\h Y} \h I=\mathcal{L}_{\breve Y} \breve I=0$.
  Then,  the proof described in section \ref{sec:3} can be generalised to show that the vector fields
 \be
 \h a^h_{\h X}\equiv \ii_{\h X} F~,~~~\h a^h_{\h Y}\equiv\ii_{\h Y} F~,~~~\breve a^h_{\breve X}\equiv\ii_{\breve X}F~,~~~\breve a^h_{\breve Y}\equiv\ii_{\breve Y}F~,
 \label{biktvf}
 \ee
 are tangent to $\mathscr{A}^*_{\asd}$ and horizontal as it has already been indicated-- we denote the corresponding vector fields on $\mathscr{M}^*_{\asd}$ with $\alpha_{\h X}$, $\alpha_{\h Y}$, $\alpha_{\breve X}$ and $\alpha_{\breve Y}$, respectively. In particular, the vector fields (\ref{biktvf}) are tangent to $\mathscr{A}^*_{\asd}$ because the exterior derivatives $d_A \h a^h_{\h X}$, $d_A\h a^h_{\h Y}$, $d_A\breve a^h_{\breve X}$ and $d_A\breve a^h_{\breve Y}$ are all anti-self-dual.  As $\h I$ and $\breve I$ induce the same orientation on $M^4$, for the vector fields $\h a^h_{\h X}$  and  $\h a^h_{\h Y}$  this property  follows because $d_A \h a^h_{\h X}$ and $d_A\h a^h_{\h Y}$ are $(1,1)$-forms with respect to $\h I$ and $\h\omega$-traceless. Similarly, for the vector fields $\breve a^h_{\breve X}$ and $\breve a^h_{\breve Y}$ this property follows because $d_A\breve a^h_{\breve X}$ and $d_A\breve a^h_{\breve Y}$ are $(1,1)$-forms with respect to $\breve I$ and $\breve\omega$-traceless. However, the horizontality conditions will differ as
 \be
 D_A^i\ii_{\h X} F_i+\h \theta^i\ii_{\h X} F_i=0~,~~~D_A^i\ii_{\breve X} F_i+\breve \theta^i\ii_{\breve X} F_i=D_A^i\ii_{\breve X} F_i-\h \theta^i\ii_{\breve X} F_i=0~,
 \label{hbikt}
\ee
 because $\breve \theta=-\h \theta$ and similarly for $\h Y$ and $\breve Y$.
The use of different horizontality conditions is not of concern here, as it was for the metric and torsion on the moduli space $\mathscr{M}^*_{\asd}$. The vector fields $\h X$, $\h Y$, $\breve X$ and $\breve Y$ are expected to induce different vector fields on $\mathscr{A}^*_{\asd}$ and on $\mathscr{M}^*_{\asd}$.

We have demonstrated that the vector fields\footnote{Note that the vector field $\alpha_{\h X}$ commutes with $\alpha_{\h Y}$ and, similarly, $\alpha_{\breve X}$ commutes with  $\alpha_{\breve Y}$.} $\alpha_{\h X}$, $\alpha_{\h Y}$, $\alpha_{\breve X}$ and $\alpha_{\breve Y}$ on $\mathscr{M}^*_{\asd}$ are Killing.
Moreover, $\alpha_{\h X}$ and  $\alpha_{\h Y}$ are holomorphic with respect to the complex structure $\h {\mathcal{I}}$, while $\alpha_{\breve X}$ and $\alpha_{\breve Y}$ are holomorphic with respect to the complex structure $\breve {\mathcal{I}}$, of $\mathscr{M}^*_{\asd}$.  These follow as a direct generalisation of  the results of section \ref{sec:hol}.  Thus, all these four vector fields leave also invariant the torsion $\mathcal{H}$ of $\mathscr{M}^*_{\asd}$.

Furthermore, it follows from the analysis in section \ref{ccvf} that the vector fields $\alpha_{\h X}$ and  $\alpha_{\h Y}$ are $\h{\mathcal{D}}$-covariantly constant on $\mathscr{M}^*_{\asd}$ provided that
\be
\h X^\flat\wedge \h\theta~,~~~\h Y^\flat\wedge \h\theta~,
\label{11biktforms}
\ee
are $(1,1)$-forms on $M^4$ with respect to the complex structure $\h I$.    Similarly, the vector fields $\alpha_{\breve X}$ and  $\alpha_{\breve Y}$ are $\breve{\mathcal{D}}$-covariantly constant on $\mathscr{M}^*_{\asd}$ provided that
\be
\breve X^\flat\wedge \breve \theta=-\breve X^\flat\wedge \h \theta~,~~\breve Y^\flat\wedge \h \theta~,
\label{11biktforms2}
\ee
are $(1,1)$-forms on $M^4$ with respect to the complex structure $\breve I$.

\subsection{ Bi-KT invariant vector fields on $\mathscr{M}^*_{\asd}$}\label{sec:bikt2}

So far, we have explored the holomorphic properties of the vector fields $\alpha_{\h X}$ ($\alpha_{\h Y}$) and $\alpha_{\breve X}$ ($\alpha_{\breve Y}$) with respect to the complex structures $\h{\mathcal{I}}$ and $\breve{\mathcal{I}}$ of $\mathscr{M}^*_{\asd}$, respectively.  Here we shall establish how the complex structure $\h{\mathcal{I}}$ transforms under the action of  $\alpha_{\breve X}$ vector field. This will establish how all four vector fields act on the bi-KT structure of $\mathscr{M}^*_{\asd}$

To perform such a calculation, it is required to examine the properties of the horizontal lift $\h a^h_{\breve X}$ of $\alpha_{\breve X}$.  This means the horizontal lift of $\alpha_{\breve X}$ with respect to the $\h I$ complex structure that satisfies the condition $D_A^i \h a^h_{{\breve X}i}+\h\theta^i\h a^h_{{\breve X}i}=0$. Notice the key sign difference in the condition that the horizontal lift $\breve a_{\breve X}^h=\ii_{\breve X} F$ of $\alpha_{\breve X}$ satisfies in  (\ref{hbikt}).
As $\h a_{\breve X}^h$ and $\breve a_{\breve X}^h$ represent the same vector $\alpha_{\breve X}$, they  should be related  by a gauge transformation.  Therefore, we set
 \be
 \h a_{\breve X}^h=\ii_{\breve X} F- d_A\eta~,
 \ee
 where $\eta$ satisfies the condition
 \be
 \mathcal{O}\eta=2 F_{ij}\breve X^i \h \theta^{j}~.
 \ee
 This  always has a solution since the operator $\mathcal{O}$ is invertible.

 Next, we shall demonstrate that
\be
\h\Theta (\h a^h, \h a^h_{{}_{\breve X}})=-\h\Theta (\h a^h_{{}_{\breve X}}, a^h)=\ii_{\breve X} \h a^h+ \h a^h\cdot \eta~,
\label{TF2}
\ee
where as usual $\h a^h\cdot \eta$ denotes the action of the vector field as in (\ref{actvf}).
This is a generalisation of the result of the lemma proven in section \ref{sec:key}.

Indeed, as $\h a_{\breve X}^h$ is horizontal, we have
\be
D_A^i \h a^h_{{\breve X}i}+\h\theta^i\h a^h_{{\breve X}i}=0\Longrightarrow D_A^i ({\breve X^j} F_{ji}- D_{Ai}\eta)+\h\theta^i ({\breve X^j} F_{ji}- D_{Ai}\eta))=0~.
\ee
To continue, we act with the tangent vector $a$ on the equation above.  The calculation for the terms containing $\ii_{\breve X} F$ is the same as that presented in section \ref{sec:key}  for $\ii_X F$; the only requirements are that $X$ is Killing and $\mathcal{L}_X\theta=0$ that  are satisfied\footnote{$\breve X$ is Killing because it is $\breve\nabla$-covariantly constant.  Moreover, $\mathcal{L}_{\breve X}\h\theta=0$ because $\breve X$ is holomorphic with respect to $\breve I$ and $\h \theta=-\breve\theta$.}  by $\breve X$. Moreover,
\be
 -a\cdot (D_A^2 \eta+ \theta^i D_{Ai}\eta)=- D_A^2 (a\cdot \eta)- \theta^i D_{Ai} (a\cdot \eta)-2 [a^i, D_{Ai}\eta]_{\mathfrak{g}}-[D_A^ia_i+\theta^i a_i, \eta]_{\mathfrak{g}}~.
 \label{heta}
 \ee
 Adding the expression (\ref{hconxf}), after replacing $X$ with $\breve X$, which gives the contribution of the $\ii_{\breve X}F$ term,  to (\ref{heta}) and  imposing the horizontality condition $\h a^h$ on the tangent vector $a$, we deduce that
 \be
 \mathcal{O} (\ii_{\breve X} \h a^h+ \h a^h\cdot \eta)= 2[\h a^{hi}, \breve X^jF_{ji}- D_{Ai}\eta]_{\mathfrak{g}}=2[\h a^{hi}, \h a^h_{{\breve X}i}]_{\mathfrak{g}}~.
 \ee
Comparing this to the equation (\ref{diftheta}) that determines $\h\Theta$, we  derive (\ref{TF2}) as the operator $\mathcal{O}$ is invertible.

As $\alpha_{\breve X}$ is Killing, it suffices to compute the Lie derivative of
 $\h\Omega$ along $\alpha_{\breve X}$. This  computation is similar to that  we have already presented in (\ref{holx}). In particular, we have that
\begin{align}
\mathcal{L}_{\alpha_{{}_{\breve X}}} \h\Omega(\alpha_1,& \alpha_2)=\ii_{\alpha_{{}_{\breve X}}} d\h\Omega(\alpha_1, \alpha_2)+d \ii_{\alpha_{{}_{\breve X}}} \h\Omega(\alpha_1, \alpha_2)
\cr
&
= \int_{M^{4}} \h\omega\wedge \Big(-\langle d_A\Theta (\ii_{\breve X} F-d_A\eta, \h a_1^h)\wedge \h a_2^h\rangle_{\mathfrak{g}}-\langle d_A\Theta ( \h a_2^h, \ii_{\breve X} F-d_A\eta)\wedge \h a_1^h\rangle_{\mathfrak{g}}
\cr
&\qquad -\langle d_A\Theta ( \h a_1^h, \h a_2^h)\wedge (\ii_{\breve X} F-d_A\eta)\rangle_{\mathfrak{g}}+ \h a_1^h \cdot \langle (\ii_{\breve X} F-d_A\eta)\wedge \h a_2^h\rangle_{\mathfrak{g}}
\cr
&\qquad
- \h a_2^h \cdot \langle (\ii_{\breve X}F-d_A\eta)\wedge \h a_1^h\rangle_{\mathfrak{g}}
-\langle (\ii_{\breve X}F-d_A\eta)\wedge \lsq a_1^h, a_2^h\rsq^h \rangle_{\mathfrak{g}}\Big)
\cr
&
=\int_{M^{4}} \h\omega\wedge \Big(\langle d_A(\ii_{\breve X}\h a_1^h+\h a_1^h\cdot \eta)\wedge \h a_2^h\rangle_{\mathfrak{g}}-\langle d_A (\ii_{\breve X} \h a_2^h+\h a_2^h \eta)\wedge \h a_1^h\rangle_{\mathfrak{g}}
\cr
&
\qquad-\langle d_A\Theta (\h a_1^h,\h a_2^h)\wedge (\ii_{\breve X} F-d_A\eta)\rangle_{\mathfrak{g}}
+   \langle \big(\ii_{\breve X}d_A \h a_1^h-[\h a_1^h, \eta]_{\mathfrak{g}}- d_A(\h a_1^h \cdot\eta)\big)\wedge \h a_2^h\rangle_{\mathfrak{g}}
\cr
&\qquad
-
 \langle \big(\ii_{\breve X} d_A \h a_2^h- [\h a_2^h, \eta]_{\mathfrak{g}}- d_A(\h a_2^h \cdot\eta)\big)  \wedge \h a_1^h\rangle_{\mathfrak{g}}
-\langle (\ii_{\breve X}F-d_A\eta)\wedge d_A\Theta(\h a_1^h, \h a_2^h)  \rangle_{\mathfrak{g}}\Big)
\cr
&
= \int_{M^{4}} \h\omega\wedge \big(\langle \mathcal{L}_{\breve X} \h a_1^h\wedge \h a^h_2\rangle_{\mathfrak{g}}+\langle  \h a_1^h\wedge \mathcal{L}_{\breve X} \h a^h_2\rangle_{\mathfrak{g}}\big)=-\int_{M^{4}} \mathcal{L}_{\breve X} \h\omega\wedge \langle \h a_1^h\wedge \h a^h_2\rangle_{\mathfrak{g}}~.
\label{holx2}
\end{align}
Therefore, the Lie derivative of the $\h\Omega$ Hermitian form of $\mathscr{M}^*_{\asd}$ along the vector field $\alpha_{\breve X}$ is determined in terms of the Lie derivative of the Hermitian form $\h\omega$ of $M^4$ along the vector field $\breve X$. A similar conclusion holds for the Lie derivatives $\mathcal{L}_{\alpha_{\breve Y}}\h\Omega$, $\mathcal{L}_{\alpha_{\h X}}\breve\Omega$ and $\mathcal{L}_{\alpha_{\h Y}}\breve\Omega$.  Thus, if $\mathcal{L}_{\h X} \breve\omega=0$, then $\mathcal{L}_{\alpha_{\h X}}\breve\Omega=0$ and so $\alpha_{\h X}$ will be holomorphic with respect the complex structure $\breve{\mathcal{I}}$.

It remains to examine the commutator $\lsq \alpha_{\h X},  \alpha_{\breve X}\rsq$.  For this, it suffices to compute
\begin{align}
\lsq \h a^h_{\h X} , \h a^h_{\breve X}\rsq^h&=\lsq \ii_{\h X} F, \ii_{\breve X} F-d_A\eta\rsq^h=(\ii_{\breve X} d_A \ii_{\h X} F-\ii_{\h X} d_A \ii_{\breve X} F-d_A (\h a^h_{\h X}\cdot\eta))^h
\cr
&
=(\mathcal {L}_{\breve X} \ii_{\h X} F -\ii_{\h X} \mathcal {L}_{\breve X} F-d_A \ii_{\breve X} \ii_{\h X} F -d_A (\h a^h_{\h X}\cdot\eta))^h
\cr
&
=-\big(\ii_{[\h X,\breve X]} F+ d_A (\ii_{\breve X} \ii_{\h X} F+\h a^h_{\h X}\cdot\eta)\big)^h= -\big(\ii_{[\h X,\breve X]} F\big)^h~.
\label{commutxy2}
\end{align}
Notice that $\ii_{[\h X,\breve X]} F$ is tangent to $\mathscr{A}^*_{\asd}$ as $d_A\ii_{[\h X,\breve X]} F=\mathcal{L}_{[\h X,\breve X]} F$ is anti-self-dual and both $\h X$ and $\breve X$ are Killing vector fields. As $\h \theta=-\breve \theta$, one concludes that $\mathcal{L}_{\h X} \h\theta=\mathcal{L}_{\breve X} \h\theta=0$ and so there is an $\eta=\eta(\h X, \breve X)$ such that
\be
\lsq \h a^h_{\h X} , \h a^h_{\breve X}\rsq^h=-\big(\ii_{[\h X,\breve X]} F-d_A\eta(\h X, \breve X)\big)~.
\label{commutxy3}
\ee
This determines the commutator of these two vector fields as $\lsq \alpha_{\h X},  \alpha_{\breve X}\rsq=-\alpha_{[\h X, \breve X]}$. In particular, if $\h X$ and $\breve X$ commute, then $a^h_{\h X}$ and  $\h a^h_{\breve X}$ also commute.

\section{Instantons on HKT and bi-HKT  manifolds}

\subsection{ HKT and bi-HKT manifolds with $\h\nabla$-covariantly constant vectors}

\subsubsection{Summary of HKT and bi-HKT geometry}

HKT manifolds $M^{4k}$ are hyper-complex manifolds\footnote{These are manifolds that admit three complex structure $(\h I_r; r=1,2,3)$ that satisfy the algebra of imaginary unit quaternions, i.e. $\h I_r \h I_s=-\delta_{rs} \bf{1}+\epsilon_{rs}{}^t \h I_t$.} equipped with metric $g$, which is Hermitian with respect to all three complex structures $I_r$, $g(\h I_rX, \h I_r Y)=g(X,Y)$,  and a connection $\h\nabla$ with skew-symmetric torsion $H$ such that the complex structures are covariantly constant, $\h\nabla \h I_r=0$. The HKT structure on $M^{4k}$ is strong provided that $H$ is closed, $dH=0$; for a recent review see \cite{pw} and references therein.

As an HKT manifold admits a KT structure with respect to each of the complex structures $\h I_r$, the torsion $H$ of $\h\nabla$ can be written as in the KT case in three different ways, i.e.
\be
H=-d_{\h I_r} \h\omega_r~,~~~\mathrm{for}~~r=1,2,3,
\label{hkttorsion}
\ee
where
\be
\h\omega_{rij}=g_{ik} \h I_r{}^k{}_j~,
\ee
is the Hermitian form of  the $\h I_r$ complex structure.  The integrability of the complex structures implies that $H$ is $(2,1)\oplus (1,2)$-form with respect to each complex structure $\h I_r$. Another key property of HKT manifolds is that the three Lee forms $\h\theta_r$, $\h\theta_{ri}=D^j\omega_{rjk} \h I_r{}^k{}_j$, each associated with the complex structures $\h I_r$ are equal
\be
\h\theta\equiv \h\theta_1=\h\theta_2=\h\theta_3~.
\label{eqlee}
\ee
A brief proof of this can be found in \cite{pw}.

A bi-HKT  manifold $M^{4k}$ admits two HKT structures associated with the hyper-complex structures $(\h I_r; r=1,2,3)$ and $(\breve I_r; r=1,2,3)$ such that
$\h\nabla \h I_r=0$ and $\breve\nabla \breve I_r=0$, where $\h\nabla$ has torsion $H$ while $\breve\nabla$ has torsion $-H$.  The hyper-complex structures
$(\h I_r; r=1,2,3)$ and $(\breve I_r; r=1,2,3)$ may or may not commute. If they commute, the bi-HKT structure is called commuting. It is also strong if $dH=0$.  It follows that $H$ can be written in three different ways as
\be
H=-d_{\h I_r} \h\omega_r=d_{\breve I_r} \breve \omega_r~,~~~\mathrm{for}~~r=1,2,3
\ee
where $\breve \omega_r$ is the Hermitian form of $\breve I_r$.  For applications to the moduli space of anti-self-dual instantons, the two HKT structures should induce the same orientation on $M^4$, i.e. the bi-HKT structure must be {\sl oriented}. For oriented bi-HKT structures, the hyper-complex structures $(\h I_r; r=1,2,3)$ and $(\breve I_r; r=1,2,3)$ do not commute.  This also implies that $\h\theta=-\breve \theta$.

\subsubsection{ Geometry of HKT manifolds with $\h\nabla$-covariantly constant vectors}\label{sec:hktpb}

Before, we proceed to investigate the geometry of instanton moduli spaces on HKT and bi-HKT manifolds, it is useful to describe the geometry of HKT manifolds that admit $\h\nabla$-covariantly constant vector fields. A related investigation has also been carried out in \cite{gp1}, see also \cite{Verbitsky}.

   Suppose a   manifold $M^{d}$ with metric $g$ and closed torsion $H$, $dH=0$, admits $\h\nabla$-covariantly constant vector field $(\h X_\alpha; \alpha=1, \dots, \ell)$, whose Lie bracket algebra $\mathfrak{k}$ closes.  As $g(\h X_\alpha, \h X_\beta)$ are constant, after possibly a transformation, $\h X_\alpha$ can be chosen as  orthogonal and of length 1, i.e. $g(\h X_\alpha, \h X_\beta)=\delta_{\alpha\beta}$. Next define\footnote{The forms $\lambda$ can also be identified with $X^\flat$. We have changed notation to simplify the formulae that follow and emphasise the interpretation of $\lambda$ as a (non-abelian) connection.} the 1-forms
   \be
   \lambda^\beta(\cdot )= \delta^{\beta\alpha} g(\h X_\alpha, \cdot)~.
   \ee
   Clearly $(\lambda^\alpha; \alpha=1, \cdots, \ell) $ are no-where vanishing and can be used as a (co-)frame on the manifold. In fact, $(\lambda^\alpha; \alpha=1, \cdots, \ell) $ can be viewed as  a principal bundle connection\footnote{The Lie algebra $\mathfrak{k}$ acts freely on $M^d$,  $\lambda$ takes values in the Lie algebra $\mathfrak{k}$ and has all the properties of a connection including the equivariance property ${\mathcal L}_{X_\alpha} \lambda^\beta=H_{\alpha\gamma}{}^\beta \lambda^\gamma$. But, again, the language of foliations may be more appropriate to describe the geometry, even though the modelling of the geometry on principal bundles may be more intuitive.}  on $M^d$. Indeed, the metric $g$ and torsion $H$ in the orthonormal (co-)frame $(\lambda^\alpha, \mathbf{e}^\ma; \alpha=1, \cdots, \ell, \ma=1,\cdots d-\ell) $ of $M^{d}$ can be written\footnote{The expressions for $g$ and $H$ are the most general possible in the chosen frame.  In particular, the component $H_{\alpha\beta \mc}$ of $H$ vanishes because the Lie algebra of vector fields $\h X_\alpha$ closes and the property of the  commutator of two such vector fields to be given by (\ref{zwcom}).} as
  \begin{align}
g&\equiv \delta_{\alpha\beta} \lambda^\alpha \lambda^\beta+\delta_{\ma\mb} \mathbf{e}^\ma \mathbf{e}^\mb=\delta_{\alpha\beta} \lambda^\alpha \lambda^\beta+ g^\perp~,~~~
\cr
H&\equiv {1\over3!} H_{\alpha\beta\gamma} \lambda^\alpha\wedge \lambda^\beta\wedge \lambda^\gamma+ {1\over2} H_{\mb\mc\alpha}\mathbf{e}^\mb\wedge \mathbf{e}^\mc\wedge \lambda^\alpha+{1\over 3!} H_{\ma\mb\mc}  \mathbf{e}^\ma\wedge \mathbf{e}^\mb\wedge \mathbf{e}^\mc
\cr
&= CS(\lambda)+ H^\perp~,
\label{decompgh}
\end{align}
 where  $CS(\lambda)$ is given by
 \be
CS(\lambda)\equiv{1\over3} \delta_{\alpha\beta} \lambda^\alpha\wedge d\lambda^\beta+{2\over3} \delta_{\alpha\beta} \lambda^\alpha\wedge G^\beta~,
\ee
and should be thought as the Chern-Simon form of $\lambda$. The components $H_{\alpha\beta\gamma}$ are constants, which follows from the Bianchi identity (\ref{bia1}), and are identified with the structure constants of $\mathfrak{k}$ Lie algebra.  The curvature of $\lambda$ is
\be
G^\alpha\equiv d\lambda^\alpha-{1\over2} H^\alpha{}_{\beta\gamma} \lambda^\beta\wedge \lambda^\gamma={1\over2} H^\alpha{}_{\mb\mc}\,\mathbf{e}^\mb\wedge \mathbf{e}^\mc~.
\label{curvlambda}
\ee
The horizontal subspaces are those orthogonal to the tangent space of the orbits of $\mathfrak{k}$ in $M^{d}$, spanned by $\h X_\alpha$, with respect to the metric $g$.  Furthermore, $g^\perp$ and $H^\perp$ have the properties
\be
\ii_{\h X_\alpha} g^\perp=\mathcal{L}_{\h X_\alpha} g^\perp=0~,~~~\ii_{\h X_\alpha} H^\perp=\mathcal{L}_{\h X_\alpha} H^\perp=0~.
\ee
Thus they can ``descent'' to a metric and 3-form on the space of orbits $B^{d-\ell}$.

 Before we proceed to describe the HKT structure, let $\h\Omega$ be the components\footnote{We use the same symbol $\Omega$ to denote the components of the frame connection and the Hermitian form on the moduli space of connections. It should be clear from the context to which one of the two we refer to in the text.} of the frame connection  of $\h\nabla$ in the frame $(\lambda^\alpha, \mathbf{e}^\ma; \alpha=1,\dots, \ell, \ma=1,\dots d-\ell)$. As $\lambda^\alpha$ are $\h\nabla$-covariantly constant, one finds that
 \be
 \h\Omega_A{}^\alpha {}_B=\Omega_A{}^\alpha{}_B+\frac{1}{2} H^\alpha{}_{AB}=0~,
 \ee
 where $\Omega$ are the components of the frame connection of the Levi-Civita connection $D$ and $A=(\alpha,\ma)$.
 This can be solved to express some of the components of $\Omega$ in terms of $H$ as
 \be
 \Omega_{\alpha\beta\gamma}=\frac{1}{2} H_{\alpha\beta\gamma}~,~~~\Omega_{\mb\alpha \mc}=-\frac{1}{2} H_{\alpha \mb\mc}~,~~~\Omega_{\alpha \mc\beta}=\Omega_{\mc\alpha\beta}=0~.
 \label{omh}
\ee
A further simplification is possible upon a more careful choice of the frame $(\lambda^\alpha, \mathbf{e}^\ma; \alpha=1,\dots, \ell, \ma=1,\dots d-\ell)$. For this notice that $\ii_{\h X_\alpha} g^\perp=0$ and ${\mathcal L}_{\h X_\alpha} g^\perp=0$. Thus, $g^\perp$ is orthogonal to the  directions spanned by the vector fields $\h X_\alpha$ and invariant. So we can always choose the frame $(\mathbf{e}^\ma; \ma=1,\dots d-\ell)$ such that
\be
\mathcal{L}_{\h X_\alpha} \mathbf{e}^\ma=0~,~~~\ii_{\h X_\alpha} \mathbf{e}^\ma=0~.
\label{framechoice}
\ee
This implies that $\ii_{\h X_\alpha} d \mathbf{e}^\ma=0$ and in turn
 the torsion free condition for the Levi-Civita frame connection implies that
\be
\Omega_{\alpha \mb\mc}=-\Omega_{\mb\mc\alpha }=-\frac{1}{2} H_{\alpha \mb\mc}~.
\label{extracond}
\ee
The last equality follows from (\ref{omh}).

The description of the geometry that we have presented so far applies to all manifolds $M^d$ that admit a connection with skew-symmetric torsion $\h\nabla$ and $\h\nabla$-covariantly constant vector fields. Returning to strong HKT manifolds first notice that if $\h X$ is a $\h\nabla$-covariantly constant vector field, then
$\h Y_r= -I_r \h X$ are also $\h\nabla$-covariantly constant and linearly independent vector fields.  The vector fields $(\h X, \h Y_r)$ are orthogonal and $\h Y_r$ have length 1,  if $\h X$ has length 1. If $Z$ is another $\h\nabla$-covariantly constant vector field, which is linearly independent from $(\h X, \h Y_r)$, then $\h W_r=-I_r \h Z$ are also linearly independent from $(\h X, \h Y_r, \h Z)$.  Continuing in this way, one can establish that the Lie algebra $\mathfrak{k}$ of $\h\nabla$-covariantly constant vector fields for HKT manifolds has dimension $\ell=4p$.  Denoting these vector fields with $(\h X_\alpha; \alpha=1,\cdots, 4p)$ and the associated 1-forms with $(\lambda^\alpha; \alpha=1,\cdots, 4p)$ and introducing an orthonormal basis $(\lambda^\alpha, \mathbf{e}^\ma; \alpha=1,\cdots, 4p, \ma=1,\cdots, 4k-4p)$ on $M^{4k}$ , the metric and torsion can be decomposed as (\ref{decompgh}).  Furthermore,
the Hermitian forms $\h \omega_r$ can be arranged as
\be
\h \omega_r\equiv {1\over2}\h {\omega}_{r\alpha\beta} \lambda^\alpha\wedge \lambda^\beta+{1\over2} \h {\omega}_{r\ma\mb} \mathbf{e}^\ma\wedge \mathbf{e}^\mb={1\over2}\h \omega_{r\alpha\beta} \lambda^\alpha\wedge \lambda^\beta+ \h \omega_r^\perp~,
\label{complexl}
\ee
where $\omega_{r\alpha\beta}$ are constants as a consequence of $\h\nabla \h \omega_r=0$ and $\h \omega_r^\perp$ is again orthogonal  to all directions spanned by the vector fields generated by $\mathfrak {k}$ on $M^{4k}$.   The components of the complex structures on $M^{4k}$ in this basis can be denoted with $\h I_r=(\h I_r{}^\alpha{}_\beta, \h I_r{}^\ma{}_\mb)$. Clearly, the Lie algebra $\mathfrak{k}$ of $\h\nabla$-covariantly constant vector fields admits a bi-invariant metric, which in the chosen basis has components $(\delta_{\alpha\beta})$, a (constant) hyper-complex structure $(\h{\tilde I}_r=(\h I_r{}^\alpha{}_\beta); r=1,2,3)$ and the structure constants of the Lie algebra are $H_{\alpha\beta\gamma}$.

 As for KT geometries, the integrability of the complex structures $\h I_r$ on $M^{4k}$ implies that $H$ must be a $(2,1)\oplus (1,2)$ form on $M^{4k}$ with respect to each $\h I_r$. In terms of the frame $(\lambda^\alpha, \mathbf{e}^\ma; \alpha=1,\cdots, 4p, \ma=1,\cdots, 4k-4p)$ , this condition decomposes as
\begin{align}
& H_{\delta \alpha\beta} \h I_r{}^\delta{}_\gamma+ H_{\delta \gamma\alpha} \h I_r{}^\delta{}_\beta+H_{\delta \beta\gamma} \h I_r{}^\delta{}_\alpha-  H_{\alpha'\beta'\gamma'} \h I_r{}^{\alpha'}{}_\alpha \h I_r{}^{\beta'}{}_\beta \h I_r{}^{\gamma'}{}_\gamma=0~,
\cr
&G^\alpha_{\mc\ma} \h I_r{}^\mc{}_\mb-G^\alpha_{\mc\mb} \h I_r{}^\mc{}_\ma+ \h I_r{}^\alpha{}_\beta (G^\beta_{\mc\md} \h I_r{}^\mc{}_\ma \h I_r{}^\md{}_\mb-G^\beta_{\ma\mb})=0~,
\cr
& H_{\md\ma\mb} \h I_r{}^\md{}_\mc+H_{\md\mc\ma} \h I_r{}^\md{}_\mb+H_{\md\mb\mc} \h I_r{}^\md{}_\ma-H_{\ma'\mb'\mc'} \h I_r{}^{\ma'}{}_\ma  \h I_r{}^{\mb'}{}_\mb \h I_r{}^{\mc'}{}_\mc=0~.
\label{nijenhuis}
\end{align}
The first condition implies that the structure constants of $\mathfrak{k}$ are $(2,1)\oplus (1,2)$ forms with respect to the (constant) hyper-complex structure $(\h{\tilde I_r}; r=1,2,3)$.  The last condition implies the same condition on  $H^\perp$ with respect to $\h I^\perp_r=(\h I_r{}^\ma{}_\mb)$.

The Lie derivative of the Hermitian forms $\h\omega_r$ on $M^{4k}$ along the vector fields $\h V_a$ can be written  in the frame $(\lambda^\alpha, \mathbf{e}^\ma; \alpha=1,\cdots, 4p, \ma=1,\cdots, 4k-4p)$ as
\begin{align}
&{\mathcal L}_{\h X_\alpha} \h\omega_{r \beta\gamma }=H_{\delta\alpha\beta} \h I_r{}^\delta{}_ \gamma-H_{\delta\alpha\gamma} \h I_r{}^\delta{}_\beta~,
\cr
&{\mathcal L}_{\h X_\alpha}\h \omega_{r \mb\mc}=-G_{\alpha \md \mb} \h I_r{}^\md{}_\mc+G_{\alpha \md \mc} \h I_r{}^\md{}_\mb~,
\label{complexlie}
\end{align}
where $G_\alpha=\delta_{\alpha\beta} G^\beta$.
Notice that the middle condition for the integrability of the complex structure is weaker than the second condition in (\ref{complexlie}).  This is significant as this allows the construction of examples of HKT manifolds for which one of their Hermitian forms is  not invariant under the action of $\mathfrak{k}$.

Finally,  the closure of $H$, $dH=0$, on $M^{4k}$ can be reexpressed as
\be
 dH^\perp+ \delta_{\alpha\beta} G^\alpha\wedge G^\beta=0~.
 \label{torsion4}
 \ee
   This in particular implies that the first Pontryagin class of $M^{4k}$ viewed as a principal bundle vanishes.

\subsection{Instantons on HKT manifolds}

\subsubsection{The HKT and bi-HKT geometry of instanton moduli space}\label{sec:hktbihkt}

It has been demonstrated by Moraru and Verbitsky   \cite{moraru} that the instanton moduli over an HKT manifold $M^4$ admits a strong HKT structure. This statement can be proved as a consequence of the results we have derived so far. Because of this, we shall proceed with a brief description of the proof.

First, on $\mathscr{A}$, we can define a metric and three Hermitian forms as
\begin{align}
( a_1, a_2)_{\mathscr{A}}&\equiv \int_{M^4} d^4x \sqrt{g} g^{-1} \langle a_1, a_2\rangle_{\mathfrak{g}}~,
\cr
\h\Omega_{\mathscr{A}\, r}(a_1, a_2)&\equiv \int_{M^4} \h\omega_r\wedge \langle a_1\wedge  a_2\rangle_{\mathfrak{g}}~.
\end{align}
The latter are associated with the complex structures $\h{\mathcal{I}}_r a =-\ii_{\h I_r} a$ on $\mathscr{A}$, which define a hyper-complex structure on $\mathscr{A}$.

The metric and Hermitian forms above can be restricted on $\mathscr{M}^*_{\asd}$ as follows. First,  the tangent vectors $a$ are  taken to be tangent to $\mathscr{A}^*_{\asd}$. This implies that for each complex structure $I_r$, $d_A a$ must be a (1,1)-from on $M^4$ and $\omega_r\llcorner d_A a=0$.  However, as all complex structures $I_r$ are compatible with the orientation of $M^4$, these conditions simply imply that $d_A a$ is anti-self-dual 2-form on $M^4$  and so it is tangent to $\mathscr{A}^*_{\asd}$.

To define the metric and Hermitian forms on $\mathscr{M}^*_{\asd}$, we have to further restrict the tangent vectors of $\mathscr{A}^*_{\asd}$ to satisfy a horizontality condition like for example that in (\ref{horizon}).  As a priori such a condition depends on the complex structure. Thus there are potentially three such choices,  one for each complex structure $\h I_r$.  However, all these three choices lead to the same horizontality condition, as  for the HKT structures all three Lee forms are equal (\ref{eqlee}).  As a result, the horizontality condition on the tangent vectors is the same for all complex structures.  Thus, we can define
\begin{align}
(\alpha_1, \alpha_2)_{\mathscr{M}^*_{\asd}}&=\int_{M^4} d^4x \sqrt{g} g^{-1} \langle a^h_1, a^h_2\rangle_{\mathfrak{g}}~,
\cr
\h\Omega_{\mathscr{M}^*_{\asd}\,r}(\alpha_1, \alpha_2)&=\int_{M^4} \h\omega_r\wedge \langle a^h_1\wedge  a^h_2\rangle_{\mathfrak{g}}~.
\end{align}
As for each choice of a complex structure $\h I_r$ on the HKT manifold $M^4$, we have shown that $\mathscr{M}^*_{\asd}$ admits a strong KT structure, it remains to show that the 3-form torsion $\mathcal{H}$ on $\mathscr{M}^*_{\asd}$ does not depend on the choice of complex structure used to determine it.  For this note that $\mathcal{H}$ can be written as
\be
\mathcal {H}(\alpha_1, \alpha_2, \alpha_3)= \,\int_{M^4} H\wedge \Big(\langle \h\Theta (\h a_1^h, \h a_2^h),   \h a^h_3\rangle_{\mathfrak{g}} + \mathrm{cyclic~ in~} (\h a^h_1, \h a^h_2, \h a^h_3)\Big)~.
\ee
We have already demonstrated that the choice of the horizontal vectors $\h a_1^h$, $\h a_2^h$ and $\h a^h_3$ does not depend on the choice of the complex structure $\h I_r$ on $M^4$. Moreover,  $\h\Theta$ does not depend on the choice of the complex structure $\h I_r$ on $M^4$ because it obeys the equation
\be
\mathcal{O}\h\Theta(\h a_1^h, \h a_2^h)=2 g^{-1}[\h a_1^h, \h a_2^h]_{\mathfrak{g}}~,
\ee
and the operator $\mathcal{O}$  as well as the tangent vectors $\h a_1^h$ and $\h a_2^h$  do not depend on the choice of complex structure. Note that $\Theta$ is a (1,1)-form on $\mathscr{M}^*_{\asd}$ with respect all three complex structures $\h{\mathcal{I}}_r$. Finally, as $H$ does not depend on the choice of complex structure $\h I_r$ because of (\ref{hkttorsion}), we conclude that $\mathcal {H}$ is independent from the choice of complex structure that it is used to determine it.  Therefore,  $\mathscr{M}^*_{\asd}$ admits a strong HKT structure.

This result can be adapted to manifolds $M^4$ that admit an oriented bi-HKT structure. In particular, if $M^4$ is a strong bi-HKT manifold with hyper-complex structures $(\h I_r; r=1,2,3)$ and $(\breve I_r; r=1,2,3)$, $\h\nabla \h I_r=0$ and $\breve \nabla \breve I_r=0$, such that $(\h I_r; r=1,2,3)$ and $(\breve I_r; r=1,2,3)$ induce the same orientation on $M^4$,  then $\mathscr{M}^*_{\asd}$ is also  a strong bi-HKT manifold. The proof of this uses a combination of arguments that include those above developed to prove  the strong HKT structure on  $\mathscr{M}^*_{\asd}$  as well as those used in section \ref{sec:bkt} to prove that $\mathscr{M}^*_{\asd}$ is a strong oriented bi-KT manifold.  In particular, if $M^4$ admits a strong bi-HKT structure with respect to the hyper-complex structures $(\h I_r; r=1,2,3)$ and $(\breve I_r; r=1,2,3)$,  then from the results above in this section $\mathscr{M}^*_{\asd}$ admits a strong HKT structure induced from that on $M^4$ with hyper-complex structure $(\h I_r; r=1,2,3)$.  $\mathscr{M}^*_{\asd}$ will also admit another strong HKT with respect to the $(\breve I_r; r=1,2,3)$ hyper-complex structure on $M^4$. It remains to show that these two HKT structures on $\mathscr{M}^*_{\asd}$ can be combined to a bi-HKT structure. Indeed, as $\h I_r$ and $\breve I_r$ induce the same orientation on $M^4$, the tangent vectors of $\mathscr{A}^*_{\asd}$ are independent from the choice of complex structures. However, there are two choices of horizontality condition for the tangent vectors.  One is
\be
D_A^i\h a_i+\theta^i\h a_i=0~,
\ee
as in (\ref{ha}), and the other is
\be
D_A^i\breve a_i-\theta^i\breve a_i=0~,
\ee
  as in (\ref{baha}), where $\theta\equiv \h \theta=-\breve\theta$.  But again, it has been demonstrated in section \ref{sec:bkt} that they induce the same metric on $\mathscr{A}^*_{\asd}$.  Moreover, if the first horizontality condition induces a torsion 3-form $\mathcal{H}$ on $\mathscr{M}^*_{\asd}$, then the second horizontality condition  induces the   torsion 3-form $-\mathcal{H}$ on $\mathscr{M}^*_{\asd}$.  Thus if the hyper-complex structure $(\h{\mathcal{I}}_r; r=1,2,3)$ on $\mathscr{M}^*_{\asd}$ is covariantly constant with respect to the connection $\h{\mathcal {D}}$ with torsion $\mathcal{H}$ on $\mathscr{M}^*_{\asd}$, then the hyper-complex structure  $(\breve{\mathcal{I}}_r; r=1,2,3)$ will be covariantly constant with respect to the connection $\breve{\mathcal D}$ with torsion $-\mathcal{H}$. This demonstrates that $\mathscr{M}^*_{\asd}$ admits a bi-HKT structure.

\subsubsection{Covariantly constant vector fields on $\mathscr{M}^*_{\asd}(M^4)$}

Suppose that $M^4$ is an HKT manifold that admits $\h\nabla$-covariantly constant vector fields  $\h X$, which are holomorphic with respect to {\sl  one} of the complex structures $\h I_r$ of $M^4$, but not necessarily tri-holomorphic.  As they are holomorphic with respect to one of the complex structure on $M^4$, there are tangent to $\mathscr{A}^*_\asd(M^4)$.  Moreover, they induce Killing  vector fields $\alpha_{\h X}$ on $\mathscr{M}^*_\asd(M^4)$, which are holomorphic with respect to one of the complex structures $\h{\mathcal{I}}_r$ of $\mathscr{M}^*_\asd(M^4)$.  The main task of this section is  to derive the formulae that are necessary to determine commutators of the associated vectors fields on $\mathscr{A}^*_\asd(M^4)$ as well as the Lie derivative of the Hermitian forms on $\mathscr{A}^*_\asd(M^4)$ with respect to these vector fields.

To begin, it follows  from (\ref{commutxy}) and (\ref{holx}) that
\begin{align}
\lsq \alpha_{\h X}, \alpha_{\h X'}\rsq &= -\alpha_{[\h X, \h X']}~,
\cr
{\cal L}_{\alpha_{\h X}} \h\Omega_s(\alpha_1, \alpha_2) &=-\int_{M^{4}} \mathcal{L}_{\h X} \h\omega_s\wedge \langle \h a_1^h\wedge \h a^h_2\rangle_{\mathfrak{g}}~,
\label{hatcom}
 \end{align}
  where the last step in (\ref{holx}) has not been carried out because $\h X$ may not be holomorphic with respect to the $\h I_s$  complex structures on $M^4$.  Furthermore, if $\h X$ is holomorphic with respect to $\h I_r$ complex structure, then the vector field $\alpha_{\h X}$ is $\h{\mathcal{D}}$-covariantly constant provided that $\h X^\flat\wedge \h \theta$ is an (1,1)-form with respect to  $\h I_r$, where the 1-form $\h X^\flat(\cdot)=g(\h X, \cdot)$.

For $M^4$ an oriented bi-HKT manifold,  which admits   $\h\nabla$- ($\breve \nabla$-) covariantly constant vector fields  $\h X$ ($\breve X$), which are holomorphic with respect to {\sl  one} of the complex structures $\h I_r$ ($\breve I_r$) of $M^4$,  similar results hold.  In particular, the commutators $\lsq \alpha_{\h X}, \alpha_{\h X'}\rsq$ and the Lie derivatives ${\cal L}_{\alpha_{\h X}} \h\Omega_r$ are given as in (\ref{hatcom}). The same expressions hold for the commutators of $\breve X$ vectors fields and the Lie derivatives of $\breve\Omega_r$ with respect to $\breve X$  after replacing in (\ref{hatcom}) $\h X$, $\h X'$, $\h \Omega_r$
 and $\h\omega_r$ with  $\breve X$, $\breve X'$, $\breve \Omega_r$
 and $\breve\omega_r$, respectively. Similarly, if the vector field $\breve X$ is holomorphic with respect to $\breve I_r$, then the vector field $\alpha_{\breve X}$ is $\breve{\mathcal{D}}$-covariantly constant provided that $\breve X^\flat\wedge \h \theta$ is an  (1,1)-form with respect to $\breve I_r$, where the 1-form $\breve X^\flat(\cdot)=g(\breve X, \cdot)$.

  It remains to state the commutators of $\h X$ and $\breve X$ vector fields as well the rest of Lie derivatives of the Hermitian forms on $\mathscr{M}^*_\asd(M^4)$.
  It follows from (\ref{commutxy3}) and (\ref{holx2}) that
  \begin{align}
  \lsq \alpha_{\h X},  \alpha_{\breve X}\rsq&=-\alpha_{[\h X, \breve X]}~,
  \cr
  \mathcal{L}_{\alpha_{{}_{\breve X}}} \h\Omega_s(\alpha_1, \alpha_2)&=-\int_{M^{4}} \mathcal{L}_{\breve X} \h\omega_s\wedge \langle \h a_1^h\wedge \h a^h_2\rangle_{\mathfrak{g}}~,
  \label{hbcom}
  \end{align}
  and similarly for $\mathcal{L}_{\alpha_{{}_{\h X}}} \breve\Omega_s$.  The above formulae can be easily adapted to the case that $M^4$ admits an oriented $\h\nabla$-HKT and a $\breve\nabla$-KT structure, and vice-versa,  instead of a bi-HKT structure.

\subsection{The bi-HKT geometry and symmetries of $\mathscr{M}^*_{\asd}(M^4)$ }\label{sec:hktvf}

\subsubsection{The bi-HKT geometry and symmetries in four dimensions}

All compact  HKT manifolds with $H\not=0$ in  four dimensions are locally isometric to $S^3\times S^1$. Thus all examples of instanton moduli spaces with an HKT structure are the instanton moduli spaces of  identifications of $S^3\times S^1$ with a discrete group.  It turns out that both $S^3\times S^1$ and $\RP^3\times S^1$ admit oriented bi-HKT structures. These for $S^3\times S^1$   have been explained in much detail in \cite{pw}. Here, we shall state  those that are relevant for the applications to the geometry of $\mathscr{M}^*_{\asd}(S^3\times S^1)$.  At the end, we shall describe the oriented bi-HKT structures on  $\RP^3\times S^1$.

Viewing $S^3\times S^1$ as the group manifold $SU(2)\times U(1)$, the left- and right- -invariant vector fields $(L_\alpha; \alpha=0,1,2,3)$ and
$(R_\alpha; a=0,1,2,3)$, respectively, satisfy the Lie brackets
\be
[L_r, L_s]=-\epsilon_{rs}{}^t L_t~,~~~[R_r, R_s]=\epsilon_{rs}{}^t R_t~,
\label{commutL}
\ee
where $V_0\equiv L_0=R_0$ commutes with all the other vector fields and it is tangent to $S^1\times \{p\}$ at every $p\in S^3$.
The associated dual 1-forms  $(V^0, L^r; r=1,2,3)$ and
$(V^0, R^r; r=1,2,3)$, where $V^0\equiv L^0=R^0$,  satisfy the exterior relations,
\be
dV^0=0~,~~~dL^r=\frac{1}{2} \epsilon^r{}_{st} L^s\wedge L^t~,~~~ dR^r=-\frac{1}{2} \epsilon^r{}_{st} R^s\wedge R^t~.
\ee
The metric $g$ and torsion $H$ on $S^3\times S^1$ can be written as
\be
g=\delta_{\alpha\beta} L^\alpha L^\beta=\delta_{\alpha\beta} R^\alpha R^\beta~,~~~H=L^1\wedge L^2\wedge L^3=R^1\wedge R^2\wedge R^3~,
\ee
i.e. both $g$ and $H$ are bi-invariant. Of course $g$ is not unique as it can be scaled with an arbitrary positive constant but the chosen normalisation is useful in calculations.  The vector fields $L_\alpha$ ($R_\alpha$) are covariantly constant with respect to the connection $\h\nabla$ ($\breve \nabla$), i.e.
\be
\h\nabla L_\alpha=0~,~~~\breve\nabla R_\alpha=0~,
\ee
with torsion $H$ and $-H$, respectively. These are the parallelizable connections associated with the left and right group actions. In fact, $V_0=L_0=R_0$ is covariantly constant with respect to the Levi-Civita connection.

Choosing the orientation on $S^3\times S^1$ as
\be
V^0\wedge L^1\wedge L^2\wedge L^3=V^0\wedge R^1\wedge R^2\wedge R^3
 \ee
 there are two compatible HKT structures. One such HKT structure is associated with the Hermitian forms
\be
\h\omega_1=L^0\wedge L^1+L^2\wedge L^3~,~~\h\omega_2=L^0\wedge L^2-L^1\wedge L^3~,~~~\h\omega_3=L^0\wedge L^3+L^1\wedge L^2~,
\ee
and the other with the Hermitian forms
\be
\breve\omega_1=R^0\wedge R^1+R^2\wedge R^3~,~~\breve\omega_2=R^0\wedge R^2-R^1\wedge R^3~,~~~\breve\omega_3=R^0\wedge R^3+R^1\wedge R^2~.
\ee
Both of  these are self-dual with respect to the orientation chosen and the two HKT structures do not commute. The  complex structures of the first HKT structure are $\h\nabla$-covariantly constant while those of the second are  $\breve\nabla$-covariantly constant.    Therefore,  $S^3\times S^1$ is an oriented bi-HKT manifold.

The Lee forms of the above two HKT structures are
\be
\h\theta=-\breve \theta=V^0~.
\ee
Furthermore, notice\footnote{This easily follows as a result of the computation  $\h I_r(L_0)=\h I_r{}^i{}_j L_0^j\partial_i= L_s \h I_r{}^s{}_0=-L_s \delta^s{}_r =-L_r$.} that
\be
L_r=-\h I_r(L_0)~,~~~
R_r=-\breve I_r (R_0)~.
\ee
Therefore, all the $\h\nabla$- and $\breve\nabla$-covariantly constant vector fields are related to $V_0=L_0=R_0$ upon acting on it with the complex structures of $S^3\times S^1$.

The manifold $\RP^3\times S^1$ can be constructed from $S^3\times S^1$ after identifying the antipodal points of $S^3$, $\RP^3=S^3/\bZ_2$. Again $\RP^3\times S^1$ can be viewed as the group manifold $SO(3)\times SO(2)$. As $SO(3)\times SO(2)$ has the same Lie algebra as $SU(2)\times U(1)$, the above geometric construction, see also appendix A, can be carried out in the same way to argue that $\RP^3\times S^1$ admits an oriented bi-HKT structure.

\subsubsection{The geometry of $\mathscr{M}^*_{\asd}(S^3\times S^1)$ and $\mathscr{M}^*_{\asd}(\RP^3\times S^1)$} \label{sec:gms3s1}

As $S^3\times S^1$ admits an oriented bi-HKT structure, from the results of previous sections, one concludes that $\mathscr{M}^*_{\asd}(S^3\times S^1)$  is a bi-HKT manifold. It remains to investigate the symmetries of $\mathscr{M}^*_{\asd}(S^3\times S^1)$.  These have originally been explored in \cite{witten}. Here, we shall derive the same results for the commutators of vector fields as well as for the Lie derivatives of the Hermitian form  on the moduli space, up to differences in notation, using the general theory we have developed  in the previous sections. Turning to the properties of vector field on $\mathscr{M}^*_{\asd}(S^3\times S^1)$, we have that  the vector fields
\be
\h a_{V_0}^h\equiv \ii_{V_0} F~,~~\h a^h_{L_r}\equiv \ii_{L_r} F~,
\label{v0lrf}
\ee
 are tangent to $\mathscr{A}^*_{\asd}(S^3\times S^1)$ and as indicated horizontal.  To prove this, first observe that by construction $V_0$ and $L_r$ are $\h\nabla$-covariantly constant. From the results of section \ref{sec:3} and the independence of the anti-self-duality condition from the choice of a compatible complex structure, it suffices to show that they are also holomorphic with respect to one of the complex structures $\h I_r$. Recall that a $\h\nabla$-covariant constant vector field $X$ is holomorphic with respect to a complex structure $\h I$, iff $\ii_X H$  is a (1,1)-form. This is indeed the case for $(V_0, L_r)$ as $\ii_{V_0} H=0$ and so $V_0$ is tri-holomorphic. Moreover, $\ii_{L_r} H$ is (1,1) with respect to the complex structure $\h I_r$.  This confirms that (\ref{v0lrf}) are tangent to $\mathscr{A}^*_{\asd}(S^3\times S^1)$ and horizontal.

 To prove that the vector fields $\alpha_{V_0}$ and $\alpha_{L_r}$ on $\mathscr{M}^*_{\asd}(S^3\times S^1)$ constructed from those in (\ref{v0lrf}) are $\h{\mathcal{D}}$-covariantly constant, it suffices to demonstrate from the results of section \ref{sec:3} that $V^0\wedge \theta$ and $L^r\wedge \theta$ are  (1,1)-forms with respect to anyone  of the complex structures $\h I_r$.  This follows from
\be
V^0\wedge \theta=V^0\wedge V^0=0~,~~~L^r\wedge \theta= \ii_{\h I_r} V^0\wedge V^0~.
\ee
In particular in the latter case $L^r\wedge \theta$ is (1,1)-form with respect to the complex structure $\h I_r$.  This completes the proof that $\alpha_{V_0}$ and $\alpha_{L_r}$ are
 $\h{\mathcal{D}}$-covariantly constant.

Using the results of section \ref{sec:3}, we can also compute the exterior derivative of the dual 1-forms to the vector fields (\ref{v0lrf}).  Indeed from (\ref{dF}), one finds that
\begin{align}
d\mathcal{F}_{L_r}(\alpha_2, \alpha_3)
& =\int_{M^4} d^{4}x \sqrt{g} \big(  (L_r\wedge V_0)^{ij} \langle a_{2i}^h, a_{3j}^h\rangle_{\mathfrak{g}}- \langle\h \Theta ( a^h_2, a_3^h),   F(L_r, V_0) \rangle_{\mathfrak{g}}\big)
\cr
d\mathcal{F}_{V_0}(\alpha_2, \alpha_3)&=0~.
\label{dFhkt}
\end{align}
As the exterior derivative of $\mathcal{F}_{V_0}$ vanishes, this confirms that the vector field $\alpha_{V_0}$ is covariantly constant with respect to the Levi-Civita connection of $\mathscr{M}^*_{\asd}(S^3\times S^1)$. The exterior derivatives  $d\mathcal{F}_{L_r}$ do not a priori vanish. Moreover, they are (1,1)-forms with respect to $\h{\mathcal{I}}_r$ but not with respect to $\h{\mathcal{I}}_s$ for $s\not=r$. So $\ii_{L_r} F$ is a holomorphic vector field with respect to $\h{\mathcal{I}}_r$ but not with respect to $\h{\mathcal{I}}_s$ for $s\not=r$. To find the Lie derivative of $\h{\mathcal{I}}_s$ along the vector field $\alpha_{L_r}$, it suffices to compute the Lie derivative of $\Omega_s$ along $\alpha_{L_r}$ as all these vector fields are Killing. Using the formula (\ref{hatcom}) for $\h X=L_r$ and (\ref{commutL}), we  find that
\begin{align}
\mathcal{L}_{\alpha_{L_r}} \h\Omega_s(\alpha_1, \alpha_2)
=-\epsilon_{rs}{}^t \h\Omega_t(\alpha_1, \alpha_2)~.
\label{holxhkt}
\end{align}
Thus
\be
\mathcal{L}_{\alpha_{L_r}} \h{\mathcal{I}}_s=\epsilon_{rs}{}^t \h{\mathcal{I}}_t~,~~~\mathcal{L}_{\alpha_{V_0}} \h{\mathcal{I}}_s=0~,
\ee
where we have added that $\alpha_{V_0}$ is tri-holomorphic.
It remains to compute the commutator of the vector fields (\ref{v0lrf}). This can  easily be done using the formula (\ref{commutxy}) and (\ref{commutL}). In particular, we find that
 \be
 \lsq \alpha_{V_0}, \alpha_{L_r}\rsq=0~,~~~\lsq \alpha_{L_r}, \alpha_{L_s}\rsq=\epsilon_{rs}{}^t \alpha_{L_t}~.
 \ee
 This completes the analysis for the one of the HKT structures on $\mathscr{M}^*_{\asd}(S^3\times S^1)$.

The investigation of the HKT structure on $\mathscr{M}^*_{\asd}(S^3\times S^1)$ induced from the other HKT structure on $S^3\times S^1$, which is associated to the hyper-complex structure $ (\breve I_r; r=1,2,3)$, can be made in a similar way.  We shall not present details. Instead, we shall state the relevant formulae. The vector fields $\alpha_{R_r}$ on $\mathscr{M}^*_{\asd}(S^3\times S^1)$, associated the vector fields
\be
\breve a^h_{R_r}\equiv \ii_{R_r} F~,
\label{rrf}
\ee
on $\mathscr{A}^*_{\asd}(S^3\times S^1)$,  are all $\breve {\mathcal{D}}$-covariantly constant, where  $\breve {\mathcal{D}}$ is the HKT connection on $\mathscr{M}^*_{\asd}(S^3\times S^1)$ with torsion $-\mathcal{H}$ -- as $\alpha_{R_0} =\alpha_{V_0} $, its properties  have already been explored.

The exterior derivatives of the 1-form dual to the vector fields (\ref{rrf}) are
\begin{align}
d\mathcal{F}_{R_r}(\alpha_2, \alpha_3)
 =\int_{M^4} d^{4}x \sqrt{g} \big(  (R_r\wedge V_0)^{ij} \langle a_{2i}^h, a_{3j}^h\rangle_{\mathfrak{g}}- \langle\breve \Theta ( a^h_2, a_3^h),   F(R_r, V_0) \rangle_{\mathfrak{g}}\big)~.
\label{dFhkt2}
\end{align}
Furthermore, the Lie derivative of the complex structures $\breve{\mathcal{I}}_s$ are
\be
\mathcal{L}_{\alpha_{R_r}} \breve{\mathcal{I}}_s=-\epsilon_{rs}{}^t \breve{\mathcal{I}}_t~,~~~\mathcal{L}_{\alpha_{V_0}} \breve{\mathcal{I}}_s=0~.
\ee
The commutators of the vector fields (\ref{rrf}) are
 \be
 \lsq \alpha_{V_0}, \alpha_{R_r}\rsq=0~,~~~\lsq \alpha_{R_r}, \alpha_{R_s}\rsq=-\epsilon_{rs}{}^t \alpha_{R_t}~.
 \ee
 Of course the vectors fields $\alpha_{L_r}$ and $\alpha_{R_s}$ also commute, $\lsq\alpha_{L_r}, \alpha_{R_s}\rsq=0$.

 It remains to investigate the action of the vector fields $\alpha_{L_r}$ ($\alpha_{R_r}$) on the complex structures $\breve {\mathcal{I}}$ ($\h{\mathcal {I}}$) of $\mathscr{M}^*_{\asd}(S^3\times S^1)$.  As $\mathcal{L}_{L_r} \breve I_s=0$ and $\mathcal{L}_{R_r} \h I_s=0$, an application of (\ref{hbcom}) reveals that
 \be
\lsq \alpha_{L_r}, \alpha_{R_s}\rsq=0~,~~~ \mathcal{L}_{\alpha_{L_r}} \breve {\mathcal{I}}_s=0~,~~~\mathcal{L}_{\alpha_{R_r}} \h {\mathcal{I}}_s=0~.
 \ee
 This completes the description of the symmetries of the  bi-HKT structure on $\mathscr{M}^*_{\asd}(S^3\times S^1)$.

The description of the symmetries of the oriented bi-HKT structure of $\mathscr{M}^*_{\asd}(\RP^3\times S^1)$ is similar to that of $\mathscr{M}^*_{\asd}(S^3\times S^1)$.  A $\bZ_2$ identification, which is the difference between the $S^3\times S^1$ and $\RP^3\times S^1$, is not expected to alter the local geometric formulae.  Though of course $S^3\times S^1$ and $\RP^3\times S^1$ are different as topological spaces.  Because of this, we shall not repeat the formulae.

\subsection{Instanton moduli spaces  and QKT structures  }\label{sec:qkt}

\subsubsection{Summary of the geometric conditions}

In what follows to simplify the analysis and notation,  it is useful to view $\mathscr{M}^*_{\asd}(S^3\times S^1)$ and $\mathscr{M}^*_{\asd}(\RP^3\times S^1)$ as (another ordinary) $4k$-dimensional  strong bi-HKT manifold.  For this, we relabel the tensors that characterise the geometry of $\mathscr{M}^*_{\asd}(M^3\times S^1)$, for either $M^3=S^3$ or $M^3=\RP^3$,  as follows
\begin{align}
&\mathcal{G}\leftrightarrow g~,~~~\mathcal {H}\leftrightarrow H~,~~~\h{\mathcal{D}}\leftrightarrow \h\nabla~,~~~ \breve{\mathcal{D}}\leftrightarrow \breve\nabla~,~~~\h\Omega_r\leftrightarrow\h\omega_r~,~~~\breve \Omega_r\leftrightarrow\breve\omega_r~.
\cr
&
\h{\mathcal{I}}_r\leftrightarrow \h{I}_r~,~~~\breve{\mathcal{I}}_r\leftrightarrow \breve{I}_r~,~~~\alpha_{V_0}\leftrightarrow V_0~,~~~\mathcal{F}_{\alpha_{V_0}}\leftrightarrow \lambda^0=\h\lambda^0=\breve\lambda^0~,~~~\mathcal{F}_{\alpha_{L_r}}\leftrightarrow \h\lambda^r~,~~~\mathcal{F}_{\alpha_{R_r}}\leftrightarrow \breve\lambda^r~,
\cr
&
\alpha_{L_r}\leftrightarrow \h V_r~,~~~\alpha_{R_r}\leftrightarrow \breve V_r~.
\label{hktcorr}
\end{align}
In terms of these, $\mathscr{M}^*_{\asd}(M^3\times S^1)$ is an oriented bi-HKT manifold with metric $g$, torsion $H$, and non-commuting hyper-complex structures $(\h I_r; r=1,2,3)$ and $(\breve I_r; r=1,2,3)$,  which are covariantly constant with respect to the connections (with torsion) $\h\nabla$ and $\breve \nabla$, respectively.  Moreover,  $\mathscr{M}^*_{\asd}(M^3\times S^1)$ admits $\h\nabla$- and $\breve\nabla$- covariantly constant vector fields
\be
 \h\nabla \h V_a= \breve \nabla \breve V_a=0~,~~~a=0,1,2,3~,
 \ee
 with $\h V_0=\breve V_0=V_0$, and so $V_0$ is covariantly constant with respect to the Levi-Civita connection of $\mathscr{M}^*_{\asd}(M^3\times S^1)$,  $DV_0=0$.  Furthermore, $V_0$ commutes with all other vector fields $\h V_r$ and $\breve V_s$ and leaves the complex structures invariant,
 \be
 [V_0, \h V_r]=[V_0, \breve V_r]=0~,~~~\mathcal{L}_{V_0} \h I_r=\mathcal{L}_{V_0} \breve I_r=0~,
 \ee
 while
\begin{align}
[\h V_r, \h V_s]&=\epsilon_{rs}{}^t \h V_t~,~~~[\breve V_r, \breve V_s]=-\epsilon_{rs}{}^t \breve V_t~,~~~r,s,t=1,2,3
\cr
[\h V_r, \breve V_s]&=0~,
\label{vectorlie}
\end{align}
and
\begin{align}
{\mathcal L}_{\h V_r} \h I_s=\epsilon_{rs}{}^t \h I_t~,~~~
{\mathcal L}_{\breve V_r} \breve I_s=-\epsilon_{rs}{}^t \breve I_t~.
\label{vectorlie2}
\end{align}
Therefore, the Lie algebra of $\h\nabla$- and $\breve\nabla$-covariantly constant vector fields can be identified with $\mathfrak{so}(4)\oplus \mathfrak{so}(2)$.
Note that the vector fields $\h I_r \h V_a$ are also $\h\nabla$-covariantly constant and as $\h V_a$ span all such vector fields, one can set $\h V^i_r= -(\h I_r)^i{}_j V^j_0$ and similarly for $\breve V_r$.  Observe that $\h V_a$ are orthogonal as a consequence of the Hermiticity of $g$ and can be chosen to have length 1 provided, after an appropriate constant re-scaling, $V_0$ is chosen to have length 1.

\subsection{The bi-HKT structure on $\mathscr{M}^*_{\asd}(M^4)$ and QKT manifolds }\label{sec:mqkt}

\subsubsection{QKT geometry from the HKT structure on  $\mathscr{M}^*_{\asd}(M^4)$}

To begin, let us examine in more detail the consequence on the geometry of $\mathscr{M}^*_{\asd}(M^4)$, for $M^4$ either $S^3\times S^1$ or $\RP^3\times S^1$,  arising from the presence of the $\h\nabla$-covariantly constant vector fields $(\h V_\alpha; \alpha=0,1,2,3)$. The Lie algebra of these vector fields is $\mathfrak{su}(2)\oplus\mathfrak{u}(1)$. As $d\lambda^0=0$ and the 1-form $\lambda^0$ is covariantly constant with respect to the
Levi-Civita connection of $\mathscr{M}^*_{\asd}(M^4)$, up to an identification with a discrete group,
\be
\mathscr{M}^*_{\asd}(M^4)= S^1\times  {\mathscr {P}}^{4k-1}(M^4)~,
\ee
where the geometry of $\mathscr {P}^{4k-1}(M^4)$ is modelled after that of a principal bundle $SU(2)$-bundle over $\mathscr{B}^{4k-4}$ with fibre $SU(2)$.  Introducing a coordinate $\tau$ along $S^1$, then $\lambda^0=d\tau$ and $V_0=\partial_\tau$.
Therefore,  the moduli space $\mathscr{M}^*_{\asd}(M^4)$ splits,  as a Riemannian manifold with skew-symmetric torsion, because the metric $g$ decomposes as a sum of that on $S^1$ and that on ${\mathscr {P}}^{4k-1}(M^4)$ and  $\ii_{V_0} H=0$.  However, it does not split as an HKT manifold because the Hermitian forms do not decompose in an appropriate  way.

As the hypercomplex structure is invariant under $V_0$, i.e. $V_0$ is a tri-holomorphic vector field, it remains to determine the implications of the action
of $\h V_r$ on $\h I_s$ given in (\ref{vectorlie2}).  For this, we use the results of  section \ref{sec:hktpb} on HKT manifolds that admit $\h\nabla$-covariantly constant vector fields and compare (\ref{vectorlie2}) with (\ref{complexlie}). It can always be arranged such that the first condition in (\ref{complexlie}), which is along the fibre directions of $\mathscr{M}^*_{\asd}(M^4)$ viewed as a principal $K$ bundle, $K=SU(2)\times U(1)$ or $SO(3)\times SO(2)$, is satisfied. This is because  restriction of the HKT structure of $\mathscr{M}^*_{\asd}(M^4)$ along the $K$ fibre induces the standard left-invariant HKT structure on $M^4$ -- the restriction of the torsion 3-form $H$ on the fibres $H$ is the volume form of $M^3$ as a consequence of the decomposition for $H$ in (\ref{decompgh}).

On comparing (\ref{vectorlie2}) with (\ref{complexlie}) along the horizontal directions, we find  $\lambda$ must be a $\mathfrak{sp}(k-1)\oplus \mathfrak{sp}(1)$ instanton on $\mathscr{B}^{4k-4}$.  In particular, the curvature\footnote{Note that we have not computed $G$ from a first principles calculation on the moduli space $\mathscr{M}^*_{\asd}(M^4)$ as we have done for the metric $\mathcal{G}$, the torsion $\mathcal{H}$ and 2-forms $d\mathcal{F}_X$.  Nevertheless, we are able to carry out the computation using the restrictions imposed on $G$ by the HKT structure of $\mathscr{M}^*_{\asd}(M^4)$.}
  $G$ of $\lambda$, defined in (\ref{curvlambda}),  decomposes as
\be
G^r= G^r_{\mathfrak{sp}(k-1)}\oplus G^r_{\mathfrak{sp}(1)}~,~~~
\ee
(with $G^0=0$).  This is a consequence of representation theory and in particular the decomposition of $\Lambda^2\bR^{4k}$ in  $Sp(2)$ representations.
The second equation in (\ref{complexlie}) does not depend on $G^r_{\mathfrak{sp}(k-1)}$. This is because the component $G^r_{\mathfrak{sp}(k-1)}$ commutes with the horizontal components $(\h I_r{}^\ma{}_\mb)$ of the complex structures. Next set
\be
( {G}^r_{\mathfrak{sp}(1)})_{\ma\mb}=A^{rs} \h \omega_{s\ma\mb}~,
\ee
for some constants $A^{rs}$. These are constants as a consequence of the Bianchi identity (\ref{bia1}) and the restriction of the holonomy of $\h \nabla$ to lie in $Sp(k)$. Furthermore, the second condition of (\ref{complexlie}) reproduces (\ref{vectorlie2}) provided that $A^{rs}=1/2\delta^{rs}$.  Thus
\be
( {G}^r_{\mathfrak{sp}(1)})_{\ma\mb}=\frac{1}{2} \delta^{rs} \h \omega_{s\ma\mb}~.
\label{sp1comp}
\ee
This specifies the $\mathfrak{sp}(1)$ component of the curvature $G$ in terms of the horizontal components of the Hermitian forms on $\mathscr{M}^*_{\asd}(M^4)$.

It remains to identify the geometry of the base space $\mathscr{B}^{4k-4}$. The base space $\mathscr{B}^{4k-4}$ has metric $g^\perp$, skew-symmetric torsion $H^\perp$ and a ``hyper-complex structure'' associated to the Hermitian forms $\h \omega_r^\perp=1/2\h\omega_{r\ma\mb} \mathbf{e}^\ma\wedge \mathbf{e}^\mb$.  Though the latter do not ``project down'' to tensors on $\mathscr{B}^{4k-4}$ as they are not invariant along the fibres of $\mathscr{M}^*_{\asd}(M^4)$.  Instead, they are sections of a twisted bundle of 2-forms on $\mathscr{B}^{4k-4}$ associated with the 3-dimensional representation of $\mathfrak{sp}(1)$.  Picking a local section of the principal bundle $\mathscr{M}^*_{\asd}(M^4)$ with base space $\mathscr{B}^{4k-4}$ and pulling back the frame connection $\h\nabla$  evaluated on $\h I_r^\perp$, one finds that
\begin{align}
\h\nabla_J \h \omega_{r\ma\mb}&=\partial_J \h \omega_{r\ma\mb}- \mathbf{e}_J^\mc \h \Omega^\perp_\mc{}^\md{}_\ma\, \h \omega_{r\md\mb}+\mathbf{e}_J^\mc \h \Omega^\perp_\mc{}^\md{}_\mb\, \h \omega_{r\md\ma}-\lambda_J^\alpha \h\Omega_\alpha{}^\md{}_\ma\, \h \omega_{r\md\mb}+\lambda_J^\alpha \h\Omega_\alpha{}^\md{}_\mb\, \h \omega_{r\md\ma}
\cr
&=\partial_J \h \omega_{r\ma\mb}- \mathbf{e}_J^\mc \h \Omega^\perp_\mc{}^\md{}_\ma\, \h \omega_{r\md\mb}+\mathbf{e}_J^\mc \h \Omega^\perp_\mc{}^\md{}_\mb\, \h \omega_{r\md\ma}+ \lambda_J^\alpha ( {G}_{\mathfrak{sp}(1)})_\alpha{}^\mc{}_\ma \h I_{r\mc\mb}-\lambda_J^\alpha ( {G}_{\mathfrak{sp}(1)})_\alpha{}^\mc{}_\mb \h I_{r\mc\ma}
\cr
&=\partial_J \h \omega_{r\ma\mb}- \mathbf{e}_J^\mc \h \Omega^\perp_\mc{}^\md{}_\ma\, \h \omega_{r\md\mb}+\mathbf{e}_J^\mc \h \Omega^\perp_\mc{}^\md{}_\mb\, \h \omega_{r\md\ma}-\lambda_J^s \epsilon_{sr}{}^t\, \h\omega_{t\ma\mb}
\cr
&=\h\nabla^\perp_J \h \omega_{r\ma\mb}-\lambda_J^s \epsilon_{sr}{}^t\, \h\omega_{t\ma\mb}=0~,
\label{qktcond}
\end{align}
where $(y^J; J=1, \dots, 4k-4)$ are some coordinates on $\mathscr{B}^{4k-4}$, $\h\nabla^\perp$ is  the connection on $\mathscr{B}^{4k-4}$ with metric $g^\perp$ and skew-symmetric torsion $H^\perp$. To prove the above equation, we have used the relation between some components of the torsion $H$ and the curvature $G$ given in (\ref{curvlambda}), the some relations between components of torsion $H$ and components of the frame connection  in (\ref{omh}),  (\ref{extracond}) and
(\ref{sp1comp}).
Manifolds that satisfy the last line in (\ref{qktcond}) are known as Quaternionic-Kaehler with torsion (QKT) \cite{QKT}.  When the torsion $H^\perp$ vanishes, one recovers the Quaternionic-Kaehler manifolds of Riemannian geometry \cite{salamon}.

To summarise, the geometry of the moduli space of connections  $\mathscr{M}^*_{\asd}(M^4)$, for $M^4$ either $S^3\times S^1$ or $\RP^3\times S^1$,  can be modelled as that of a principal bundle with fibre the HKT group manifold $K$, for $K=SU(2)\times U(1)$ or $S(3)\times SO(2)$, and base space a Quaternionic-Kaehler manifold with torsion $\mathscr{B}^{4k-4}$. In fact up to an identification with a discrete group $\mathscr{M}^*_{\asd}(M^4)=S^1\times \mathscr{P}^*_{\asd}(M^4)$.   The torsion $H^\perp$ of $\mathscr{B}^{4k-4}$ is a $(2,1)\oplus (1,2)$-form with respect to the (twisted almost) hyper-complex structure $(\h I^\perp_r; r=1,2,3)$.  The principal bundle is equipped with a connection $\lambda$, which is a $\mathfrak{sp}(k-1)\oplus sp(1)$ instanton over $\mathscr{B}^{4k-4}$.  The condition (\ref{torsion4}) also implies that the first Pontryagin class of $\mathscr{M}^*_{\asd}(M^4)$ over $\mathscr{B}^{4k-4}$ vanishes.

\subsubsection{An additional QKT structure}

The description of the geometry of $\mathscr{M}^*_{\asd}(M^4)$ in terms of a QKT structure associated to the  $\h\nabla$-HKT structure and explored in the previous section, it can be repeated for the  $\breve \nabla$-HKT structure of $\mathscr{M}^*_{\asd}(M^4)$. The details are very similar and we shall not repeat the analysis.

 Further progress on the geometry will depend on the   understanding of the action of the $\mathfrak{so}(4)\oplus \mathfrak{so}(2)$, generated by both $\h V$ and $\breve V$ vector fields on $\mathscr{M}^*_{\asd}(M^4)$.  As $\mathscr{M}^*_{\asd}(M^4)= S^1\times  {\mathscr {P}}^{4k-1}(M^4)$ and $V_0$ is tangent to $S^1$ subspace, it remains to understand the action of $\mathfrak{so}(4)$ on ${\mathscr {P}}^{4k-1}(M^4)$.  For $M^3=S^3$, unlike the individual action of the two $\mathfrak{su}(2)$ subalgebras of $\mathfrak{so}(4)$ that act freely on ${\mathscr {P}}^{4k-1}(S^3)$, $\mathfrak{so}(4)$ is expected to have fixed points and the principal bundle model for the geometry of ${\mathscr {P}}^{4k-1}(S^3\times S^1)$ cannot be used. A indication for this is that $\mathfrak{so}(4)$ has fixed points acting on $S^3$ in  $M^4=S^3\times S^1$ as $S^3=SO(4)/SO(3)$, i.e. the isotropy group of any point in $S^3$ is $SO(3)$.  A further refinement of the geometry of $\mathscr{M}^*_{\asd}(S^3\times S^1)$ will depend on the nature of the orbits of $\mathfrak{so}(4)$ in $\mathscr{M}^*_{\asd}(S^3\times S^1)$.  In particular, it will depend on whether the two  $\mathfrak{su}(2)$ subalgebras have the same or different orbits in $\mathscr{M}^*_{\asd}(S^3\times S^1)$.

 \subsection{Examples with an HKT and a KT structure on $\mathscr{M}^*_{\asd}$}

At it has already been mentioned   $S^3\times S^1$  is the only non-K\"ahler compact  four-dimensional manifold, up to an identification with a discrete group, that admits a hyper-complex structure.  As a result to find moduli spaces $\mathscr{M}^*_{\asd}(M^4)$ with a HKT structure, one has to focus on manifolds $M^4$ that arise from  identifications of $S^3\times S^1$ with a discrete group.   As HKT and KT structures on $M^4$ that are compatible with a given orientation of $M^4$ induce HKT and KT structures on $\mathscr{M}^*_{\asd}(M^4)$, we shall focus on the description of the {\sl oriented} HKT and KT structures on $M^4$.

To describe the class of examples constructed in \cite{witten}, we use the parameterisation  of $SU(2)$  in terms of the $2\times 2$ complex matrices as
\be
U(a,b)=
\begin{pmatrix}
a& b\\ -\bar b& \bar a
\end{pmatrix}~,~~~\det\,U=a\bar a+b\bar b=1~.
\label{su2m}
\ee
Then a class of examples can be constructed as $M_n^4= S^3/\bZ_n\times S^1$, where the generator $D(e^{\frac{2i\pi}{n}})$ of $\bZ_n$, $n\in \b N$, acts on $S^3$ as
\be
U(a,b)\rightarrow  U(a,b) D(e^{\frac{2i\pi}{n}})~,
\label{dtransf}
\ee
with $D(e^{\frac{2i\pi}{n}})=U(e^{\frac{2i\pi}{n}}, 0)$.  Note that for $n=2$, $S^3/\bZ_2=\RP^3$ that we have already considered.   All the $M_n^4$ manifolds are orientable.   The $\bZ_n$ action is free and leaves the right-invariant 1-forms of $SU(2)$ invariant. For $n>2$, this is also the case for the left-invariant form $L^3$ but $L^1$ and $L^2$ transform as doublets. The transformation (\ref{dtransf}) is an isometry and leaves the torsion $H$  of $S^3\times S^1$ invariant.  It also leaves invariant the oriented right-invariant HKT structure on $S^3\times S^1$ as well as the oriented left-invariant KT structure associated with the Hermitian form $\h\omega_3= V^0\wedge L^3+ L^1\wedge L^2$.  But it ``breaks'' the symmetries generated by the $L_1$ and $L_2$ $\h\nabla$-parallel vector fields.  Thus $ S^3/\bZ_n\times S^1$ admits an oriented right-invariant HKT structure and an oriented left-invariant KT structure. These in turn induce a $\h\nabla$-KT and an $\breve \nabla$-HKT structure on $\mathscr{M}^*_{\asd}(S^3/\bZ_n\times S^1)$.

One can show that, up to an identification with a  discrete group,  $\mathscr{M}^*_{\asd}(S^3/\bZ_n\times S^1)=S^1\times \mathscr{P}$.  Moreover,  the geometry of $\mathscr{M}^*_{\asd}(S^3/\bZ_n\times S^1)$ can be modelled on that of a principal fibration $S^3\times S^1$ fibration with base space $\mathscr{B}$ a QKT manifold. But in this case, the QKT structure on $\mathscr{B}$ is induced by the $\breve {\mathcal{D}}$ connection.  Moreover,
$\mathscr{M}^*_{\asd}(S^3/\bZ_n\times S^1)$ admits an additional symmetry generated by the vector field $\alpha_{L_3}$.

It turns out that the above class   of examples is the most general one  based on circle fibrations over (the lens spaces) $S^3/\bZ_n$.  It is known \cite{thornton} that if one considers the circle fibrations  $M_{n,m}^4= S^3\times_{\bZ_n} S^1$, where the generator of $\bZ_n$ acts on $S^3\times S^1$ as
\be
(U(a,b), z)\rightarrow \big( U(a,b) D(e^{\frac{2i\pi}{n}}), e^{-\frac{2im\pi}{n}} z\big)~,
\label{actionnm}
\ee
 $\bar z z=1$, $z\in S^1$,  and $m\in \bN$ with $0\leq m <n$, $n>0$, there is a $p$ such that $M_{n,m}^4=S^3/\bZ_p\times S^1$ and so these do not constitute new examples. Nevertheless, $M_{2,1}=S^3\times_{\bZ_2} S^1$ requires a mention as it can be identified with the group manifold $U(2)$. Topologically, $M_{2,1}=U(2)=S^3\times S^1$ but, as we have relied on the product group structure $SU(2)\times U(1)$ on $S^3\times S^1$ to construct the HKT structures on $S^3\times S^1$, one may wonder how much of our construction depends on the choice of group action on $S^3\times S^1$. In particular, one can ask the question whether the HKT structures constructed on $S^3\times S^1$ using the group structures $SU(2)\times U(1)$ and $U(2)$ coincide. In appendix A, we prove that they do up to a diffeomorphism.

Finally, one can consider the left-invariant KT-geometry on $S^3\times S^1$ given by the metric and Hermitian form
\be
g=(V^0)^2+(L^3)^2+ \nu^2\big((L^1)^2+(L^2)^2\big)~,~~~\h\omega=V^0\wedge L^3+\nu^2 L^1\wedge L^2~,
\ee
where $\nu^2$ is a positive real number. For $\nu^2=1$, one recovers the bi-HKT structure on $S^3\times S^1$ that we have already described. While if $\nu^2\not=1$, this reduces to a KT structure -- it is easy to see that the chosen metric is not Hermitian with respect to the other two left-invariant complex structures on $S^3\times S^1$. The metric above is that of the sum of the metric on $S^1$ with the metric of a squashed 3-sphere.  The associated complex structure is integrable.  The 3-form torsion $H=L^1\wedge L^2\wedge L^3$ and the Lee form is $\h\theta=\nu^{-2} V^0$.  The group of isometries of the background is $U(2)\times SO(2)$, where $SO(2)$ is the isometry of $S^1$ generated by $V_0$ and $U(2)=SU(2)\times_{\bbZ_2} U(1)$ with $SU(2)$ generated by the left action and $U(1)$ generated by $L_3$. Both vector fields $V_0$ and $L_3$ are holomorphic and $\h\nabla$-covariantly constant. It turns out that the associated vector fields $\alpha_{V_0}$ and $\alpha_{L_3}$ are also holomorphic and $\h{\mathcal{D}}$-covariantly constant
on $\mathscr{M}^*_{\asd}$.  Moreover, $d\alpha_{V_0}=0$ and so up to an action with a discrete group $\mathscr{M}^*_{\asd}=S^1\times \mathscr{P}_{\asd}$, with $\mathscr{P}_{\asd}$ a circle bundle over some base space $\mathscr{B}_\asd$ -- $\mathscr{M}^*_{\asd}$ is a holomorphic $T^2$ fibration over $\mathscr{B}_\asd$
and so $\mathscr{B}_\asd$ admits a KT structure. A similar construction can be made by replacing $S^3$ with $\RP^3$.



\section{Applications and concluding remarks}

\subsection{Geometries with skew-symmetric torsion and moduli spaces}

One of the remarkable results that have arisen in the investigation of the geometry of moduli spaces of Hermitian-Einstein connections   is that the torsion $\mathcal{H}$ of the KT structure on $\mathscr{M}^*_{\HE}(M^{2n})$  is a closed 3-form.  This is regardless on whether the KT structure on $M^{2n}$ is strong or not. The explanation for this lies in the dependence of $\mathcal{H}$ on the torsion $H$ of the underlying manifold  $M^{2n}$. As $H$ is a $(2,1)\oplus (1,2)$-form, it lies in the $(n^3-n^2)$-dimensional representation of $U(n)$ over the reals. This decomposes into a $(n^3-n^2-2n)$-dimensional irreducible representation and a $2n$-dimensional irreducible representation. The latter is determined by the Lee form $\theta$ of the Hermitian geometry of $M^{2n}$. It is straightforward to observe from  (\ref{ttorsion}) that the only dependence of $\mathcal{H}$ on the torsion $H$ of $M^{2n}$ is via the $2n$-dimensional representation, i.e. it only depends on the Lee form $\theta$.  As a result, the closure of $\mathcal{H}$ imposes a rather weak condition on the geometry of $M^{2n}$, the co-closure of $\theta$ (Gauduchon condition), which is always satisfied after an appropriate conformal rescaling of the Hermitian metric of $M^{2n}$.

Every Hermitian manifold admits a KT structure.   Thus there are many examples of KT manifolds. However, it is less the case for manifolds, $M^{2n}$,  with a strong KT for $n>2$.  The strong condition imposes an additional restriction on the Hermitian structure of $M^{2n}$. It is known that not all Hermitian manifolds admit a strong KT structure \cite{verbitsky2}.  Nevertheless,  many examples can be constructed with standard complex geometry techniques. The $n=2$ case is special as every compact 4-dimensional Hermitian manifold admits a strong KT structure.  This is  a consequence of the Gauduchon theorem. Explicit examples of manifolds that admit a strong KT structure include group manifolds with a bi-invariant metric and torsion \cite{Spindel, OP}. The latter is constructed from the structure constants of the group manifold.  This construction does not necessarily extend to homogeneous spaces, which can admit a KT structure, but the torsion is not necessarily closed. Many  strong KT manifolds are known as well as methods of constructing them.  Such constructions include  nilmanifold, fibrations, blow ups along holomorphic submanifolds  and many other techniques,  see \cite{fino1, ugarte, grantcharov2, fino2, fino3, swann}.

The strong bi-KT structure is even more restrictive than strong KT. It has been extensively investigated in connection with generalised K\"ahler geometry. Examples of bi-KT manifolds include group manifolds. In fact, every strong KT group manifold with a bi-invariant metric and torsion of \cite{Spindel, OP} admits a strong bi-KT structure\footnote{For such group manifold $G$ with a bi-KT (or bi-HKT) structure, one can always consider $G/D$, with $D$ an appropriate discrete subgroup, and reduce the bi-KT  (bi-HKT) structure on $G$  to a KT (KT and HKT or HKT) one on $G/D$.}. One of the KT structures is invariant under the left action of the group and the other is invariant under the right action of the group.
Far more progress has been made in  four-dimensional generalised K\"ahler manifolds, where   there is a classification \cite{grantcharov, hitchin, apostolov}, for both commuting and non-commuting complex structures.  However to our knowledge, a topological characterisation of manifolds with a generalised K\"ahler structure  is still an open problem.

Similarly, manifolds with a strong HKT structure\footnote{Not all hyper-complex manifolds admit an HKT structure \cite{fino4, dotti}.} include the group manifold examples of \cite{Spindel, OP}. Compact examples can be constructed using fibrations, like those of \cite{Verbitsky, gp1}, but the torsion may not be closed $dH\not=0$.  There are also nilmanifold examples, see e.g. \cite{dottifino, barberisfino}.  The same applies for bi-HKT structures.  In four dimensions, the only known examples of a compact manifold with a bi-HKT structure are $S^3\times S^1$ and $\RP^3\times S^1$.

In view of these results, it is remarkable that the moduli spaces for Hermitian-Einstein connections, $\mathscr{M}^*_{\HE}(M^{2n})$, provide another source of examples of strong KT, bi-KT and HKT manifolds.  Though of course, the description of their geometry and other properties is not explicit.
The case of pursuing  examples of strong HKT manifolds using the instanton moduli space $\mathscr{M}^*_{\asd}(S^3\times S^1)$ has been made in \cite{witten}, where their singularities and the difficulties of resolving them are also explored.    As we have seen in section \ref{sec:mqkt}, the geometry of $\mathscr{M}^*_{\asd}(S^3\times S^1)$ can be modelled after that of a $S^3\times S^1$ principal fibration over a QKT manifold.  As a result, the moduli spaces $\mathscr{M}^*_{\asd}(S^3\times S^1)$ will produce a significant number of QKT manifolds as well. Though such QKT manifolds are not expected to satisfy the strong condition. However, it may be possible to use the underlying QKT geometry to describe $\mathscr{M}^*_{\asd}(S^3\times S^1)$, especially those of low dimension.

\subsection{Sigma models on moduli spaces and their symmetries}

There is an extensive literature, see for example \cite{Curtrightfreedman, LAGDF, Curtrightzachos, Howesierra, JGCHMR, Hullwitten, Buscher, Braaten, Hullreview, HoweGP1}, on the geometry of the target spaces of 2-dimensional sigma models with $(p,q)$  worldsheet supersymmetry following the early work of \cite{Zumino2} on four-dimensional supersymmetric sigma models.  There is also the recent concise review \cite{pw}, where the relationship between worldsheet supersymmetry in sigma models and the geometry of their target spaces is explained.  So we shall use this relationship without providing further explanation.

The (1,1) superfields $\phi$ are maps from the $(1,1)$-worldsheet superspace $\Xi^{1,1}$ with Grassmannian even (odd) coordinates $(u,v)$ ($(\vartheta^+, \vartheta^-)$) into the sigma model target space,   which here is taken to be either $\mathscr{M}^*_{\HE}(M^{2n})$ or $\mathscr{M}^*_{\asd}(M^{4})$.  The action of $(1,1)$-supersymmetric sigma models written in terms of $(1,1)$ superfields $\phi$ is
\be
I= \int du dv d^2\vartheta   (g+b)_{ij} D_+\phi^i D_-\phi^j~,
\label{sigmaact}
\ee
where $D_+$ and $D_-$ are superspace derivatives with $D^2_+=i \partial_u$ and $D^2_-=i \partial_v$ and $H=db$.

 This action is invariant under the worldsheet translation and $(1,1)$-supersymmetry transformations
\be
\delta_{\h T}\phi^i= D_+\h \epsilon D_+\phi^i+2 i \h \epsilon \partial_u \phi^i~,~~~\delta_{\breve T}\phi^i= D_-\breve \epsilon D_-\phi^i+2 i \breve \epsilon \partial_v \phi^i~,
\label{susytrans}
\ee
where $\h \epsilon= \h \epsilon(u, \vartheta^+) $ and $\breve \epsilon=\breve \epsilon(v, \vartheta^-)$ are the infinitesimal parameters.

In what follows for $(p,q)$-supersymmetric sigma models, we denote the  $q$-extended and the $p$-extended supersymmetry transformations with the complex structure associated with them as either
\be
\delta_{\h {\mathcal{I}}} \phi^i=\h \eta_- \h {\mathcal{I}}^i{}_j D_+\phi^j~,~~~\mathrm {or}~~~\delta_{\breve {\mathcal{I}}} \phi^i=\breve \eta_+ \breve {\mathcal{I}}^i{}_j D_-\phi^j~,
\ee
respectively, where $\h \eta_-=\h \eta_-(u, \vartheta^+)$ and $\breve \eta_+=\eta_+(v, \vartheta^-)$ are the infinitesimal parameters. Similarly, we denote the symmetries of the action generated by the Killing vector fields $ V$  that also leave $H$ invariant  with
\be
\delta_{ \h V} \phi^i=\h\zeta\, \h V^i~,~~~\delta_{ \breve V} \phi^i=\breve\zeta\, \breve V^i~,
\ee
where $\h \zeta=\h \zeta(u, \vartheta^+)$ and $\breve\zeta=\breve \zeta(v, \vartheta^-)$  are the infinitesimal parameters. Notice that the parameters of the transformations are allowed to depend appropriately as indicated on the Grassmannian even and odd coordinates of $\Xi^{1,1}$.  This is because the action (\ref{sigmaact}) is classically invariant under superconformal transformations. However, as the superconformal symmetry is not expected to persist in the quantum theory for all such theories, we shall consider  the transformations for which their parameter depends only on the Grassmannian odd coordinates $\vartheta^\pm$ as it has been indicated, i.e. $\h\epsilon=\h\epsilon(\vartheta^+)$, $\h\eta=\h\eta(\vartheta^+)$ and $\h\zeta=\h\zeta(\vartheta^+)$ and similarly for $\breve\epsilon$, $\breve \eta$ and $\breve\zeta$.  This subset of symmetries is expected to be symmetries of the quantum theory as well. We shall  comment later on the sigma models   that are  superconformal.

As all $\mathscr{M}^*_{\HE}(M^{2n})$ admit a strong KT structure, they can be probed by $(p,2)$-supersymmetric 2-dimensional (string) sigma models with either $p=0$ or $p=1$. Here, we shall focus on sigma models with $p,q\not=0$ -- the investigation of the properties of $(0,q)$-supersymmetric sigma models with target space either $\mathscr{M}^*_{\HE}(M^{2n})$ or $\mathscr{M}^*_{\asd}(M^{4})$ is similar to that of $(1,q)$-supersymmetric sigma models. The extended supersymmetry of $(1,2)$-supersymmeric sigma model with target space the KT manifold $\mathscr{M}^*_{\HE}(M^{2n})$ is generated by the complex structure $\h {\mathcal{I}}$. Moreover, for every  holomorphic and $\h\nabla$-covariantly constant vector fields $\h X$ on  $M^{2n}$, there will be a corresponding holomorphic Killing vector field $ \alpha_{\h X}$ on $\mathscr{M}^*_{\HE}(M^{2n})$ that will generate a symmetry for the sigma model action. The algebra of these variations is schematically
\begin{align}
&[\delta_{\h T}, \delta_{\h {\mathcal{I}}}]= \delta_{\h {\mathcal{I}}}~,~~~[\delta_{ \h T}, \delta_{ \alpha_{\h X}}]=\delta_{ \alpha_{\h X}}~,~~~[\delta_{\h {\mathcal{I}}}, \delta_{\h {\mathcal{I}}}]=\delta_{\h T}+ \delta_{\h {\mathcal{I}}}~,~~~[\delta_{\h {\mathcal{I}}}, \delta_{ \alpha_{\h X}}]=\delta_{\alpha_{\h Y}}~,~~~
\cr
&[\delta_{ \alpha_{\h X}}, \delta_{ \alpha_{\h X'}}]=-\delta_{\alpha_{[\h X, \h X']}}~,
\label{ktmalg}
\end{align}
where $\delta_{\h T}$ is given in (\ref{susytrans}), $\h Y=-\h I\h X$ and we have used (\ref{commutxy}) for the last commutator.  Moreover, if $\alpha_{\h X}$ and $\alpha_{\h X'}$
are $\h{\mathcal{D}}$-covariantly constant, then $[\delta_{ \alpha_{\h X}}, \delta_{ \alpha_{\h X'}}]=0$.

Next, let us turn to investigate the symmetries of $(2,2)$-supersymmetric sigma models with target spaces bi-KT instanton moduli spaces $\mathscr{M}^*_{\asd}(M^{4})$.  As the complex structures $\h{\mathcal{I}}$ and  $\breve{\mathcal{I}}$ do not commute, the closure of the supersymmetry algebra is on-shell \cite{JGCHMR}, i.e. it closes up to terms containing the field equations of the sigma model action (\ref{sigmaact}).  Also, the action will be invariant under all vector fields $\alpha_{\h X}$ ($\alpha_{\breve X}$) on $\mathscr{M}^*_{\asd}(M^{4})$ generated by the  $\h\nabla$-($\breve \nabla$-)-covariantly constant and $\h I$- ($\breve I$-) holomorphic vector fields $\h X$ ($\breve X$)  of $M^4$.  The algebra of  $\delta_{\h {\mathcal{I}}}$ and $\delta_{ \alpha_{\h X}}$ variations is as in (\ref{ktmalg}).  The algebra of  $\delta_{\breve {\mathcal{I}}}$ and $\delta_{ \alpha_{\breve X}}$ variations are given as in (\ref{ktmalg}) with  $\h {\mathcal{I}}$, $\alpha_{\h X}$ and $\h T$ replaced with $\breve {\mathcal{I}}$, $\alpha_{\breve X}$ and $\breve T$, respectively.  Again if $\alpha_{\breve X}$ and $\alpha_{\breve X'}$ are $\breve{\mathcal{D}}$-covariantly constant, then $[\alpha_{\breve X},\alpha_{\breve X'}]=0$.

It remains to determine how the symmetries of the sigma model action generated by the $\alpha_{\breve X}$ vector fields act on the $(1,2)$ supersymmetry transformations of the fields. The commutator of such transformations, in particular that of symmetries generated by $\alpha_{\breve X}$ with that of the extended
supersymmetry transformation generated by $\h{\mathcal{I}}$, is determined by (\ref{holx2}). The result will depend on the properties of the underlying manifold $M^4$ and in particular the Lie derivative, $\mathcal{L}_{\breve X} \h I$.   Similarly, the commutator of symmetries generated by $\alpha_{\h X}$ with the supersymmetry transformation generated by $\breve{\mathcal{I}}$ can be expressed in terms of $\mathcal{L}_{\h X} \breve I$. These results can be summarised schematically using the variations on the sigma model fields $\phi$ as
\begin{align}
&[\delta_{\h T}, \delta_{\h {\mathcal{I}}}]= \delta_{\h {\mathcal{I}}}~,~~~[\delta_{ \h T}, \delta_{ \alpha_{\h X}}]=\delta_{ \alpha_{\h X}}~,~~~[\delta_{\h {\mathcal{I}}}, \delta_{\h {\mathcal{I}}}]=\delta_{\h T}+ \delta_{\h {\mathcal{I}}}~,~~~[\delta_{\h {\mathcal{I}}}, \delta_{ \alpha_{\h X}}]=\delta_{\alpha_{\h Y}}~,~~~
\cr
&[\delta_{ \alpha_{\h X}}, \delta_{ \alpha_{\h X'}}]=-\delta_{\alpha_{[\h X, \h X']}}~,
\cr
&[\delta_{\breve T}, \delta_{\breve {\mathcal{I}}}]= \delta_{\breve {\mathcal{I}}}~,~~~[\delta_{ \breve T}, \delta_{ \alpha_{\breve X}}]=\delta_{ \alpha_{\breve X}}~,~~~[\delta_{\breve {\mathcal{I}}}, \delta_{\breve {\mathcal{I}}}]=\delta_{\breve T}+ \delta_{\breve {\mathcal{I}}}~,~~~[\delta_{\breve {\mathcal{I}}}, \delta_{ \alpha_{\breve X}}]=\delta_{\alpha_{\breve Y}}~,~~~
\cr
&[\delta_{ \alpha_{\breve X}}, \delta_{ \alpha_{\breve X'}}]=-\delta_{\alpha_{[\breve X, \breve X']}}~,
\cr
&
[\delta_{\h {\mathcal{I}}}, \delta_{\breve {\mathcal{I}}}]=0~~\mathrm{(on~shell)}~,~~~[\delta_{\h {\mathcal{I}}}, \delta_{ \alpha_{\breve X}}]=\delta_{\mathcal{L}_{\alpha_{\breve X}} \h{\mathcal {I}}}~,~~~[\delta_{\breve {\mathcal{I}}}, \delta_{ \alpha_{\h X}}]=\delta_{\mathcal{L}_{\alpha_{\h X}} \breve{\mathcal {I}}}~,~~~
\cr
&[\delta_{ \alpha_{\h X}}, \delta_{ \alpha_{\breve X'}}]=-\delta_{\alpha_{[\h X, \breve X']}}~,
\end{align}
where we have used (\ref{holx2}) and (\ref{commutxy3}) for the last commutator and $\delta_{\breve T}$ denotes the variation on the sigma model fields induced by the right-handed worldsheet translations. If the Lie derivatives $\mathcal{L}_{\breve X} \h I$ and $\mathcal{L}_{\h X} \breve I$ vanish, the commutators will vanish as well.

 As $\mathscr{M}^*_{\asd}(S^3\times S^1)$ is a bi-HKT manifold, sigma models with target space $\mathscr{M}^*_{\asd}(S^3\times S^1)$ can admit $(4,4)$ worldsheet supersymmetry. From the results in section \ref{sec:gms3s1}, it is straightforward to establish  the commutator algebra of variations on the sigma model fields generated by the complex structures on $\mathscr{M}^*_{\asd}(S^3\times S^1)$,  and the $\h{\mathcal{D}}$- and $\breve{\mathcal{D}}$-covariantly constant vector fields.  This can be written schematically as
\begin{align}
&[\delta_{\h T}, \delta_{\h {\mathcal{I}}_r}]= \delta_{\h {\mathcal{I}}_r}~,~~~[\delta_{ \h T}, \delta_{ \alpha_{V_0}}]=\delta_{ \alpha_{V_0}}~,~~~[\delta_{ \h T}, \delta_{ \alpha_{L_r}}]=\delta_{ \alpha_{L_r}}~,~~~[\delta_{\h {\mathcal{I}}_r}, \delta_{\h {\mathcal{I}}_s}]=\delta_{rs} \delta_{\h T}+\epsilon_{rs}{}^t \delta_{\h {\mathcal{I}}_t}~,~~~
\cr
&
[\delta_{\h {\mathcal{I}}_r}, \delta_{ \alpha_{ L_s}}]=\epsilon_{rs}{}^t \delta_{\h {\mathcal{I}}_t}+\delta_{rs} \delta_{ \alpha_{ V_0}}+\epsilon_{rs}{}^t \delta_{ \alpha_{ L_t}} ~,~~~[\delta_{\h {\mathcal{I}}_r}, \delta_{ \alpha_{ V_0}}]=\delta_{ \alpha_{ L_r}}~,~~~[\delta_{ \alpha_{ L_r}}, \delta_{ \alpha_{ L_s}}]=\epsilon_{rs}{}^t \delta_{ \alpha_{ L_t}} ~,
\cr
&[\delta_{\breve T}, \delta_{\breve {\mathcal{I}}_r}]= \delta_{\breve {\mathcal{I}}_r}~,~~~[\delta_{ \breve T}, \delta_{ \alpha_{V_0}}]=\delta_{ \alpha_{V_0}}~,~~~[\delta_{ \breve T}, \delta_{ \alpha_{R_r}}]=\delta_{ \alpha_{R_r}}~,~~~[\delta_{\breve {\mathcal{I}}_r}, \delta_{\breve {\mathcal{I}}_s}]=\delta_{rs} \delta_{\breve T}+\epsilon_{rs}{}^t \delta_{\breve {\mathcal{I}}_t}~,~~~
\cr
&
[\delta_{\breve {\mathcal{I}}_r}, \delta_{ \alpha_{ R_s}}]=\epsilon_{rs}{}^t \delta_{\breve {\mathcal{I}}_t}+\delta_{rs} \delta_{ \alpha_{ V_0}}+\epsilon_{rs}{}^t \delta_{ \alpha_{ R_t}} ~,~~~[\delta_{\breve {\mathcal{I}}_r}, \delta_{ \alpha_{ V_0}}]=\delta_{ \alpha_{ R_r}}~,~~~
[\delta_{ \alpha_{R_r}}, \delta_{ \alpha_{R_s}}]=-\epsilon_{rs}{}^t \delta_{ \alpha_{R_t}}~,
\cr
&
[\delta_{\h {\mathcal{I}}_r}, \delta_{\breve {\mathcal{I}}_s}]=0~~\mathrm{(on~shell)}~,~~~[\delta_{\h {\mathcal{I}}_r}, \delta_{ \alpha_{ R_s}}]=0~,~~~[\delta_{\breve {\mathcal{I}}_r}, \delta_{ \alpha_{L_s}}]=0~,~~~
\cr
&[\delta_{ \alpha_{L_r}}, \delta_{ \alpha_{R_s}}]=0~,~~~[\delta_{\h {\mathcal{I}}_r}, \delta_{ \alpha_{V_0}}]=[\delta_{\breve {\mathcal{I}}_r}, \delta_{ \alpha_{V_0}}]=[ \delta_{ \alpha_{ L_r}}, \delta_{ \alpha_{V_0}}]=[ \delta_{ \alpha_{R_r}}, \delta_{ \alpha_{V_0}}]=0~.
\label{largesuper}
\end{align}
It is clear that the vector fields $\alpha_{L_r}$ and $\alpha_{R_s}$ on the moduli space $\mathscr{M}^*_{\asd}(S^3\times S^1)$ generate the algebra
 $\mathfrak{so}(3)\oplus\mathfrak{so}(3)$  of transformations that leave the sigma model action invariant. However, these transformations   act non-trivially, with rotations,  on the worldsheet supersymmetries of the sigma model.  A similar algebra of symmetries can be constructed for the sigma model with target space the bi-HKT manifold $\mathscr{M}^*_{\asd}(\RP^3\times S^1)$.

These results can be easily adapted to sigma models with target space $\mathscr{M}^*_{\asd}(S^3/\bZ_n\times S^1)$, $n>2$.  Such sigma models can admit $(2,4)$ worldsheet supersymmetry generated by the complex structures $(\h{\mathcal I}_r, \breve {\mathcal I}_3)$.  They are also invariant under the action of the vector fields $\alpha_{L_r}$ and $\alpha_{R_3}$. The algebra of these symmetries can be easily derived from (\ref{largesuper}) by appropriately restricting the latter to the generators of symmetries of $\mathscr{M}^*_{\asd}(S^3/\bZ_n\times S^1)$.

\subsection{Conformal invariance and moduli spaces}

All sigma models with target space the moduli spaces $\mathscr{M}^*_{\HE}(M^{2n})$  are invariant, at the classical level,  under worldsheet $(p,q)$-superconformal transformations.  However, this is not expected to persist at the quantum level for such generic\footnote{There is the possibility that one such moduli space $\mathscr{M}^*_{\HE}(M^{2n})$ has some additional structure, which is special to the choice of $M^{2n}$ and of the vector bundle over it,  that makes the corresponding sigma model superconformal.} sigma models unless the theory is either $(4,q)$ or $(p, 4)$ supersymmetric.  With this amount of supersymmetry, there are arguments, including superspace ones,  that make the sigma models power-counting ultraviolet finite.  This was first established for $(4,4)$-supersymmetric sigma models in  \cite{LAGDF, Hull4, Galperin} and then for $(p,4)$-supersymmetric sigma models in \cite{HoweGP1, howeGP2, Sokatchev, Becchi}. However, although this is sufficient to prove that the sigma models are scale invariant, it is not enough to prove that they are conformally invariant. For an in depth  discussion of scale vs conformal invariance in the context of sigma models see \cite{townsend}.  As it has been stressed in \cite{pw, witten},   conformal invariance requires some additional global information about the theory. In the original work of Polchinski on the relation between scale and conformal invariance  \cite{Polchinski}, this is the assumption that the theory must have a discrete spectrum of operator dimensions. In the examination of this problem for sigma models in \cite{pw},  using the Perelman's functional \cite{Perelman}, this assumption is the compactness, or at least the geodesic completeness, of the sigma model target space.  In  \cite{witten} three separate arguments have been stated in support of the superconformal invariance of  $(4,4)$-supersymmetric sigma models with target space the moduli spaces $\mathscr{M}^*_{\asd}(S^3\times S^1)$.

Next let us turn to examine the algebra of symmetries of the $(4,4)$-supersymmetric sigma model with target space $\mathscr{M}^*_{\asd}(S^3\times S^1)$. Such a model is expected to exhibit two copies of the large $N=4$ superconformal algebra of \cite{sevrin}.  One copy is associated with the ``left" $(4,0)$ worldsheet supersymmetries  and the other with the ``right"  $(0,4)$ worldsheet supersymmetries of the theory.  Let us focus on the ``right" large $N=4$ superconformal algebra and work at the classical level. The conserved currents of the theory written in terms of $(1,1)$ superfields $\phi$ are
\begin{align}
&\h T=g_{ij} D_+\phi^i \h\nabla_+D_+\phi^j-\frac{1}{3} H_{ijk} D_+\phi^i D_+\phi^j D_+\phi^k~,~~~\h J_r =\h{\omega}_{rij} D_+\phi^i D_+\phi^j~,~~~
\cr
&
\h W^0=\lambda^0_i D_+\phi^i~,~~~\h W^r=\lambda^r_iD_+\phi^i~,
\end{align}
where we have used  (\ref{hktcorr}) to translate the notation used for the HKT structure on $\mathscr{M}^*_{\asd}(S^3\times S^1)$ to that of ``ordinary" HKT manifolds.
Written in component fields, each of these currents decomposes into two conserved currents. $\h T$ contains the spin 2 energy-momentum tensor as well as the spin $3/2$ $(0,1)$ supersymmetry  current of the theory. One of the components of $\h J_r$, for $r=1,2,3$  are the spin $3/2$ extended $(0,q)$ supersymmetry currents, $q=2,3,4$ and the other component current generates the spin 1 R-type of symmetry transformations with Lie algebra $\mathfrak{so}(3)$.  Finally, the currents $\h W^0$ and $\h W^r$ have a spin 1/2 component and a spin 1 component currents.  The latter generate the $\mathfrak{so}(3)\oplus \mathfrak{u}(1)$ algebra of isometries. The $\mathfrak{so}(4)$ Kac-Moody symmetry of the $N=4$ large superconformal algebra is generated by the above R-type of symmetry associated with the spin 1 component of $\h J_r$ and the $\mathfrak{so}(3)$ transformations of the Killing vector fields associated with the spin 1 component of $\h W^r$. This current content of the $(4,4)$-supersymmetric sigma model with target space $\mathscr{M}^*_{\asd}(S^3\times S^1)$  is reminiscent of that of the realisation of the large $N=4$ superconformal algebra in terms of the currents of a $SU(2)$
WZW model in \cite{sevrin}. This is in agreement with the results of \cite{witten} that the symmetry of $(4,4)$-supersymmetric sigma model with target space $\mathscr{M}^*_{\asd}(S^3\times S^1)$ is two copies of the large $N=4$ superconformal algebra.

\vskip1cm
 \noindent {\it {Acknowledgements}}

 I thank Edward Witten for correspondence and for rekindling my interest in the geometry of moduli spaces.

\newpage

\setcounter{section}{0}

\appendix{Group and HKT structures  on $S^3\times S^1$ } \label{ap-a}


We shall prove that the HKT structures on $S^3\times S^1$ are independent from the  group structure used to construct them. As a group manifold $S^3\times S^1$ can either be identified with  $SU(2)\times U(1)$ or with $U(2)$. For the latter notice  that $U(2)$ is topologically $S^3\times S^1$. Indeed,
\be
SU(2)\rightarrow U(2)\xrightarrow{\det} U(1)~,
\label{seq}
\ee
and so $U(2)$ is an $SU(2)$ principal fibration over $S^1$.  However,  all $SU(2)$ fibrations over $S^1$ are topologically trivial, which proves the assertion.  Moreover, $U(2)$ is the semi-direct product of
$SU(2)$ and $U(1)$, $U(2)=U(1)\ltimes SU(2)$, with group multiplication given by
\be
(z_1, U_1) (z_2, U_2)=\big(z_1 z_2, U_1 \zeta(z_1) U_2 \zeta^{-1}(z_1) \big)~,
\ee
where $\zeta: U(1)\rightarrow U(2)$ with
\be
\zeta(z)=\begin{pmatrix}
z& 0\\0 &  1
\end{pmatrix}~,
\ee
is a group homomorphism  and  an inverse of the map $\det$ in (\ref{seq}), i.e. it is splitting the sequence (\ref{seq}), $z_1, z_2\in U(1)$ and $U_1, U_2\in SU(2)$ as in (\ref{su2m}).

The inverse of the element $(z, U(a,b))$ of $U(2)$ is $(\bar z, \zeta(\bar z) U(\bar a, - b) \zeta(\bar z)^{-1})$.  The left-invariant frame $(\tilde L^0, \tilde L^r)$ on $U(2)$ can now be computed using
\be
(\bar z, \zeta(\bar z) U(\bar a, - b) \zeta(\bar z)^{-1}) (d z, d U)=\bar z dz+ \zeta(\bar z)  U(\bar a, - b) d U(a,b) \zeta(\bar z)^{-1}=i\tilde L^0+ \frac{i}{2}\sigma_r \tilde L^r~,
\ee
where $U(\bar a, - b) d U(a,b)$ gives the left-invariant 1-forms of $SU(2)$ and $\sigma^r$ are the Pauli matrices. A direct computation reveals that
\begin{align}
\tilde L^0=d\tau~,~~~\tilde L^3= L^3~,~~~\tilde L^2=\cos\tau L^2+\sin\tau L^1~,~~~\tilde L^1=\cos\tau L^1-\sin\tau L^2~,
\end{align}
where we have set $z=e^{i\tau}$ and $(L^1, L^2, L^3)$ is the left-invariant frame on $SU(2)$. Because of the $\tau$ dependence of the $\tilde L^1$ and  $\tilde L^2$, this frame does not obey the standard exterior relations expected from the $\mathfrak{su}(2)\oplus \mathfrak{u}(1)$ Lie algebra of $U(2)$ but this can be easily corrected by redefining $\tilde L^3$ as
\be
\tilde L^3\rightarrow \tilde L^3+\tilde L^0~.
\ee
In terms of the new $\tilde L^3$, one has that
\be
d \tilde L^0=0~,~~~d \tilde L^3=\tilde L^1\wedge \tilde L^3~,~~~d\tilde L^2=-\tilde L^1\wedge \tilde L^3~,~~~d\tilde L^1=\tilde L^2\wedge \tilde L^3~,~~
\ee
i.e. these are the same equations as those derived using the product group structure $SU(2)\times U(1)$ on $S^3\times S^1$, but of course, unlike the product case,  this left-invariant frame depends non-trivially on the coordinate $\tau$ in addition to those of $SU(2)$.  One can proceed to construct the metric $\tilde g$, torsion $\tilde H$ and Hermitian forms $\h{\tilde\omega}_r$  as in the product group structure case but now using the $(\tilde L^0, \tilde L^r)$ frame instead of $(L^0, L^r)$ of the product group structure. Clearly, all these tensors will depend on $\tau$ and so it remains to see whether such dependence can be eliminated with a diffeomorphism.  For this write $a=x e^{i\varphi}$ and $b=y e^{i\psi}$, $0\leq\varphi, \psi<2\pi$, $x,y>0$ with $x^2+y^2=1$, and after a short computation, one finds that
\begin{align}
&2\tilde L^3=d\tau+d\varphi-d\psi+(x^2-y^2) (d\varphi+d\psi)~,
\cr
&2\tilde L^1=\sin(\psi-\varphi-\tau) (xdy-ydx)+\cos(\psi-\varphi-\tau) xy (d\psi+d\varphi)
\cr
&
2\tilde L^3=\cos (\psi-\varphi-\tau) (xdy-y dx)-\sin(\psi-\varphi-\tau) xy (d\psi+d\varphi)~.
\label{u2frame}
\end{align}
The left-invariant frame $(L^1, L^2, L^3)$ of $SU(2)$ can be recovered from the above expression by setting $\tau=0$. (The left-invariant frame of $\RP^3$ is as that above but with $0\leq\varphi, \psi<\pi$.)  Performing the coordinate transformation
\bea
\psi'-\varphi'=\psi-\varphi-\tau~,~~~\psi'+\varphi'=\psi+\varphi~,~~~\tau'=\tau~,
\label{thetransf}
\eea
the left-invariant frame (\ref{u2frame}) we have computed using the $U(2)$ group structure on $S^3\times S^1$ coincides with that we have computed using the produce $SU(2)\times U(1)$ structure on $S^3\times S^1$.  As a result, the metric $g$, torsion $H$ and Hermitian forms $\h\omega_r$ of the left-invariant  HKT structure on $S^3\times S^1$ computed using the two group structures coincide up to a diffeomorphism and so they should be considered as identical. This is expected to be the case for each individual HKT structure on $S^3\times S^1$ constructed using the two different group actions.

It remains to investigate whether this is also the case for the oriented bi-HKT structure on $S^3\times S^1$. For this we have to compute the right-invariant frame
$(\tilde R^0, \tilde R^r)$. For this observe that
\begin{align}
&d \big(z, U(a,b)\big) \big(\bar z, \zeta(\bar z) U^{-1}(a,b) \zeta^{-1}(\bar z)\big)=-\big(z, U(a,b)\big) d\big(\bar z, \zeta(\bar z) U(\bar a,-b) \zeta^{-1}(\bar z)\big)
\cr
&=-\big(z, U^{-1}(\bar a,-b)\big) d\big(\bar z, U(\bar a,-\bar z b) \big)=-\big(z,\zeta(z) U^{-1}(\bar a,-\bar z b)\zeta^{-1}(z)\big) d\big(\bar z, U(\bar a,-\bar z b)\big )~.
\end{align}
Therefore, the right-invariant frame can be computed from the left-invariant frame upon replacing in the latter
\be
\tau\rightarrow -\tau~,~~~\varphi\rightarrow -\varphi~,~~~\psi\rightarrow \psi-\tau+\pi
\ee
and adding an overall sign. This implies that
\begin{align}
&-2\tilde R^3=-d\varphi-d\psi+(x^2-y^2) (d\varphi+d\psi)~,
\cr
&-2\tilde R^1=\sin(\psi+\varphi+\pi) (xdy-ydx)+\cos(\psi+\varphi+\pi) xy (d\psi-d\varphi-d\tau)~,
\cr
&
-2\tilde R^3=\cos (\psi+\varphi+\pi) (xdy-y dx)-\sin(\psi+\varphi+\pi) xy (d\psi-d\varphi-d\tau)~.
\end{align}
Again, the transformation (\ref{thetransf}) eliminates all the $\tau$ dependence from the right-invariant frame.  Thus the oriented bi-HKT invariant structure on $S^3\times S^1$ is independent from the group structure on $S^3\times S^1$ used to construct it.

\bibliographystyle{unsrt}

\end{document}